\documentclass[twocolumn]{autart}    % Enable this line and disable the
\usepackage[comma,numbers,square,sort&compress]{natbib}

\usepackage{amsmath}
\usepackage{amssymb}
\usepackage{float}
\usepackage{graphicx}
\usepackage{subfigure}
\usepackage{color}
\usepackage{verbatim}
\usepackage{tikz,times}
\usepackage{enumitem}

\usepackage{savesym}
\savesymbol{AND}
\savesymbol{OR}
\savesymbol{NOT}
\savesymbol{TO}
\savesymbol{COMMENT}
\savesymbol{BODY}
\savesymbol{IF}
\savesymbol{ELSE}
\savesymbol{ELSIF}
\savesymbol{FOR}
\savesymbol{WHILE}

\usepackage{algorithm}
\usepackage{algorithmic}

\DeclareMathOperator{\diag}{diag}

\newcommand{\norm}[1]{\left\lVert#1\right\rVert}

\newtheorem{theorem}{Theorem}
\newtheorem{proposition}{\textbf{Proposition}}
\newtheorem{lemma}{\textbf{Lemma}}
\newtheorem{corollary}{\textbf{Corollary}}

\theoremstyle{definition}

\newtheorem{assumption}{\textbf{Assumption}}

\theoremstyle{remark}
\newtheorem{remark}{Remark}

\begin{document}

\begin{frontmatter}

\title{Distributed control under compromised measurements: \\Resilient estimation, attack detection,
	and 
	vehicle platooning
%	\thanksref{The work is supported by the Knut \& Alice Wallenberg Foundation, and   Swedish Research Council.}
	} % Title, preferably not more

\thanks[footnoteinfo]{This paper was not presented at any IFAC
meeting. \\Corresponding author: Xingkang He}

\author[kth]{Xingkang~He}\ead{xingkang@kth.se},    % Add the
\author[Ehsan]{Ehsan Hashemi}\ead{ehashemi@uwaterloo.ca},               % e-mail address
\author[kth]{Karl H. Johansson}\ead{kallej@kth.se}  % (ead) as shown

\address[kth]{Division of Decision and Control Systems, School of Electrical Engineering and Computer Science. \\KTH Royal Institute of Technology, Sweden }  % Please supply
\address[Ehsan]{Department of Mechanical and Mechatronics Engineering, University of Waterloo, Waterloo, ON, Canada}             % full addresses
%\address[Baiae]{The White House, Baiae}        % here.

                             % keyword list or with the
                                          % help of the Automatica
                                          % keyword wizard

\begin{abstract}                          % Abstract of not more than 200 words.
		We study how to design a secure observer-based distributed controller 	such 	that  a group of vehicles   can achieve accurate state estimates and formation control even if the measurements of a  subset of   vehicle sensors are  compromised by a malicious attacker.	
		We propose an architecture  consisting of a resilient observer, an  attack detector, and an observer-based distributed controller.
		The distributed detector is able to update three  sets of vehicle sensors: the ones surely under attack, surely attack-free, and suspected to be under attack.		   
		The adaptive observer saturates the measurement innovation through a preset static or time-varying threshold, such that the potentially compromised measurements have limited influence on the estimation.  
		Essential properties of the proposed architecture include: 
		1) The detector is fault-free, and  the attacked and attack-free vehicle sensors can be identified in finite time;		 
		2) The observer   guarantees both real-time error bounds and asymptotic error bounds, with tighter bounds  when  more attacked or attack-free vehicle sensors are identified by the detector;
		3) The  distributed controller ensures  closed-loop stability.
		The effectiveness of the proposed methods is  evaluated through simulations by an application to vehicle platooning.
\end{abstract}

\end{frontmatter}

%%%%%%%%%%%%%%%%%%%%%%%%%%%%%%%%%%%%%%%%%%%%%%%%%%%%%%%%%%%%%%%%%%%%%%%%%%%%%%%%
%%%%%%%%%%%%%%%%%%%%%%%%%%%%%%%%%%%%%%%%%%%%%%%%%%%%%%%%%%%%%%%%%%%%%%%%%%%%%%%%
\section{Introduction}
\subsection*{Motivations and related work}
Networked control systems (NCS) are ubiquitous. 
The performance of NCS significantly depends on widely deployed sensors  which might be compromised due to 
%communication ranges, adversaries, malicious behaviors of agents, or sensor attacks
the presence of  malicious attackers
\cite{shoukry2018smt,baras2019trust}. The attackers can strategically manipulate the sensor measurements in order to affect stability and performance of NCS. Attack detection, state estimation, and  system control  are three major components in the design of secure NCS in malicious environments.
% that are not equipped with large redundancy in measurements.

To detect whether systems are under attack and identify attacked components, quite a few  detection methods are proposed. 
Attack detection and identification  for linear descriptor systems are studied in \cite{pasqualetti2013attack}. 
Methods of attack detection and    correction for noise-free linear systems are proposed in \cite{RN18}. 
To detect the Byzantine adversaries with quantized false alarm rates, \cite{baras2019trust} study a trust-aware consensus algorithm.   In \cite{gallo2020distributed,RN25},     distributed detectors are designed for false data injection (FDI) attacks in communications.
Detection and mitigation  methods are proposed by \cite{deghat2019detection} for   distributed observers under  a class of bias injection attacks. A joint detection and estimation problem is investigated in \cite{forti2018distributed} with the knowledge of some attack statistics. 
There are some methods for multi-observer based detector design  \cite{chowdhury2020observer,kim2018detection,yang2020multi}. However, the computational complexity of these methods substantially 
  increases as the number of sensors is increasing. 
Thus, designing single-observer based detectors without relying on the knowledge of  attack signals  needs more investigations.
Moreover, most existing methods focus on detecting the attacked sensors, but few results are given    for the identification of  attack-free   sensors.

There are two major approaches in the literature for handling state estimation under sensor attacks.
The first approach is based on solving optimization problems  \cite{shoukry2018smt,RN89,fawzi2014secure,pajic2017attack,shoukry2017secure,RN92,RN13}. This approach  needs  a large number of computational resources    in enumerating all sensor combinations in order to find the attacked sensor set. Thus, it is not suitable to large-scale sensor networks if the resources are constrained. The second approach is to use robust techniques  in handling potentially compromised data, such as  discarding a few largest and  smallest elements \cite{su2019finite,ren2020secure,mitra2019byzantine,mitra2019resilient}, using the signum information of  measurement innovations \cite{lee2020fully}, and saturating the innovation which reaches a threshold \cite{chen2018resilient,he2020secure_journal}.
This approach is more suitable in online estimation since it needs very less computational resources than the first approach. However, there are few results in this direction, especially for dynamical systems under FDI sensor attacks.

Some resilient distributed control strategies have been proposed
to achieve  formation control of a group of   vehicles or robots in malicious environments. 
There are strategies on how to   handle  different attacks, such as replay attack on  control commands \cite{zhu2013distributed}, denial-of-service (DoS) attack on measurement and control channels \cite{RN65}, FDI attack in the transmission from controller to actuator \cite{RN1}, attack on   network topology of multi-agent systems \cite{feng2017distributed},
and   stealthy  integrity attacks \cite{weerakkody2016graph}. 
%For  multi-agent systems subject to attacks on   network topology,  resilient consensus tracking control strategy is studied in \cite{feng2017distributed}. 
%To handling deception attack on  observers,
% event-triggered  control strategy is investigated \cite{ding2016observer}.  
% $H_{\infty}$ control strategy is proposed  in \cite{modares2019resilient} for leader-follower systems with  attacks on actuators and sensors. 
 However,   there is no  unified architecture integrating resilient estimation, attack detection and distributed control.

\subsection*{Contributions}
In this paper, 
%different from  \cite{shoukry2018smt,ren2020secure,fawzi2014secure,pajic2017attack,an2019distributed,su2019finite,ding2017distributed,mitra2019byzantine,zhu2013distributed,weerakkody2016graph,feng2017distributed,modares2019resilient},
we propose an architecture  comprising of a resilient observer, an online attack detector, and a  distributed controller, such that a group of vehicles can achieve accurate state estimates and formation control even if the measurements of  a   subset of  the vehicle sensors are  compromised by a malicious attacker.
The main contributions of this paper are summarized as follows:
% The problem of designing a decentralized Coop-erative Adaptive Cruise Control (CACC) resilient to Denial-of-Service (DoS) attacks while satisfying performance require-ments is studied. Finally, the effectiveness of the proposed approach is shown in a numerical example
%	To this end, by using available secure radar and stereo camera, which are available in ADS and advanced driver-assistance systems, this paper provides a secure design algorithm such that a group of autonomous vehicles achieve practical platooning under potentially attacked GPS data in an unknown vehicle.

\begin{enumerate}[label=\roman*)]
	\item  We   propose an adaptive resilient observer, designed by saturating the measurement innovation through a preset static or time-varying threshold, such that the potentially compromised measurements have limited influence to the estimation (Algorithm \ref{alg:obser}).  
	%	with three update candidates, one of which is selected depending on the vehicle position in the network and whether the vehicle is attack-free from the detector (Algorithm \ref{alg:obser}).	
	%	 for each vehicle with the potentially compromised measurement   
	%	The key observer candidate   is designed by saturating the measurement innovation through a preset static or time-varying threshold, such that the attacked measurements have limited influence to the estimation.  
	Some essential properties are found: i) The observer is able to provide an upper bound of the estimation error  at each time (Proposition \ref{lem_detec22});
	ii)  If the observer threshold is static and satisfied with some explicit design principle (Proposition \ref{prop_feasibility}), the estimation error is asymptotically upper bounded (Theorem \ref{thm_estimation2}); and iii)  If the observer threshold is time-varying and computed adaptively, the estimation error is also asymptotically upper bounded (Theorem \ref{thm_estimation22}) and the   bound is tighter than that of the static threshold.

	%	by applying a saturation method to  an augmented and 
	%	The observer is able to provide an upper bound of the estimation error 
	%	 while providing an upper bound of the estimation error at each time.  Under mild conditions, we prove the state estimation error of each vehicle  is asymptotically uniformly bounded against the system uncertainties and the attacked measurements (Theorem \ref{thm_estimation2}).
	
	\item 	We develop an online distributed attack detector 
	with the potentially compromised sensor measurements and the observer's estimates. 	 
%	In contrast to recent works \cite{pasqualetti2013attack,baras2019trust,gallo2020distributed,deghat2019detection,forti2018distributed}, 
	The designed detector is able to update three  sets of vehicle sensors: the ones surely under attack, surely attack-free, and suspected to be under attack (Algorithm \ref{alg:detec}). Some properties are found: i) The detector is  fault-free (Lemma \ref{prop_ass}), which differs from the existing   results with false alarms (e.g., \cite{baras2019trust}); and ii) If some condition holds,  all  attacked and attack-free vehicle sensors are identified in finite time (Theorem \ref{prop_detect});
	%	%	The detector leverages the constrains in relative state of two adjacent vehicles, and the local  detection method and a local innovation method
	%	The detector is able to provide three detection sets: a  set of attacked vehicles, a set of secured vehicles, and a set of suspiciously attacked vehicles.	
	%	The three sets are updated through the local detected results and the neighbor vehicle communication 

	%	 leverages two detection approaches (Algorithm \ref{alg:detec}),
	
	%	 leveraging a shared trust scheme and two detection approaches (Algorithm \ref{alg:detec}), through which   the estimates of the  secured vehicle set and the attacked vehicle set are shared among neighbor vehicles. 	
	%    The detection scheme leverages two detectors, where one finds out a pair of suspicious adjacent vehicles if their  relative measurements are out of range,  and   the other detects an attacked vehicle when its  local update innovation is larger than the detector threshold. 

	\item  We design a distributed controller (Algorithm \ref{alg:control}) to achieve the formation control of the vehicles.
	%	regulate the speed of all vehicles to a desired one while maintaining a safe distance between any two adjacent vehicles. 	
	We find that if 
	the controller parameters satisfy some graph-related conditions, the overall performance function  is asymptotically upper bounded in the presence of noise and tending to zero in the absence of noise (Theorem \ref{thm_control} and Corollary \ref{coro_overall1}), which ensures the closed-loop stability of the proposed architecture. 
	%	 These attack resilience results are the extensions of string stability \cite{swaroop1996string,besselink2017string} in the case of vehicles  under sensor attacks.
	%	Also, the most general set of the control parameters for achieving formation control is provided (Theorem \ref{thm_control}).

	%	 which is initiated when the time is larger than a preset threshold for  steady transient. Moreover, we provide the design principle of  control parameters, such that the practical platooning  is achieved, i.e., 	 with small errors, all vehicles reach consensus in  speed and keeping fixed desired distance between two neighbors. 
	
	%	\item 	We apply the unified framework to  the vehicle platooning and find mild conditions such that the platooning is achieved (Theorem  \ref{thm_control2}), which has not been well studied in \cite{biron2017resilient,zhang2020distributed,feng2019secure,petrillo2018collaborative, pirani2018resilient}.
	%	The efficiency of the framework in the vehicle platooning is also evaluated through some  numerical simulations.

	%	We apply the general framework to  the vehicle platooning under GPS spoofing attacks, 
	
	%	 establish a secure vehicle platooning  framework, comprising of the proposed observer and controller, as well as an online attack detection scheme, which enable the observer gain to be adaptively adjusted by finding out the sets of attacked and attacked-free vehicles.  
	%		\item 	The resilience of the framework is analyzed. 
	%	\item We apply the develop the methods to the vehicle platooning...
\end{enumerate}

%	The observer is quite different from the Luenberger observer widely used in the  existing works for secure state estimation, e.g., \cite{chowdhury2020observer,modares2019resilient}. 
%To the best knowledge of the authors, the proposed architecture is the first unified architecture integrating resilient estimation, attack detection and distributed control.
		The proposed observer is able to handle more typical sensor attacks than   \cite{forti2018distributed,deghat2019detection}, such as random attack, DoS attack, bias injection attack, and replay attack. 
	The proposed detector is  based on one observer, which requires less computational resources than the detectors based on multiple observers \cite{chowdhury2020observer,kim2018detection,yang2020multi}. Although  \cite{mitra2019byzantine} study a wider range of attacks than this paper,
  we remove the requirements of  graph robustness. Moreover, the sufficiently large communication times between two updates \cite{an2019distributed} is not required.
Note that in comparison with our recent work  \cite{he2020secure_journal}, 
the current paper studies a different problem, and  uses potentially compromised measurements with new  approaches.

\subsection*{Outline}
The remainder of the paper is organized as follows: Section  \ref{sec_formulation} is on the problem formulation, followed by an overview of the proposed distributed observer-based control architecture in Section \ref{sec:structure}.  Section \ref{sec:observer} designs  a resilient observer for each vehicle, based on which 
Section \ref{sec:detection} studies the attack detection problem.  In Section \ref{sec:control}, 
a distributed controller is proposed to close the loop. After   simulations of vehicle platooning in Section \ref{sec:simulation}, the paper is concluded in   Section \ref{sec_conclusion}. 
The main proofs are   given in Appendix.

\textbf{Notations:} 
	  $\mathbb{R}^{n\times m}$ denotes the set of  real-valued  matrices with $n$ rows and $m$ columns, and  $\mathbb{R}^n$   the set of $n$-dimensional real-valued vectors. Without specific explanation, the scalars and matrices in this paper are real-valued. Denote $\mathbb{N}^+$ the set of positive integers and $\mathbb{N}=\mathbb{N}^+\cup 0.$
%The matrices and scalars in this paper are all real-valued.
The matrix	$I_{n}$ stands for the $n$-dimensional square identity matrix.  
%$\textbf{1}_N$ stands for the $N$-dimensional vector with all elements being one. 
%$\diag\{\cdot\}$ and $\blockdiag\{\cdot\}$   represent the diagonalization operators of scalar elements and block elements, respectively.  
The superscript ``$\sf T$" represents the transpose. 
The operator	$\diag\{\cdot\}$   represents the diagonalization.  
We denote   the Kronecker product of $A$ and $B$ by 	$A\otimes B$. The vector norm $\norm{x}$ is the 2-norm of a vector $x$. 
The matrix norm $\norm{A}$ is the induced 2-norm, i.e., $\norm{A}=\sup_{x\neq 0}\norm{Ax}/\norm{x}$.  
The notations	$\lambda_{\min}(A)$ and $\lambda_{\max}(A)$ are the  minimal and maximal eigenvalues of a real-valued symmetric matrix  $A$, respectively. The notation $a=(a_i)_{i=1,2,\dots,n}$ is a vector consisting of elements $a_1,\dots,a_n$. Let $\mathbb I_{i\in \mathcal{C}}$ be an indicator function, which equals 1 if $i\in \mathcal{C}$; otherwise, it is  0.
The function $\lceil \cdot\rceil$ stands for the ceiling function.
%$|\mathcal{\hat  S}^a|$ is the cardinality of the set $\mathcal{\hat  S}^a.$
%	A continuous function $\chi: [0, a) \rightarrow [0,\infty)$ is said to be of class
%	$\mathcal{K}$ if it is strictly increasing and $\chi(0)=0$. If, additionally, $a=\infty$ and 
%	$\chi(r)=\infty$ as $r\rightarrow \infty,$ it is of class $\mathcal{K}_{\infty}$. A continuous function $\xi:  [0, a)\times [0,\infty) \rightarrow [0,\infty)$ is said to be of class
%	$\mathcal{KL}$ if for each fixed $s$, the function $\xi(\cdot,s)$ is of class $\mathcal{K}$ and, for each fixed $r$, $\xi(r,\cdot )$ is decreasing and satisfies $\xi(r,s)\rightarrow 0$ as $s\rightarrow \infty.$

\section{Problem Formulation}\label{sec_formulation}
In this section, we first motivate the problem through a vehicle platooning example, and then formulate the problem. 
\begin{figure}[t]
	\centering
	%		\subfigure[Vehicle communication  with $L=1$ and vehicle 2 under GPS spoofing attacks]{
	%			\includegraphics[scale=0.5]{vehicle1_small.pdf}}
	%		\subfigure[Vehicle communication topology with $L=2$ and vehicles 1 and 3 under GPS spoofing attacks]{
	\includegraphics[scale=0.5]{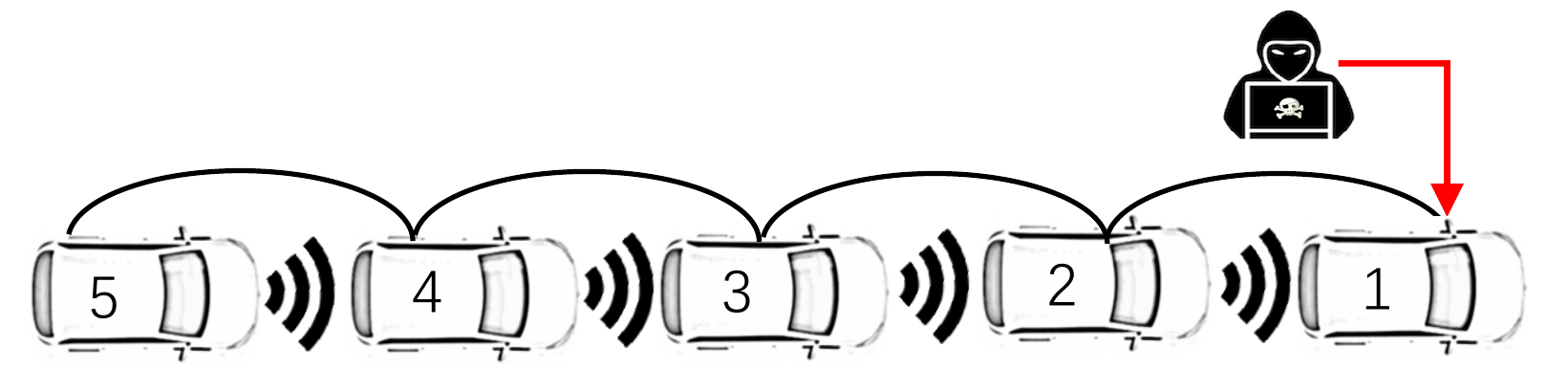}
	%			}
	\caption{ Platoon of five vehicles. The position and velocity measurements of  vehicle 1 are compromised by a malicious attacker. Each vehicle is able to exchange messages with other vehicles nearby   through   wireless communication.
		%			Communication topology of five vehicles with $L=1$ in (a) and $L=2$ in (b): Each vehicle is able to measure the relative state between itself and its front vehicle by radar or camera.  In (a), vehicle $i$, $i=2,3,4$ has $2L$ neighbors, but  only vehicle $3$ has $2L$ neighbors in (b).  For two  neighbor vehicles connected  by wireless communications (e.g., WiFi), they can exchange their messages. 
		%			A detailed message transmission in this paper is provided in Fig. \ref{fig:diagram}. 
	}
	\label{fig:Model_Description}
\end{figure}
%In this section, we formulate the problem of interest, which is motivated by the example in Fig. \ref{fig:Model_Description}, where five autonomous vehicles, localized by GPS, are running in a highway to achieve vehicle platooning (i.e., consensus in vehicle speed and keeping  desired distance between two neighboring vehicles) under the case that the GPS data of some vehicles is compromised due to spoofing attacks. The problem is how to employ the unreliable measurements to design distributed controller such that the vehicle platooning is achieved against the attacks and system uncertainties.

\subsection{Motivating example}
Consider the five-vehicle platooning in Fig. \ref{fig:Model_Description}. The aim is to control the speed of all vehicles to a desired value while maintaining a safe distance between any two adjacent vehicles.  Each vehicle is able to obtain its position and velocity measurements through a GPS receiver or a similar sensor, and the relative position and velocity measurements to its front vehicle through a sensor like a camera or radar. All  vehicles collaborate in the platoon  by using their local   measurements, 
and vehicle-to-vehicle communication. 

Suppose there is a malicious attacker, which aims to affect the platoon by compromising the position and velocity measurements of vehicle 1. Such attack could be a spoofing attack on a GPS receiver. 
By  using  the compromised   measurements, vehicle 1 is unable to  control its velocity to the desired value.  Consequently, the platoon is not able to maintain a proper formation.  The data redundancy resulting from the absolute and   relative measurements of the follower vehicles, however, provides an opportunity for designing resilient estimation and control algorithms. The algorithms are expected to mitigate such sensor attacks in order to achieve   vehicle platooning.

\subsection{System model} 
Consider $N\geq 3$ vehicles, which are labeled  from the leader to the tail by $1,2,\dots,N$.  We study the second-order   vehicle model: for $i=1,2,\dots,N$,
\begin{align}\label{eq_local_state}
\begin{split}
%		x_1(t+1)&=Ax_1(t)+d_{1}(t),  \\
x_i(t+1)&=Ax_i(t)+[0,Tu_i(t)]^{\sf T}+d_{i}(t),  
\end{split}
\end{align}
where $x_i(t)=(s_i(t),v_i(t))^{\sf T}\in\mathbb{R}^2$ is the state of vehicle~$i$ consisting of position $s_i(t)$ and velocity $v_i(t)$, $u_i(t)\in\mathbb{R}$  the control input,
$d_{i}(t)\in\mathbb{R}^2$  the process noise,   all at time $t$. Moreover,  $A=\left(\begin{smallmatrix}
1&T\\
0&1
\end{smallmatrix}\right)$,
where $T>0$  is the  time step. 			
%%			Without loss of generality, we assume that the order of these vehicles from the leader to the tail is $1,2,\dots,N$.  
%	
%%	The leader vehicle, i.e., vehicle $1,$  satisfies the following second-order dynamics:
%%	%Then we can rewrite \eqref{eq_system} as follows:
%%	\begin{align}\label{eq_local_state}
%%	x_1(t+1)=Ax_1(t)+d_{1}(t),
%%	\end{align}
%%	and  vehicle $i$, $i=2,3,\dots,N$, satisfies
%%	\begin{align}\label{eq_local_state2}
%%	x_i(t+1)=Ax_i(t)+[0,Tu_i(t)]^{\sf T}+d_{i}(t),
%%	\end{align}
%%	where 
%	
%
%	
%	
%	
%	
%	
%	
%	
Vehicle~$i$ is able to obtain its  absolute  measurements of position and velocity through sensor $i$, which is a potentially attacked sensor (e.g., a GPS receiver under spoofing attack):
\begin{equation}\label{eq_system2}
\begin{split}
y_{i,i}(t)&=x_i(t)+a_{i}(t)+n_{i,i}(t)
\end{split}
\end{equation}
where
$y_{i,i}(t)\in\mathbb{R}^2$ and $n_{i,i}(t)\in\mathbb{R}^2$ are the measurement and measurement noise, and the vector $a_{i}(t)\in\mathbb{R}^2$ represents an attack signal injected by a malicious attacker. 
Moreover, we assume each vehicle $j\in\{2,3,\dots,N\}$ has a secured sensor (e.g., an onboard radar or camera) to measure 	
%	by employing the vehicle sensors (e.g., radar and camera), vehicle $i\in\{2,3,\dots,N\}$ is able to measure
the relative state between itself and its front vehicle (i.e., vehicle~$j-1$):
\begin{equation}\label{eq_system3}
\begin{split}
y_{j-1,j}(t)&=x_j(t)-x_{j-1}(t)+n_{j-1,j}(t), 
\end{split}
\end{equation}
where $y_{j-1,j}(t)\in\mathbb{R}^2$ and $n_{j-1,j}(t)\in\mathbb{R}^2$ are the measurement and measurement noise.

	Although the relative state measurements $\{y_{j-1,j}(t)\}$ are secured, it is not possible to accurately estimate the absolute state $x_j(t)$ simply with these measurements. In the rest of the paper, we say that sensor $i$ is under attack if the unsecured sensor of vehicle $i$ is under attack.

%	Notice that  in this paper we consider the GPS spoofing attacks, thus   the relative state measurements from the vehicle sensors (e.g., radar and camera)  are secured.
%In this setting, the problem is still of interest, since it is infeasible to obtain the position and velocity of each vehicle simply based on the relative measurements.	

%
%\begin{assumption}\label{ass_graph}
%	The graph is undirected and connected.
%\end{assumption}

\subsection{Attack model}
%%	Denote $\mathcal{V}=\{1,2,\dots,N\}$ the set of all vehicles, 
%Assume a subset $y_{i,i}$, $i\in \mathcal{S}^a\subset \{1,2,\dots,N\}$ of the vehicle sensor measurements in \eqref{eq_system2} have been attacked, but we don't know which. 
%%	Denote $\mathcal{S}$ the  set of attack-free vehicle sensors, i.e., 
%%%	$\mathcal{S}$ the  set of secured vehicle sensors, and
%%	$\mathcal{S}=\{1,2,\dots,N\}-\mathcal{S}^a$, 
The   attack model is provided in the following   assumption. 
%	In other words, the malicious attacker can compromise the position and velocity data of the vehicles in the set $\mathcal{S}^a$ by injecting the attack signal $a_i$ in \eqref{eq_system2}. 
%	However, due to  physical limitations (e.g.,  energy), the size of $\mathcal{S}^a$ is inevitably constrained, which is assumed in the following.
%	Then the following assumption on the attack model is  needed in the paper.
\begin{assumption}\label{ass_attack}
	There is an unknown and time-invariant attack set $ \mathcal{S}^a\subset \{1,2,\dots,N\}$ with at most $b\geq 1$ elements, such that
	the corresponding attack signals $a_i(t)\in\mathbb{R}^2$, $i\in \mathcal{S}^a$, $t\in\mathbb{N}$, are arbitrary, and the maximum number of  attacked sensors $b$ is known to each vehicle.  For the set of attack-free vehicle sensors $\mathcal{S}:=\{1,2,\dots,N\}\setminus\mathcal{S}^a$, it holds that $a_{i}(t)\equiv 0,    i\in \mathcal{S},$ $t\in\mathbb{N}$.
	%		\begin{align*}
	%		a_{i}(t)&\in\mathbb{R}^2,   i\in \mathcal{S}^a, |\mathcal{S}^a|\leq b, \quad 
	%		a_{i}(t)=0,    i\in \mathcal{S},
	%		\end{align*}
	
	%$		|\mathcal{S}^a|\leq b$.
	%	The number of attacked vehicles satisfies
	%	\begin{align*}
	%	|\mathcal{S}^a|\leq 1.
	%	\end{align*}
	%	where $s$ is known to the system defender.
	%	Two conditions on the attack hold as follows:
	%	\begin{enumerate}
	%		\item The relative measurements of all vehicles, i.e., $y_{i,j}$, $j\in\mathcal{N}_i,i=1,\dots,N$, and the absolute measurements of  vehicle $i$, i.e., $y_{i,i}$, $2\leq i\leq N-1$,   can be compromised by a malicious attacker.
	%		\item The label difference between two attacked vehicles is larger than 1, i.e., $|i-j|>2, \forall i,j\in\mathcal{S}^a$.
	%	\end{enumerate}
\end{assumption}

Following Assumption \ref{ass_attack}, 
a subset $ \mathcal{S}^a$ of the vehicle sensor measurements in \eqref{eq_system2} can be manipulated arbitrarily, but we do not know which ones. Assumption \ref{ass_attack}  does not impose  any specific distribution or  form of $a_i(t)$, and covers many typical  sensor attacks, including random attack, DoS attack, bias injection attack, and replay attack  \cite{teixeira2015secure}.
%	and is general than the existing literature where the attack signals are supposed to satisfy some distributions or concrete requirement \cite{deghat2019detection}.

The upper bound $b$ of the number of attacked vehicle sensors is used in the observer and detector designs. The assumption on the knowledge of $b$ can be relaxed, but will result in worse performance for the same number of attacked sensors.

\subsection{Problem}\label{sec:form_problem}
%\textbf{Problem:}

In order to achieve vehicle formation control (e.g., vehicle platooning) in a malicious environment, 
it is  important to estimate the  states of all vehicles simultaneously.  For example, when a group of vehicles are required to achieve a platoon with a desired speed,  
it is necessary to estimate the  state of 
the leader vehicle for controller design. However, its absolute measurements are potentially compromised as in \eqref{eq_system2}.  
 In order to  have data redundancy for the state estimation of the leader vehicle, the secured relative measurements and accurate  estimates of the follower vehicles are necessary.
%   Another example showing the importance of   state estimation is  when the leader vehicle is changing
%%	in practical applications, such as the platooning case that when the leader vehicle 
%for another task, and the new leader is supposed to track a new state with its state estimate.
%%	There are two important reasons to estimate the  states of all vehicles: i) As  in Remark \ref{rem_relative}, the controller     needs the state estimate of the leader vehicle. Since the attacker can compromise the absolute measurements of the leader vehicle, in order to have data redundancy in the estimation for the leader vehicle, the secured relative measurements and accurate  estimates of the follower vehicles are necessary. ii) In practice, accurate vehicle state estimates  are important   when the leader vehicle is changing
%%	%	in practical applications, such as the platooning case that when the leader vehicle 
%%	for another task, and the new leader is supposed to track a new state by implementing its state estimate.

To measure the overall estimation and control performance for the system \eqref{eq_local_state}--\eqref{eq_system3}, we introduce the performance function 	$\varphi(t)$:
%	We denote
%	the  performance metric of the system \eqref{eq_local_state}--\eqref{eq_control} by 
%Since the leader-follower formation control is considered in this paper, 
%\begin{align}
%\begin{split}
%\varphi(t)=\frac{1}{N}\sum_{i=1}^N\bigg(&\norm{\hat x_i(t)-x_i(t)}\\
%&+\norm{x_{i-1}(t)-x_{i}(t)+\Delta x_{i-1,i}(t)}\bigg),
%%&+\frac{1}{N-1}\sum_{j=1}^{N-1}\norm{x_j(t)-x_{j+1}(t)+\Delta x_{j,j+1}(t)},
%\end{split}
%\end{align} 
\begin{align}\label{eq_function}
\begin{split}
\varphi(t)&=\frac{1}{N}\sum_{i=1}^N\norm{\hat x_i(t)-x_i(t)}+\norm{x_{i}(t)-x_{i}^*(t)},
%e_i(t)&=\hat x_i(t)-x_i(t)\\
%\tilde e_i(t)&=x_{i}(t)-x_{i}^*(t)\\
\end{split}
\end{align}
%\begin{align}\label{eq_function}
%\begin{split}
%\varphi(t)=\frac{1}{N}\sum_{i=1}^N\bigg(\norm{\hat x_i(t)-x_i(t)}
%+\norm{x_{i}(t)-x_{i}^*(t)}\bigg),
%%&+\frac{1}{N-1}\sum_{j=1}^{N-1}\norm{x_j(t)-x_{j+1}(t)+\Delta x_{j,j+1}(t)},
%\end{split}
%\end{align} 
where  $\hat x_i(t)$ is the estimate of $x_i(t)$ from the observer to be designed, and $x_{i}^*(t) $ is the desired  vehicle state of the formation
    satisfying
\begin{align*}
x_{i}^*(t)=\begin{cases}
x_0(t), &\text{if } i=1\\
x_{i-1}^*(t)-\Delta x_{i-1,i}(t), &\text{if } i\in\{2,3,\dots,N\},
\end{cases}
\end{align*}
where  $x_0(t)$ is the reference state of the leader vehicle, subject to $x_0(t+1)=Ax_0(t)$, and  $\Delta x_{i-1,i}(t)$ is the desired  relative state between vehicles $i-1$ and $i$, subject to $\Delta x_{i-1,i}(t+1)=A\Delta x_{i-1,i}(t)$, $i=2,3,\dots,N$. 	 For convenience, we denote $\Delta x_{0,1}(t)\equiv[0,0]^{\sf T}$.
%  We aim to design control inputs $u_i(t)$, $i=1,2,\dots,N$, such that the following formation is achieved: the leader vehicle can reach the reference state $x_0$, and the following two adjacent vehicles $j-1$ and $j$ keep the desired relative state   $\Delta x_{j-1,j}(t)$,  subject to $\Delta x_{j-1,j}(t+1)=A\Delta x_{j-1,j}(t)$, $j=2,3,\dots,N$.  

%\begin{remark}\label{rem_platoon}
	If $\Delta x_{i-1,i}(t)\equiv[0,0]^{\sf T}$, $i=1,\dots,N$, it means all  vehicles aim to reach the reference state $x_0$; if $\Delta x_{i-1,i}(t)\equiv[s_0,0]^{\sf T}$, where $s_0$ is a positive scalar, it means all vehicles are expected to have the same speed, and two nearest neighbor vehicles keep the distance $s_0$, which is a typical scenario in vehicle platooning.
%\end{remark}
% Assume $\Delta x_{i-1,i}(t)$ is  known to vehicles $i-1$ and $i.$

\begin{assumption}\label{ass_noise}
	The noise  in \eqref{eq_local_state}--\eqref{eq_system3}, and the initial estimation error  satisfy:  $\forall i\in\{1,\dots,N\}$ and $\forall j\in\{2,\dots,N\},$   
	\begin{align*}
	&\sup\norm{\hat x_{i}(0)-x_i(0)}\leq q,\quad\sup_{t\geq 0}\norm{d_i(t)}\leq \epsilon, 	\\
	&\sup_{t\geq 0}\max\{\norm{n_{i,i}(t)},\norm{n_{j-1,j}(t)}\}\leq \mu,
	\end{align*}
	where  the scalars $q>0$, and $\epsilon\geq 0, \mu\geq 0$ are   known to each vehicle.
\end{assumption}
The upper bounds $q,\epsilon,\mu$ are used in the observer and detector designs. The assumption on the knowledge of $q,\epsilon$, and $\mu$ can be relaxed, but will result in worse performance for the same noise  and  initial estimation error.

\textbf{Problem:} How to design  an  observer-based  distributed controller $u_i(t)$ for the system \eqref{eq_local_state}--\eqref{eq_system3} under Assumptions \ref{ass_attack}--\ref{ass_noise}, such that: 
\begin{enumerate}[label=\roman*)]
	\item In the presence of noise, there is a scalar  $c_0>0$,  such that 
	\begin{align*}
\limsup_{t\rightarrow\infty}\varphi(t)<c_0;
	\end{align*}
	\item In the absence of noise,  
	\begin{align*}
\limsup_{t\rightarrow\infty}\varphi(t)=0.
	\end{align*}
\end{enumerate}

	 \section{Observer-Based Distributed  Control Architecture}\label{sec:structure}
	 In this section, we first introduce the communication structure of the vehicle network, and then  propose an architecture  consisting of a resilient observer, an  attack detector, and a distributed controller.
	 % new  architecture of a distributed  observer-based controller  with an online attack detector. 
	 Moreover, the measurements of each vehicle will  be reconstructed based on vehicle-to-vehicle communication.
	 %The algorithm design and property analysis of the observer, detector, and controller  will be respectively conducted in the following sections.

	 \subsection{Communication structure of vehicle network}
\begin{figure}[t]
	 	\centering
	 	\begin{tikzpicture}[scale=0.58, transform shape,line width=1pt]
	 	\draw (-3,0) node[rectangle,draw,scale=1.2,line width=2pt] (vehicle5)   {VCS 5};
	 	\draw (0.5,0)  node[rectangle,draw,scale=1.2,line width=2pt] (vehicle4)   {VCS 4};	
	 	%	\draw (1,-1)  node[rectangle,draw,scale=1.2,line width=2pt] (vehicle5)   {Radar sensor};
	 	\draw (4,0)  node[rectangle,draw,scale=1.2,line width=2pt] (vehicle3)   {VCS 3};	
	 	\draw (7.5,0)  node[rectangle,draw,scale=1.2,line width=2pt] (vehicle2)   {VCS 2};
	 	\draw (11,0)  node[rectangle,draw,scale=1.2,line width=2pt] (vehicle1)   {VCS 1};
	 	%	\draw (6,0)  node  (output)   {};
	 	\draw (vehicle5)++(3.5,1.7)  node  (indicator)   {$\mathcal{M}_5$};
	 	\draw (vehicle4)++(3.5,1.7)  node  (indicator)   {$\mathcal{M}_4$};
	 	\draw (vehicle4)++(3.5,-1.7)  node  (indicator)   {$\mathcal{M}_2$};
	 	\draw (vehicle5)++(3.5,-1.7)  node  (indicator)   {$\mathcal{M}_3$};
	 	\draw (vehicle3)++(3.1,1.7)  node  (indicator)   {$\mathcal{M}_3$};
	 	\draw (vehicle3)++(3.5,-1.7)  node  (indicator)   {$\mathcal{M}_1$};
	 	%	\path[->] (-6,0) edge node[above] {$n_j$} (plant);
	 	%	\path[->] (0.6,2.5) edge node[left] {$d_{j,j}$} (0.6,1.3);
	 	%	\path[->] (1.3,2.5) edge node[right] {$a_{j}$} (1.3,1.3);
	 	%	\draw[->] (plant) edge node[above] {$x_j$} (GPS);
	 	%	\draw[->] (plant) edge node[below] {$x_j$} (radar);
	 	%	\path[->] (radar)+(0,1) edge node[left] {$d_{j-1,j}$} (radar);
	 	%	\draw[->] (GPS) edge node[above] {$y_{j,j}$} (detector);
	 	%	\draw[->] (radar) edge node[below] {$y_{j-1,j}$} (detector);
	 	%	\path[->,color=red] (vehicle3)+(0,2) edge node[left] {$a_{3}$} (vehicle3);
	 	%	\path[->] (detector)+(0,2) edge node[right] {$\textcircled{2}$} (detector);
	 	\path[->,color=red] (vehicle1)+(0.0,2) edge node[below,right] {$a_{1}$} ++(0.0,0.3);
	 	%	\path[->] (control)+(0,2) edge node[left] {$\textcircled{3}$} (control);
	 	%			\draw[->] (5.9,0.2) edge node[above] {$\Gamma_j,\Theta_j,\Omega_j$} (8,0.2);
	 	%	\draw (6.8,0.38) node (estimate) {$\mathcal{W}_2$};
	 	\path[->] (vehicle5) ++(0.8,0.2) edge node[above] {$\mathcal{M}_5$} +(1.9,0);
	 	\path[<-] (vehicle5) ++(0.8,-0.2) edge node[below] {$\mathcal{M}_4$} +(1.9,0);
	 	\path[->] (vehicle4) ++(0.8,0.2) edge node[above] {$\mathcal{M}_4$} +(1.9,0);
	 	\path[<-] (vehicle4) ++(0.8,-0.2) edge node[below] {$\mathcal{M}_3$} +(1.9,0);
	 	\path[->] (vehicle3) ++(0.8,0.2) edge node[above] {$\mathcal{M}_3$} +(1.9,0);
	 	\path[<-] (vehicle3) ++(0.8,-0.2) edge node[below] {$\mathcal{M}_2$} +(1.9,0);
	 	\path[->] (vehicle2) ++(0.8,0.2) edge node[above] {$\mathcal{M}_2$} +(1.9,0);
	 	\path[<-] (vehicle2) ++(0.8,-0.2) edge node[below] {$\mathcal{M}_1$} +(1.9,0);
	 	\draw[->] (vehicle5) .. controls ++(3.5,1.7) .. (vehicle3);
	 	\draw[<-] (vehicle5) .. controls ++(3.5,-1.7) .. (vehicle3);
	 	\draw[->] (vehicle3) .. controls ++(3.5,1.7) .. (vehicle1);
	 	\draw[<-] (vehicle3) .. controls ++(3.5,-1.7) ..  (vehicle1);
	 	\draw[->] (vehicle4) .. controls ++(3.5,1.7) ..  (vehicle2);
	 	\draw[<-] (vehicle4) .. controls ++(3.5,-1.7) ..  (vehicle2);
	 	%	\path [draw] (5.9,-0.2) -- (8.1,-0.2) [<-];
	 	%	\draw (6.8,-0.5) node (estimate) {$\mathcal{W}_2$};
	 	%	\draw (11,0.38) node (estimate) {$\mathcal{W}_2$};
	 	%	\path [draw] (9.9,0.2) -- (11.95,0.2) [->];
	 	%	\path [draw] (9.9,-0.2) -- (11.95,-0.2) [<-];
	 	%	\draw (11,-0.48) node (estimate) {$u_{j}$};
	 	%	\path [draw] (plant) -- (-3,-2) -- (13,-2) -- (control) [<-];
	 	\end{tikzpicture}	
	 	\caption{Communication topology of the undirected graph $\mathcal{G}$  with five vehicle control systems (VCSs) for $L=2$, where   VCS 1 is under attack  and $\mathcal{M}_j$, defined in  \eqref{message}, is the  message sent out by  VCS $j$ to its neighbors, $j=1,2,\dots,5$, and $a_1$ is  the attack signal.
	 		%		The messages from vehicle $i$ to vehicle $j$, $i\in\mathcal{N}_j$, where $y_{i-1,i},y_{i,i}$ are  the measurements in \eqref{eq_system2} and \eqref{eq_system3}, respectively, $\bar x_{i}$ the predicted state estimate of vehicle $i$, and $\Gamma_i,\Theta_i,\Omega_i$ are detection sets defined in Section \ref{sec:detection}.
	 	}	
	 	\label{fig:diagram1}
	 \end{figure}
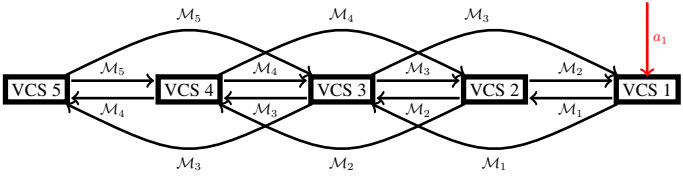

	 %In this subsection, we show the communi structure between different vehicles.	
	 %	\subsection{Communication of vehicles}
	 %	To achieve the vehicle formation control, the vehicles collaborate with each other by  communicating messages through   vehicle-to-vehicle connectivity. 
	 We model the vehicle communication topology by an undirected graph $\mathcal{G}=\{\mathcal{V},\mathcal{E}\}$, which consists of the set of nodes $\mathcal{V}=\{1,2,\dots,N\}$ and the set of edges $\mathcal{E}$. If  there is an edge $(i,j)\in \mathcal{E}$, 
	 node $i$ can exchange information with node $j$. In the case,  node $j$  is called a  neighbor of node $i$, and vice versa. 
	 Denote the neighbor set of node $i\in \mathcal{V}$ by $\mathcal{N}_{i}:=\{j\in\mathcal{V}|(i,j)\in \mathcal{E}\}$, which  in this paper is assumed to be
	 %A  bidirectional vehicle communication scheme is considered, where each vehicle is able to send its messages to other vehicles and  also receive messages from them. The graph is denoted by $\mathcal{G}=\{\mathcal{V},\mathcal{E}\}$.
	 %	We assume that the neighbor set of each vehicle is 
	 \begin{align*}
	 \mathcal{N}_i=\begin{cases}
	 \{i-L, \dots,i-1,i+1, \dots,i+L\}, &\text{if } i\in\mathcal{V}_1\\
	 \{1, \dots,i-1,i+1, \dots,i+L\}, &\text{if } i\in\mathcal{V}_{2,1}\\
	 \{i-L, \dots,i-1,i+1, \dots,L\}, &\text{if } i\in \mathcal{V}_2\setminus\mathcal{V}_{2,1},
	 \end{cases}
	 \end{align*}
	 where $L\in\mathbb{N}^+$ is a parameter indicating the neighbor range, $\mathcal{V}_{2,1}=\{1, 2,\dots,L\}$, and 
	 \begin{align}\label{sets_vehicles}
	 \begin{split}
	 \mathcal{V}_1=\{L+1,L+2,\dots,N-L\}, \quad
	 \mathcal{V}_2=\mathcal{V}\setminus\mathcal{V}_1.
	 \end{split}
	 \end{align}
	 As seen, each  vehicle $i\in\mathcal{V}_1$ has $2L$ neighbors, and each vehicle $j\in\mathcal{V}_2$ has less than $2L$ neighbors.	
	 %	each vehicle  
	 %	is able to  communicate with its $L$ front  and $L$ rear neighboring vehicles, $L\geq 1$.
	 %	
	 %	We then denote $\mathcal{V}_1$ the set of these vehicles with	$2L$ neighbors, and  denote $\mathcal{V}_2$ the  complementary of $\mathcal{V}_1$ in $\mathcal{V}$, i.e.,
	 %	Given $b\leq L\leq \lfloor \frac{N-1}{2}\rfloor$, we assume each vehicle $i$, 
	 %	$i\in \mathcal{V}_1$, can communicate with its nearest $2L$ vehicles, namely, $L$ vehicles   front  and $L$ vehicles behind, through WiFi or 5G.  
	 %	Vehicle $i\in \mathcal{V}_2$  does not have  $2L$ neighboring vehicles, but it can communicate with the vehicles within the range. For example, vehicle $1$ can communicate with vehicles $2,3,\dots,L+1$, and vehicle $N$ can communicate with vehicles $N-L+1,N-L+2,\dots,N-1$.   
	 %	%	 The communication between two vehicles is bidirectional, meaning the vehicle can send its messages to other vehicles and can also receive messages from them. 
	 %	We denote  the neighbor set of vehicle $i$ by $\mathcal{N}_i$, where 	
	 The communication topologies of five vehicle control systems (VCSs) for $L=1$ and $L=2$ are illustrated in Fig. \ref{fig:Model_Description} and Fig.~\ref{fig:diagram1}, respectively.  In the following,  we use the term `vehicle' to represent a VCS for convenience.
	 %	The communication topology of five vehicles with $L=1$ can be referred to .
	 %	 In Fig.~\ref{fig:diagram1}, vehicles 1 and 3 are under spoofing attacks, where two attack signals $a_1$ and $a_2$ are injected into the absolute measurements as in~\eqref{eq_system2}.
	 %	 the predicted state estimate and the measurements of its neighboring vehicle $j$, $j\in\mathcal{N}_i$ comprising of  absolute and relative measurements in \eqref{eq_system2} and \eqref{eq_system3}.
	 %	Note that for the leader and tail vehicles, i.e., vehicle $1$ and  vehicle $N$, to ensure the redundancy of measurement information against the possible attacks on them, we require  that vehicle $1$ can  obtain the measurements from  vehicles $2$ and $3$, and that vehicle $N$   can  obtain the measurements from vehicles $N-1$ and $N-2$.
	 Each vehicle $i\in\mathcal{V}$ is able to send  its  neighbor vehicle  $j\in\mathcal{N}_i$ a message at time $t\in\mathbb{N}^+$, denoted by $\mathcal{M}_i(t)$ (omitting the time index $t$ in the following notation):
	 %(omitting the time index $t$): for $i=1,2,\dots,N$,
	 \begin{align}\label{message}
	 %			\mathcal{M}_1&=\begin{cases}
	 %			\{y_{1,1},\bar x_{1}\}  \qquad&\text{ without detection}\\
	 %			\{y_{1,1},\bar x_{1},\mathcal{\hat S}_1,\mathcal{\hat  S}_1^a,\mathcal{\hat S}_1^s\}  \qquad  &\text{ with detection}
	 %			\end{cases}\\
	 \mathcal{M}_i&=\begin{cases}
	 \{y_{1,1},\bar x_{1},\mathcal{\hat S}_1,\mathcal{\hat  S}_1^a,\mathcal{\hat S}_1^s,\alpha_1\}  &\text{if } i=1\\
	 \{y_{i-1,i},y_{i,i},\bar x_{i},\mathcal{\hat S}_i,\mathcal{\hat  S}_i^a,\mathcal{\hat S}_i^s,\alpha_i\}   &\text{otherwise},
	 \end{cases}
	 \end{align} 
	 where $\bar x_{i}(t+1)=A\hat x_i(t)+[0,Tu_i(t)]^{\sf T}$ is the predicted value of $x_{i}(t+1)$ from the observer to be designed, $\alpha_i(t)$ denotes the estimation error bound to be specified in \eqref{real_bounds},  and
	 \begin{itemize}
	 	\item $\mathcal{\hat S}_i(t)$: the set of attack-free vehicle sensors estimated by   vehicle $i$ at time $t$, i.e., the estimate of    $\mathcal{S}$ 
	 	\item $\mathcal{\hat  S}_i^a(t)$: the set of  attacked vehicle sensors   estimated by   vehicle $i$ at time $t$,  i.e., the estimate of    $\mathcal{S}^a$
	 	\item $\mathcal{\hat S}_i^s(t)$: the set of vehicle sensors, which are suspected  to be under attack, estimated by vehicle $i$. 
	 \end{itemize}
	 Note that $\mathcal{\hat S}_i^s(t)\subseteq \mathcal{V}$ is not necessarily a subset of $\mathcal{S}^a$, since $\mathcal{\hat S}_i^s(t)$ may include some attack-free vehicle sensors.	 
	 %	$\mathcal{\hat  S}_i^a(t)$ denote the estimates of attack-free vehicle sensor set (i.e., $\mathcal{S}$  in Assumption \ref{ass_attack})  and attacked vehicle sensor set (i.e.,  $\mathcal{S}^a$), respectively,  and $\mathcal{\hat S}_i^s(t)$ is the set of vehicles suspicious under attacks, all from vehicle $i\in\mathcal{V}$  at time $t$. 
	 The three sets $ \{\mathcal{\hat S}_i(t),\mathcal{\hat  S}_i^a(t),\mathcal{\hat S}_i^s(t)\}$ are shared  between vehicles through the vehicle-to-vehicle  network $\mathcal{G}$ and   updated in a distributed manner described in Section \ref{sec:detection}. The sets are initialized as empty sets, i.e.,   $\mathcal{\hat S}_i(0)=\mathcal{\hat  S}_i^a(0)=\mathcal{\hat S}_i^s(0)=\emptyset,$  $i\in \mathcal{V}$.

	 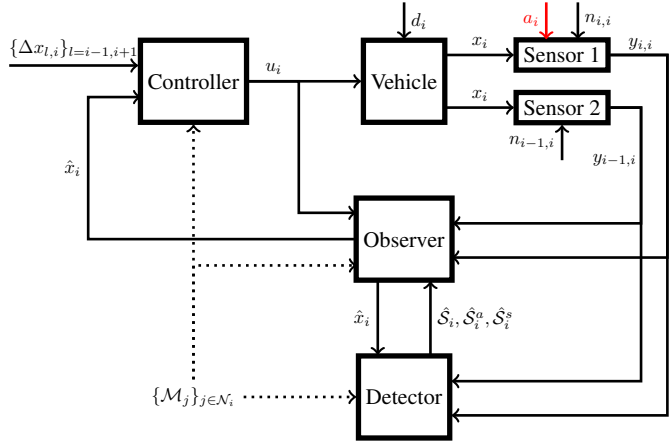
\begin{figure}[t]
	 	\centering
	 	\begin{tikzpicture}[scale=0.7, transform shape,line width=1pt]
	 	\draw (3,0)  node[rectangle,draw,scale=1.2,minimum width=0.8cm,
	 	minimum height = 1.3cm,line width=2pt] (plant)   {Vehicle};
	 	\draw (6,0.5)  node[rectangle,draw,scale=1.2,line width=2pt] (GPS)   {Sensor 1};	
	 	\draw (6,-0.5)  node[rectangle,draw,scale=1.2,line width=2pt] (radar)   {Sensor 2};	
	 	\draw (3,-3)  node[rectangle,draw,scale=1.2,minimum width=0.8cm,
	 	minimum height = 1.3cm,line width=2pt] (observer)   {Observer};
	 	\draw (-1,0)  node[rectangle,draw,scale=1.2,minimum width=0.8cm,
	 	minimum height = 1.3cm,line width=2pt] (control)   {Controller};
	 	\draw (3,-6)  node[rectangle,draw,scale=1.2,minimum width=0.8cm,
	 	minimum height = 1.3cm,line width=2pt] (detector)   {Detector};
	 	\draw[->] (plant)+(0.8,0.5) -- node[above] {$x_i$} (GPS);
	 	\draw[->] (plant)+(0.8,-0.5) -- node[above] {$x_i$} (radar);
	 	\path [draw] (observer) -- (-3,-3) -- node[left] {$\hat x_i$} (-3,-0.3) -- ++(1,0) [->];
	 	\path [draw] (control) -- node[above] {$u_i$} (1,0) -- (1,-2.5) -- ++(1.09,0) [->];
	 	\path [draw] (control) --   (plant) [->];
	 	\path [draw,->] (observer)++(-0.5,-0.8) -- node[left] {$\hat x_i$}	 ++(0,-1.4);
	 	\path [draw] (observer)++(0.5,-0.8)	 edge node[right,xshift =0cm, yshift = 0cm] {$\mathcal{\hat S}_i, \mathcal{\hat  S}_i^a,\mathcal{\hat S}_i^s$}  ++(0,-1.4) [<-];
	 	\draw (-1,-6)  node[scale=1,line width=2pt] (message)   {$\{\mathcal{M}_j\}_{j\in\mathcal{N}_i}$};
	 	\draw (7.5,0.7)   node[scale=1,line width=2pt] (relative)   {$y_{i,i}$};
	 	\draw (7.0,-1.5)  node[scale=1,line width=2pt] (absolute)   {$y_{i-1,i}$};
	 	\path [draw,dotted] (message)  -- (detector) [->];
	 	\path [draw,dotted] (message)  -- (control) [->];
	 	\path [draw,dotted] (message)  -- (-1,-3.5) -- ++(3.1,0) [->];
	 	\path[->] (plant) ++(-0,1.5) edge node[right] {$d_{i}$}    ++(0,-0.7);
	 	\path[->,color=red] (GPS)+(-0.3,1) edge node[left] {$a_{i}$}    ++(-0.3,0.3);
	 	\path[->] (GPS)+(0.3,1) edge node[right] {$n_{i,i}$}    ++(0.3,0.3);
	 	\path[->] (radar)+(0,-1) edge node[left] {$n_{i-1,i}$} (radar);
	 	\path [draw] (GPS) -- (8,0.5) -- (8,-3.35) -- ++(-4.1,0) [->];
	 	\path [draw] (GPS) -- (8,0.5) -- (8,-6.35) -- ++(-4.1,0) [->];
	 	\path [draw] (radar) -- (7.5,-0.5) -- (7.5,-2.7) -- ++(-3.58,0) [->];
	 	\path [draw] (radar) -- (7.5,-0.5) -- (7.5,-5.7) -- ++(-3.6,0) [->];
	 	\path [draw] (-4.5,0.3)  -- node[above] {$\{\Delta x_{l,i}\}_{l=i-1,i+1}$}  ++ (2.45,0) [->];
	 	%	\path [draw] (-1,1.5)  -- node[right] {$\{\Delta x_{l,i}\}_{l=i-1,i+1}$}   (control) [->];
	 	\end{tikzpicture}	
	 	\caption{Vehicle control system architecture for vehicle $i$: The  control signal for vehicle $i$ utilizes information from the other vehicles as indicated by the  dashed arrows: $\mathcal{M}_j$ is defined in  \eqref{message}, $j\in\mathcal{N}_i$. The observer, detector, and controller are designed in Sections \ref{sec:observer}, \ref{sec:detection}, and \ref{sec:control}, respectively. 
	 		%		 Here, $\hat x_j$ is the  estimate of $x_j$.
	 	}	
	 	\label{fig:diagram2}
	 \end{figure}

	 \subsection{Resilient observer-based distributed control architecture}	
	 We design an architecture for the VCS of each vehicle $i$ in Fig. \ref{fig:diagram2}.
	 The architecture   integrates the resilient observer in Section \ref{sec:observer}, the attack detector in Section \ref{sec:detection}, and the distributed controller in Section  \ref{sec:control}. The observer leverages the measurements of vehicle $i$ and neighbor vehicles. Then, the estimate $\hat x_i(t)$ from the observer is sent to the controller, which employs   $\hat x_i(t)$  as well as the estimates of neighbor vehicles to generate control signal $u_i(t)$.
	 %, which  is actuated in the vehicle $j$ to change   its state according to the system dynamics in \eqref{system_control}.
	 If the observer is inefficient, the observer-based controller would not work well.  Therefore, the key point for the observer is how to use the potentially attacked measurements and the measurements from neighbor vehicles efficiently. In Section \ref{sec:observer}, a resilient observer is proposed by leveraging a new saturation approach.  
	 %The architecture in Fig. \ref{fig:diagram} inherits the structure in Fig. \ref{fig:diagram2}, and It also adds an online attack detector in Section \ref{sec:detection}. 
%	 Different from existing results \cite{pasqualetti2013attack,baras2019trust,gallo2020distributed,deghat2019detection,forti2018distributed}, 
	 The designed detector is able to update the    three sets $\{\mathcal{\hat S}_i,\mathcal{\hat  S}_i^a,\mathcal{\hat S}_i^s\}$, and send them to the observer.
	 Then, in order to improve the estimation performance, the observer will discard the measurements of the  untrustworthy vehicles henceforth, and fully utilize the  measurements of the trustworthy vehicles.	 
%	  For the untrustworthy vehicle, its measurements will be discarded henceforth, and the measurements of the trustworthy vehicles will be fully utilized and help to  improve the estimation performance.  
	 %Compared to the architecture in Fig. \ref{fig:diagram2},  the architecture in Fig. \ref{fig:diagram} is able to gain better performance in state estimation and then formation control while providing the estimates for the secured and attacked vehicle sets. 
	 Note that the detector in Section \ref{sec:detection} ensures   consistency of the three sets in the sense that they will not conflict. In other scenarios,  if an inconsistent case occurs due to some reasons (e.g., the detection data is manipulated),  the architecture in Fig. \ref{fig:diagram2} can be   employed by  abandoning the inconsistent subsets.
	 %	\section{Co-design of observer,  detector, and controller}\label{sec_algorithm}
	 %	%Since the vehicles run in a high way, the topology of the vehicles is a line. Thus, for vehicle $2\leq i\leq N-1$, it measures   the relative states  between itself and its front vehicle $i-1$. 
	 %  and finally provide a distributed controller by employing the estimates from neighboring vehicles to achieve the vehicle platooning in \eqref{eq_platoon}.
	 %	In the following, we drop the time script (i.e., $(t)$) for notational convenience.
	 
	 %	We provide a diagram in Fig. \ref{fig:diagram}, which contains the main body of the proposed framework in dealing with the above secure vehicle localization and platooning problem.	
	 
	 \subsection{Measurement  reconstruction via vehicle communication}
	 Based on whether each vehicle has $2L$ neighbors, we split the vehicle set $\mathcal{V}$ into two subsets $\mathcal{V}_1$ and $\mathcal{V}_2$ as shown in \eqref{sets_vehicles}. In the following, we first reconstruct the measurement equation of vehicle $i\in\mathcal{V}_1$ by employing the local   measurements \eqref{eq_system2}--\eqref{eq_system3} and the messages from neighbor vehicles.
%	 To design the resilient observer, in the following, we derive an augmented measurement equation according to the local measurements \eqref{eq_system2}--\eqref{eq_system3} and the messages from neighbor vehicles.
	 %	Each vehicle $j$ can measure the relative measurement between itself and its front vehicle by using sensors like camera or radar.
	 Denote   $y_{i|j}(t)$, $j=i-1,i+1$   the absolute measurement of vehicle $i$ from the view of vehicle $j$, calculated as follows:
	 \begin{align}\label{eq_system01}
	 y_{i|j}(t)=
	 \begin{cases}
	 y_{j,j}(t)+\sum_{m=j+1}^{i}y_{m-1,m}(t), &\text{if } i>j\\
	 y_{i,i}(t),  &\text{if } i=j\\
	 y_{j,j}(t)-\sum_{m=i+1}^{j}y_{m-1,m}(t), &\text{if } i<j
	 \end{cases}
	 \end{align}
	 Substituting \eqref{eq_system2} and \eqref{eq_system3} into \eqref{eq_system01} yields  
	 $y_{i|j}(t)=x_{i}(t)+a_{j}(t)+n_{i|j}(t),$
	 where 
	 \begin{align*}
	 n_{i|j}(t)=
	 \begin{cases}
	 n_{j,j}(t)+\sum_{m=j+1}^{i}n_{m-1,m}(t), &\text{if } i>j\\
	 n_{i,i}(t),  &\text{if } i=j\\
	 n_{j,j}(t)-\sum_{m=i+1}^{j}n_{m-1,m}(t), &\text{if } i<j.
	 \end{cases}
	 \end{align*}
	 Under Assumption \ref{ass_noise}, it holds that for any $j\in\mathcal{N}_i$,
	 \begin{align}\label{new_bound_noise}
	 \norm{ n_{i|j}(t)}\leq (L+1)\mu=:\bar \mu.
	 \end{align}

	 Through the graph $\mathcal{G}$, vehicle $i\in\mathcal{V}_1$ is able to receive the absolute measurements (i.e., $\{y_{j,j}(t)\}$, $j\in\mathcal{N}_i$) and relative measurements  (i.e., $\{y_{j-1,j}(t)\}$), and then  calculate the measurements $\{y_{i|j}(t)\}_{j\in\mathcal{N}_{i}\bigcup \{i\}}$. 
	 Hence, it is feasible to reconstruct the measurement equation of vehicle $i\in\mathcal{V}_1$:
	 %	 Then  we obtain the following  the measurement equation for vehicle $i\in\mathcal{V}_1$:
	 \begin{equation}\label{eq_system5}
	 \begin{split}
	 z_i(t)&=Cx_i(t)+\boldsymbol{a}_i(t)+\boldsymbol{n}_i(t),
	 \end{split}
	 \end{equation}
	 where $C=\begin{pmatrix}
	 I_{2}&	I_{2}& \cdots &	I_{2}
	 \end{pmatrix}^{\sf T}\in\mathbb{R}^{(4L+2)\times 2},$ and
	 \begin{align*}
	 \begin{split}
	 %	z_1(t)&=[y_{1,1}^{\sf T}(t), y_{1|2}^{\sf T}(t),y_{1|3}^{\sf T}(t)]^{\sf T},\\
	 %	\boldsymbol{a}_1(t)&=[a_{1}^{\sf T}(t), a_{2}^{\sf T}(t), a_{3}^{\sf T}(t)]^{\sf T},\\
	 %	\boldsymbol{d}_1(t)&=[d_{1,1}^{\sf T}(t), d_{1|2}^{\sf T}(t),d_{1|3}^{\sf T}(t)]^{\sf T},\\
	 %  z_N(t)&=[y_{N|N-2}^{\sf T}(t), y_{N|N-1}^{\sf T}(t),y_{N,N}^{\sf T}(t)]^{\sf T},\\
	 z_i(t)&=(y_{i|i-L}^{\sf T}(t),y_{i|i-L+1}^{\sf T}(t),\dots,y_{i|i+L}^{\sf T}(t))^{\sf T}\in\mathbb{R}^{4L+2},\\
	 \boldsymbol{a}_i(t)&=(a_{i-L}^{\sf T}(t),a_{i-L+1}^{\sf T}(t),\dots,a_{i+L}^{\sf T}(t))^{\sf T}\in\mathbb{R}^{4L+2},\\
	 \boldsymbol{n}_i(t)&=(n_{i|i-L}^{\sf T}(t),n_{i|i-L+1}^{\sf T}(t),\dots,n_{i|i+L}^{\sf T}(t))^{\sf T}\in\mathbb{R}^{4L+2}.
	 %	z_N(t)&=[y_{N|N-2}^{\sf T}(t), y_{N|N-1}^{\sf T}(t),y_{N,N}^{\sf T}(t)]^{\sf T},\\
	 %	\boldsymbol{a}_N(t)&=[a_{N-2}^{\sf T}(t), a_{N-1}^{\sf T}(t),a_{N}^{\sf T}(t)]^{\sf T},\\
	 %	\boldsymbol{d}_N(t)&=[n_{N|N-2}^{\sf T}(t), n_{N|N-1}^{\sf T}(t),n_{N,N}^{\sf T}(t)]^{\sf T},\\
	 %	C&=\underbrace{\begin{pmatrix}
	 %		I_{2}&	I_{2}& \cdots &	I_{2}
	 %	\end{pmatrix}^{\sf T}}_{\text{the number is }2L+1},
	 \end{split} 
	 \end{align*}	
	 \begin{remark}
	 	The attack signal $\boldsymbol{a}_i(t)$ has at most $2b$ non-zero elements, which means at least $4L+2-2b$ elements of $z_i(t)$ are not under attack. If $L\geq b$, according to the  sparse observability \cite{shoukry2016event}, the measurement redundancy  in \eqref{eq_system5} enables us to design an effective resilient observer for vehicle $i\in\mathcal{V}_1$.
	 	%				 in the case that some vehicles are under attacks.
	 \end{remark}
	 
	 %		For  vehicle $i$,  $i\in\mathcal{V}_1$, through Appendix \ref{app_derivation}, we obtain the following  the measurement equation:
	 %		\begin{equation}\label{eq_system5}
	 %		\begin{split}
	 %		z_i(t)&=Cx_i(t)+\boldsymbol{a}_i(t)+\boldsymbol{d}_i(t),
	 %		\end{split}
	 %		\end{equation}
	 %		where $z_i(t)$, $C$, $\boldsymbol{a}_i(t)$, and $\boldsymbol{d}_i(t)$ are given in   Appendix \ref{app_derivation}. 
	 %	
	 
	 %	Since vehicle $i\in\mathcal{V}_2$ has less measurements compared to the one in $\mathcal{V}_1$, considering of the worst attack scenario where most neighbors of vehicle $i$ are under attack, we consider to use the estimates of the vehicles in $\mathcal{V}_1$ in the measurement reconstruction of vehicle $i\in\mathcal{V}_2$. 
	 %%	to ensure it can obtain stable estimates, 
	 %		Each vehicle	$i\in \mathcal{V}_2$ is able to receive the messages from its  $2L$-neighbor vehicle $j$,   $j\in \mathcal{N}_i\bigcap \mathcal{V}_1$.
	 Next, we reconstruct the  measurement equation of vehicle~$i\in\mathcal{V}_2$ by using 	 the messages from neighbor vehicles: 
	 %		Denote the prediction error by $\bar e_{j_i}=\bar  x_{j_i}-x_{j_i}$, then substituting it into \eqref{pf_rela1} and \eqref{pf_rela12} leads to the following measurement equation for $i\in\mathcal{V}_2$:
	 \begin{align}\label{eq_measure}
	 %		\boldsymbol{\hat y}_{i}=\boldsymbol{\hat C}_{i|j}x_i+\boldsymbol{\hat d}_{i|j}.
	 \hat y_{i|j}&=x_i+\hat n_{i|j}, \quad  	j\in\mathcal{N}_i\bigcap \mathcal{V}_1=:\mathcal{\hat{N}}_i,
	 \end{align}
	 where $\hat y_{i|j}$ is the absolute measurement of vehicle $i$ from the view of vehicle $j$ subject to 
	 %		$\hat y_{i|j}=\bar x_{j}-\sum_{m=i+1}^{j}y_{m-1,m}$, if $j>i$; $\hat y_{i|j}=\bar x_{j}+\sum_{m=j+1}^{i}y_{m-1,m},$ if $j<i,$
	 \begin{align*}
	 \hat y_{i|j}=
	 \begin{cases}
	 \bar x_{j}-\sum_{m=i+1}^{j}y_{m-1,m}&\text{if }j>i\\
	 \bar x_{j}+\sum_{m=j+1}^{i}y_{m-1,m}&\text{if }j<i,
	 \end{cases}
	 \end{align*}
	 and the noise $\hat n_{i|j}$ is subject to
	 \begin{align}\label{pf_d}
	 \hat n_{i|j}=
	 \begin{cases}
	 \bar  x_{j}-x_{j}-\sum_{m=i+1}^{j}n_{m-1,m} &\text{if }j>i\\
	 \bar  x_{j}-x_{j}+\sum_{m=i+1}^{j}n_{m-1,m} &\text{if }j<i.
	 \end{cases}
	 \end{align}
	As seen, vehicle $i\in\mathcal{V}_2$ uses the estimate $\bar x_j$ from neighbor vehicle $j$ and the relative measurements $\{y_{m-1,m}\}$ from neighbor vehicle $m$, where $j\in\mathcal{N}_i\bigcap \mathcal{V}_1$ and $m\in\mathcal{N}_i$.	 
	 In next section, we will design a resilient observer for vehicles $i\in\mathcal{V}_1$ and $i\in\mathcal{V}_2$ with the reconstructed measurements in \eqref{eq_system5} and \eqref{eq_measure}, respectively.
	 
	 %		Let vehicle $j_i\in\mathcal{\hat{N}}_i$ be the vehicle closest to vehicle $i$, i.e.,
	 %		\begin{align}
	 %		j_i=\min j,  j\in\mathcal{\hat{N}}_i.
	 %		\end{align}
	 %%		Since vehicle $j_i$ is able to obtain good estimates of its state by receiving and employing redundant measurements, we  aim to use the estimates of vehicle $j_i$ in the observer design for vehicle $i$, so as to provide good state estimates of vehicle  $i$. 
	 %		For $j_i>i$, recursively employing \eqref{eq_system3} yields
	 %		\begin{align}\label{pf_rela1}
	 %		x_{j_i}-\sum_{m=i+1}^{j_i}y_{m-1,m}&=x_i-\sum_{m=i+1}^{j_i}n_{m-1,m}, 
	 %		\end{align}
	 %		For $j_i<i$, recursively employing \eqref{eq_system3} yields
	 %		\begin{align}\label{pf_rela12}
	 %		x_{j_i}+\sum_{m=j_i+1}^{i}y_{m-1,m}&=x_{i}+\sum_{m=i+1}^{j_i}n_{m-1,m}, 
	 %		\end{align}
	 %		
	 %	Then the reconstruction measurement equation is 

	\section{Observer Design}\label{sec:observer}
	%		In this section, for each vehicle $i$,  $i\in\mathcal{V}$,  we will first design a local observer to estimate its velocity and position.

	%
	% Each vehicle $i$, $i=2,\dots,N-1$, is able to receive the measurements (including $y_{i,i}$ and $y_{i-1,i}$) of  vehicle $i-1$ in the front and vehicle $i+1$ in the rear, respectively. 
	
	%Thus, vehicle $3$ can communicate with vehicles $1,2,4$, vehicle $N-2$ can communicate with vehicles $N-3,N-1,N$. For convenience, we assume vehicles $3$ (res. $N-2$)  will not treat vehicle $1$ (res. $N$) as its neighbor. 
	
	%Since vehicle $i$,  $i\in\{2,\dots,N\}$, has the measurements of itself and its front vehicle, we assume that it  transmits/receives all the measurements to/from its  neighbors.
	
	%We consider the scenario of $n$ neighbor communication, where each vehicle is able to communicate with its nearest $n$ vehicles in the front and  in the rear, respectively. For the vehicles without enough neighbors (e.g., when $n\geq 1$, vehicle $N$ has no neighbors in the rear), they just communicate with their existing neighbors.
	%Since vehicle $i$,  $i\in\{2,\dots,N\}$, has the measurements of itself and its front vehicle, we assume that it  transmits all the measurements to its $N$ neighbors.

	% for the leader and tail vehicles, we have their measurement equations in the following
	%\begin{align}
	%  z_1(t)=x_1(t)+n_{1,1}(t)\\
	%    z_N(t)=x_N(t)+n_{N,N}(t)\\
	%\end{align}
	%where 
	
	%For consistent notations, we let 
	%\begin{align*}
	% y_{i|i}=y_{i,i},n_{i|i}=n_{i,i},  i\in\{2,\dots,N\}.
	%\end{align*}
	
	In this section,  we design an observer algorithm and analyze an asymptotic upper bound of the estimation error with a static observer threshold and an adaptive observer threshold, respectively.	
	%	for each vehicle $i$ based on its position in the  network $\mathcal{G}$ and   whether vehicle sensor $i$ is attack-free, i.e., $i\in \mathcal{\hat S}_i(t)$, or not.	
	%Before we move forward, the following assumption is given.
	Since the observer algorithm to be designed uses the detection results, we need the following assumption in this section.
	\begin{assumption}\label{ass_detection}
		The  sets $\mathcal{\hat S}_i(t)$ and $\mathcal{\hat  S}_i^a(t)$ introduced in \eqref{message} satisfy the following two properties:
		\begin{enumerate}[label=\roman*)]
			\item  monotonically non-decreasing, i.e., 	
			$\mathcal{\hat  S}^a_i(t_1)\subseteq\mathcal{\hat  S}^a_i(t_2), $ and $\mathcal{\hat  S}_i(t_1)\subseteq\mathcal{\hat  S}_i(t_2), $  if $t_1\leq t_2$;
			\item  no false alarm at each time, i.e.,   $\mathcal{\hat S}_i(t)$ and $\mathcal{\hat  S}_i^a(t)$ are fault-free, $t=1,2,\dots$.
		\end{enumerate} 
	\end{assumption}
	This assumption   is removed after we introduce the detector in Section \ref{sec:detection}. In other words, the integrated observer and detector in this paper satisfy 
	Assumption \ref{ass_detection}  (see  Lemma~\ref{prop_ass}).

	\subsection{Observer algorithm}
	
	%	Since the leader vehicle is control-free, i.e., $u_1(t)=0$, 	
	%	From \eqref{eq_local_state}, \eqref{eq_local_state2}, and  \eqref{eq_system5},  we have the reformulated state equation  and  measurement equation  of vehicle $i$,  $i\in\{L+1, \dots,N-L\}$ in the following
	%	\begin{align}\label{system}
	%	\begin{split}
	%	x_i(t)&=Ax_i(t-1)+[0,Tu_i(t-1)]^{\sf T}+n_{i}(t-1)\\
	%	z_i(t)&=Cx_i(t)+\boldsymbol{a}_i(t)+\boldsymbol{d}_i(t).
	%	\end{split}
	%	\end{align}
	%Denote $\eta_i^{[r]}(t)$ the $r$th element of $\eta_i(t)$, where $r=1,\dots,6$ for $i=1,\dots,N$. =
	
	%To employ the measurement innovation $\eta_i(t)$ effectively,  we design a saturation scheme as follows.
	%For each $m=\{1,2,\dots,2L+1\}$, let
	%\begin{align}\label{eq_K}
	%k_{i,j_m}(t)=\begin{cases}
	%1,\qquad\qquad\text{ if } \norm{\eta_{i,j_m}(t)}\leq\beta(t-1),\\
	%\frac{\beta(t-1)}{\norm{\eta_{i,j_m}(t)}}, \quad\text{ otherwise},
	%\end{cases}
	%\end{align}
	%where $\beta(t)\geq 0$ is a time-varying saturation parameter to be designed. 
	%Denote the measurement gain for vehicle $i\in\mathcal{V}_1$ by
	%\begin{align}\label{eq_notation}
	%\begin{split}
	%%\eta_i(t)&=[\eta_{i,j_m}^{[r]}(t)]_{m=\{1,2,\dots,2L+1\},r=\{1,2\}}, \\
	%K_i(t)&=\diag\{k_{i,j_m}(t)I_2\}_{m=\{1,2,\dots,2L+1\}}.
	%\end{split}
	%\end{align}
	%where    $	K_i(t)\in\mathbb{R}^{(4L+2)\times (4L+2)}.$
	
	%For vehicle $i\in \mathcal{V}_2$, denote $\mathcal{\hat{N}}_i=\mathcal{N}_i\bigcap \mathcal{V}_1$.  
	From  the reconstructed measurement equation \eqref{eq_system5},
	we denote  the    innovation of vehicle $i\in\mathcal{V}_1$  by $z_i(t)-C\bar x_i(t)=\eta_i(t)=[\eta_{i,m_s}(t)]_{s=\{1,2,\dots,2L+1\}}, $
	where $m_s\in\mathcal{N}_i\cup\{i\}$, $\eta_{i,m_s}(t)\in\mathbb{R}^{2},$   and $\eta_i(t)\in\mathbb{R}^{4L+2}$.
	For example, when $L=1$ and $i\in \{2,\dots,N-1\}$,    we have $m_1=i-1,m_2=i,m_3=i+1$.		
	For each vehicle $i\in\mathcal{V}$,
	given the   sets $ \{\mathcal{\hat S}_i(t),\mathcal{\hat  S}_i^a(t)\}$ from the detector, we design the following observer  by employing the measurements from \eqref{eq_system2}, \eqref{eq_system5}, and \eqref{eq_measure}:
	\begin{align}\label{update_2}
	\hat x_{i}(t)=
	\begin{cases}
	\bar x_{i}(t)+ \frac{1}{2L}C^{\sf T}K_i(t)\eta_i(t), &\text{if }i\in\mathcal{V}_1\\
	\bar x_i(t)+\frac{1}{\varpi}( y_{i,i}(t)-\bar x_i(t)), &\text{if }i\in\mathcal{V}_2\bigcap \mathcal{\hat S}_i(t),\\
	\bar x_i(t)+\frac{1}{\varpi}(\hat y_{i|j_i(t)}(t)-\bar x_i(t)), &\text{if }i\in\mathcal{V}_2\setminus\mathcal{\hat S}_i(t),
	\end{cases}
	\end{align}		
	where 
	%$\varpi\in(1,\frac{\norm{A}}{\norm{A}-1})$, and 
	%$j_i$ is given as follows  
	\begin{align}\label{obser_parameter}
	\begin{split}
	\forall \varpi&\in\left (1,\frac{\norm{A}}{\norm{A}-1}\right )\\
	j_i(t)&=\arg\min_{j\in\mathcal{\hat{N}}_i\cup \mathcal{\hat S}_i(t)} |j-i|\\
	K_i(t)&=\diag\{k_{i,m_s}(t)I_2\}_{s=\{1,2,\dots,2L+1\}},
	\end{split}
	\end{align}
	%and
	% $\eta_i(t)=z_i(t)-C\bar x_i(t)$,  $i\in\mathcal{V}_1$, and 
	%\begin{align*}
	%%\eta_i(t)=z_i(t)-C\bar x_i(t)\\
	%K_i(t)=\diag\{k_{i,m_s}(t)I_2\}_{s=\{1,2,\dots,2L+1\}},
	%\end{align*}
	where $\mathcal{\hat{N}}_i$ is introduced in \eqref{eq_measure}, and $k_{i,m_s}(t)$ is designed by leveraging the following saturation method with a   threshold $\beta_i(t)>0$ (designed in Subsections \ref{subsec:beta_static} and \ref{subsec:beta_varying}):
	%		To achieve the two points,  the observer gain $K_{i}(t)$ for vehicle $i\in\mathcal{V}_1$ is supposed to be adjusted, compared to the one in Algorithm \ref{alg:obser}. Recall the form of $K_i(t)$ in \eqref{eq_notation}: 
	%		\begin{align}\label{eq_notation2}
	%		%	K_i(t)&=:\diag\{k_{i,j_1}^{[1]}(t),k_{i,j_1}^{[2]}(t),k_{i,j_2}^{[1]}(t),k_{i,j_2}^{[2]}(t),\\
	%		%	&\qquad\qquad\qquad k_{i,j_3}^{[1]}(t),k_{i,j_3}^{[2]}(t)\}.\nonumber
	%		%K_i(t)&=\diag\{k_{i,j_m}^{[r]}(t)\}_{m=\{1,2,\dots,2L+1\},r=\{1,2\}}.
	%		K_i(t)&=\diag\{k_{i,j_m}(t)I_2\}_{m=\{1,2,\dots,2L+1\}}.
	%		\end{align}
	%		Then we adjust the design of the elements $k_{i,j_m}(t)$,   $m=1,2,\dots,2L+1$ in the following.	
	%For $s=1,2,\dots,2L+1$, let
	%	if  $j_m\in\mathcal{\hat  S}^a_i(t)$ or $j_m\in \mathcal{\hat S}_i(t)$, then 
	\begin{align}\label{eq_K2}
	k_{i,m_s}(t)=\begin{cases}
	0, &\text{if } m_s\in\mathcal{\hat  S}^a_i(t)\\
	1, &\text{if } m_s\in\mathcal{\hat S}_i(t)\\
	\min\left\{1,\frac{\beta_i(t)}{\norm{\eta_{i,m_s}(t)}}\right\}, &\text{otherwise.}
	\end{cases}
	\end{align}

	%	First, we will design an observer for vehicle $i$,  $i\in\mathcal{V}_1$,  to estimate its velocity and position by employing the measurements received from its $2L$-neighbor vehicles. Then for vehicle $i\in\mathcal{V}_2$, its estimate is updated by using the estimates of its closest neighbor within the set $\mathcal{V}_1$.  Denote the state estimate and the predicted state estimate by $\hat x_i(t)$ and $\bar x_i(t)$, respectively. 
	
	%	To alleviate the influence of the malicious attacker to the platooning performance, we are desired to remove 	 the measurements   of the surely attacked vehicles, i.e., the vehicles in the set $\mathcal{\hat  S}^a_i(t).$ Moreover, for the surely attacked-free vehicles, i.e., the vehicles in the set $\mathcal{\hat S}_i(t)$, their measurements are worth fully trusting. 
	
	\begin{remark}
		The observer \eqref{update_2} shows: i)
		For one sensor in the set $\mathcal{V}_1$, if it is attacked, i.e., $m_s\in\mathcal{\hat  S}^a_i(t),$ its measurements are no longer employed, i.e., $k_{i,m_s}(t)=0$; If it is attack-free, i.e., $m_s\in\mathcal{\hat S}_i(t),$ its measurements are fully trusted, i.e., $k_{i,m_s}(t)=1$. Otherwise, the saturation method with the threshold $\beta_i(t)$ can reduce the influence of the potentially compromised measurements.
		ii)
		   For each vehicle  $i\in\mathcal{V}_2$, if it is  attack-free (i.e., $i\in\mathcal{V}_2\bigcap \mathcal{\hat S}_i(t)$), it uses its own local measurements with full trust to update the state estimate, otherwise, it  uses the estimate of vehicle $j_i(t)$ which is either in the set $\mathcal{V}_1$ with redundant measurements or in the set of attack-free vehicle sensors $\mathcal{V}_2\bigcap\mathcal{\hat S}_i(t).$ 
	\end{remark}
	
	\begin{remark}
		The reason to find   vehicle $j_i(t)$, which is nearest to vehicle $i$, is to alleviate the influence of the noise in  relative measurements. This is seen from  \eqref{pf_d}, where $\hat n_{i|j_i}(t)$ includes  the noise of the relative measurements from vehicles $j_i$ to~$i$. 
	\end{remark}

	%
	%
	%Then through Appendix \ref{app_derivation2}, we have
	%\begin{align}\label{eq_measure}
	%%		\boldsymbol{\hat y}_{i}=\boldsymbol{\hat C}_{i|j}x_i+\boldsymbol{\hat d}_{i|j}.
	%\hat y_{i|j_i}&=x_i+\hat n_{i|j_i}, \quad j_i\in\mathcal{\hat{N}}_i,
	%\end{align}
	%where $\hat y_{i|j_i}$ and $\hat n_{i|j_i}$ are defined in \eqref{pf_y} and \eqref{pf_d}, respectively, and vehicle $j_i\in\mathcal{\hat{N}}_i$ is the vehicle closest to vehicle $i$, i.e., 
	%\begin{align*}
	%j_i=\arg\min_{j\in\mathcal{\hat{N}}_i} |j-i|.
	%\end{align*} 
	For each vehicle $i\in\mathcal{V}$, based on \eqref{eq_system5}--\eqref{eq_measure} and \eqref{update_2}--\eqref{eq_K2}, we propose a resilient observer in Algorithm \ref{alg:obser}. 
	\begin{algorithm}
						\caption{Resilient Observer}
		\label{alg:obser}
		\small
		\begin{algorithmic}[1]
%		\SetAlgoLined
%		\SetKwInOut{Input}{init.}
%		\SetKwInOut{Output}{output}
		\STATE{\textbf{Initialization}: Initial estimate $\bar x_{i}(0)$, observer parameter $\varpi\in(1,\frac{\norm{A}}{\norm{A}-1})$,  saturation parameter $\beta_i(t)$, and vehicle communication  parameter $L$}\\
		\STATE{\textbf{Output}: State estimate $\hat x_{i}(t)$}\\
		%		\KwResult{Write here the result }
		%		initialization\;
		\FOR{$t\geq 0$}
		\STATE{
			\textbf{Communications between neighboring vehicles:}  Vehicle~$i$ sends out $\mathcal{M}_i$ defined in \eqref{message};\\
			%its measurements $y_{i-1,i}(t)$, $y_{i,i}(t)$,  and the estimate $\bar x_i(t)$ to its neighboring vehicle $j\in\mathcal{N}_i$\;
			\textbf{Time update:} For each vehicle $i$, $i\in\mathcal{V}$;
\begin{align}\label{pred_equation}
\bar x_i(t)=A\hat x_i(t-1)+[0,Tu_i(t-1)]^{\sf T},
\end{align}
			where $u_i(t)$ is specifically designed by vehicle $i$;\\
			\textbf{Measurement update:} See \eqref{update_2}.
		}
		\ENDFOR
	\end{algorithmic}
	\end{algorithm}
	
	%\subsubsection{Design of time-varying $\beta$}
	%
	%
	%
	%
	%\subsubsection{Design of static $\beta$}
	
	Next, we study a real-time upper bound of the estimation error of  Algorithm \ref{alg:obser}.
	In the following a)--c) items, we define three  sequences, namely, $\rho_i(t)$, $\lambda_i(t)$, and $\tau_i(t)$, which are proved in Proposition \ref{lem_detec22}   to be the upper bounds of the estimation errors of the  three updates  \eqref{update_2}.

	\textbf{a)} For vehicle $i\in\mathcal{V}_1$,
	%, since the influence of the attacker is alleviated by the designed $K_i(t)$,  we are able to obtain an estimation error  bound  by employing the sets $\mathcal{\hat  S}^a_i(t)$ and  $\mathcal{\hat S}_i(t)$.
	we denote $\mathcal{\hat S}_{i,1}(t)$  the estimate of the set of   attack-free vehicle sensors in the  $2L$-neighborhood of vehicle sensor   $i$,  i.e.,
	\begin{align}\label{eq_normal}
	\begin{split}
	%\bar \mathcal{\hat  S}^a_i(t)&=\mathcal{\hat  S}^a_i(t)\bigcap \left(\mathcal{N}_i\bigcup \{i\}\right)\\
	\mathcal{\hat S}_{i,1}(t)&=\mathcal{\hat S}_i(t)\bigcap \left(\mathcal{N}_i\bigcup \{i\}\right).
	\end{split}
	\end{align}
	Then, for  $i\in\mathcal{V}_1$, we
	define a sequence $\{\rho_i(t)\}$ with $	\rho_i(0)=q$ in the following
	\begin{align}\label{sequ_detect}
	\rho_i(t)=\bar m_{i}(t)\norm{A}	\rho_i(t-1)+\bar Q_i(t),
	\end{align}
	where 
	\begin{align*}
	%\begin{split}
	\bar  m_{i}(t)=&1-\frac{|\mathcal{\hat S}_{i,1}(t)|+(2L+1-b-|\mathcal{\hat S}_{i,1}(t)|)\bar k_i(t)}{2L},\nonumber\\
	\bar k_i(t)=&\min\left\{1,\frac{\beta_i(t)}{\norm{A}\rho_i(t-1)+\epsilon+\bar\mu}\right \},\\
	\bar Q_i(t)=&\frac{(\epsilon+\bar\mu)(2L+1-b)+(b-| \mathcal{\hat  S}^a_i(t)|)\beta_i(t)}{2L}.\nonumber
	%\end{split}
	\end{align*}

	\textbf{b)} For vehicle $ i\in\mathcal{V}_2\bigcap\mathcal{\hat S}_i(t)$, we define a sequence $\{\lambda_i(t)\}$, as follows
	\begin{align}\label{lambada}
	\lambda_i(t)=&\frac{(\varpi-1)\norm{A}}{\varpi}\lambda_i(t-1)+\frac{\epsilon(\varpi-1)+\mu}{\varpi},
	\end{align}
	where the parameter $\varpi$ is introduced in \eqref{obser_parameter}, $\lambda_i(T_i)=\tau_i(T_i)$,  
 the sequence $ \{\tau_i(t)\}$ is to be defined in  \eqref{tau_detect}, and $T_i$ is the time after which vehicle sensor $i$ is attack-free by detection, i.e.,  
 $T_i=\min \bar t$, s.t., $i\in\mathcal{\hat S}_i(\bar t+1)$. \vskip 5pt
	
	\textbf{c)} For vehicle $ i\in\mathcal{V}_2\setminus\mathcal{\hat S}_i(t)$,   we   define a sequence $\{\tau_i(t)\}$, as follows
	\begin{align}\label{tau_detect}
	\begin{split}
	\tau_i(t)=&\frac{(\varpi-1)\norm{A}}{\varpi}\tau_i(t-1)\\
	&+\frac{\epsilon\varpi+\mu|j_i(t)-i|+\norm{A}s_i(t-1)}{\varpi},
	\end{split}
	\end{align}
	where $\tau_i(0)=q,$   $j_i(t)$ is given in \eqref{obser_parameter}, and $s_i(t-1)=\rho_{j_i}(t-1),$ $\text{if }j_i(t)\in \mathcal{V}_1$, otherwise $s_i(t-1)=\lambda_{j_i}(t-1),$
	%\begin{align*}
	%%j_i=&\arg\min_{j\in\mathcal{\hat{N}}_i\cup \mathcal{\hat S}_i(t-1)} |j-i|,\\
	%s_i(t-1)=&\begin{cases}
	%\rho_{j_i}(t-1), &\text{if }j_i\in \mathcal{V}_1\\
	%\lambda_{j_i}(t-1), &\text{otherwise},
	%\end{cases}
	%\end{align*}
	where $\rho_{j_i}(t)$ and $\lambda_{j_i}(t)$ are given in \eqref{sequ_detect} and \eqref{lambada}, respectively.
	\begin{remark}
		Although the constructions of the two sequences $\{\lambda_i(t)\}$ and $\tau_i(t)$ need each other, they are both well defined. Because, $\tau_i(t)$ starts at time $t=0$, which does not require $\lambda_i(t)$, and  $\lambda_i(t)$ starts at $t=T_i$. 
	\end{remark}	
	\begin{proposition}\label{lem_detec22}
		Consider Algorithm \ref{alg:obser} for the system \eqref{eq_local_state}--\eqref{eq_system3}  satisfying Assumptions \ref{ass_attack}--\ref{ass_detection}.  The estimation error of each vehicle $i\in\mathcal{V}$ is subject to
		\begin{align}\label{real_bounds}
		\norm{\hat x_{i}(t)-x(t)}\leq \alpha_i(t):=
		\begin{cases}
		\rho_i(t), &\text{if }  i\in\mathcal{V}_1,\\
		\lambda_i(t), &	\text{if }i\in\mathcal{V}_2\bigcap \mathcal{\hat S}_i(t),\\
		\tau_i(t), &\text{if }  i\in\mathcal{V}_2\setminus\mathcal{\hat S}_i(t),
		%			\bar \sigma_i(t),
		\end{cases}
		\end{align}
		where $\rho_i(t)$, $\lambda_i(t)$ $\tau_i(t)$ are given in \eqref{sequ_detect}, \eqref{lambada}, and \eqref{tau_detect}, respectively.
	\end{proposition}
	\begin{pf}
		See Appendix \ref{pf_lem_detec22}.
	\end{pf}
	\begin{remark}
		Based on local information and the vehicle-to-vehicle network 
		$\mathcal{G}$, vehicle $i\in\mathcal{V}$ is able to compute the  sequence $\{\alpha_i(t)\}$.  It enables  evaluation of the error bounds offline by setting $\mathcal{\hat  S}^a_i(t)\equiv\mathcal{\hat S}_i(t)\equiv\emptyset$, which reduces to the case without detection.
	\end{remark}
	
	Since the observer threshold $\beta_j(t), j\in\mathcal{V}_1,$ in \eqref{eq_K2} is essential,  we study the properties of Algorithm \ref{alg:obser} by designing $\beta_j(t)$ in a static way and in an adaptive way respectively in the following two subsections.
	
	\subsection{Observer property with static  threshold}\label{subsec:beta_static}
	%In this subsection, we analyze the observer properties by designing a static observer threshold $\beta_i$ in \eqref{eq_K2}.
	In this subsection, we design the observer threshold  $\beta_j(t)\equiv\beta_j,$ for all $j\in\mathcal{V}_1.$
	Given a scalar $\omega\in(0,1)$, denote
	\begin{align}\label{eq_beta}
	%\begin{split}
	\beta_0&=\norm{A}q+\epsilon+\bar\mu\nonumber\\
	\bar\beta_1(\omega)&=\frac{2L}{2L+1-b}\frac{\left(\omega+\norm{A}-1\right)\beta_0}{\norm{A}}\\
	%\bar\beta_1(\omega)&=\frac{2L}{2L+1-b}\frac{\left(\omega+\norm{A}-1\right)\left(\norm{A}q+\epsilon+\bar\mu\right)}{\norm{A}}\\
	%\frac{2L}{2L+1-b}\left((\omega+\norm{A}-1) q-\frac{\left(\epsilon+\bar\mu\right)\left(\omega+\norm{A}-1\right)}{\norm{A}}\right)\\
	%
	\bar\beta_2(\omega)&=\min\left\{\beta_0,\frac{2L}{b}\left(\omega q-\frac{\left(\epsilon+\bar\mu\right)\left(2L+1-b\right)}{2L}\right)\right\},\nonumber
	%\end{split}
	\end{align}
	where $\bar \mu$ is defined in \eqref{new_bound_noise}.	
	In the following theorem, we study the boundedness of the estimation error of the observer in Algorithm \ref{alg:obser} with a static observer threshold $\beta_j,$ $j\in\mathcal{V}_1$ introduced in   \eqref{eq_K2}. 
	\begin{theorem}\label{thm_estimation2}
		Consider the observer in Algorithm \ref{alg:obser} for the system \eqref{eq_local_state}--\eqref{eq_system3}  satisfying Assumptions \ref{ass_attack}--\ref{ass_detection}. Given the sets $\mathcal{\hat S}_i(T_i)$ and $\mathcal{\hat  S}^a_i(T_i)$ at time $T_i$  for any $i\in\mathcal{V}$, 	if there is a scalar $\omega\in(0,1)$, such that $0<\bar\beta_1(\omega)<\bar\beta_2(\omega)$,  then for any $ \beta_j\in (\bar\beta_1(\omega),\bar\beta_2(\omega))$ with  $ j\in\mathcal{V}_1$,  the estimation error of vehicle $i$ is asymptotically upper bounded, i.e.,
		\begin{align*}
		&\limsup\limits_{t\rightarrow \infty}\norm{\hat x_i(t)-x_i(t)}\leq 
		\begin{cases}
		\tilde\alpha_1, &\text{if }  i\in\mathcal{V}_1,\\
		\tilde\alpha_2, & \text{if }i\in\mathcal{V}_2\bigcap \mathcal{\hat S}_i(T_i),\\
		\tilde\alpha_3, &\text{if }  i\in\mathcal{V}_2\setminus\mathcal{\hat S}_i(T_i)),
		\end{cases}
		\end{align*}
		where $\bar\beta_1(\omega)$ and $\bar\beta_2(\omega)$ are defined in \eqref{eq_beta}, and
		\begin{align}\label{bounds}
		\begin{split}
		\tilde\alpha_1&=\frac{\tilde Q_i}{1-\tilde  m_i\norm{A}}\\
		\tilde\alpha_2&=\frac{\epsilon(\varpi-1)+\mu}{\varpi-(\varpi-1)\norm{A}}\\
		\tilde\alpha_3&=\frac{\epsilon\varpi+\mu|j_i^*-i|+\norm{A}\max \{\tilde\alpha_1,\tilde\alpha_2\}}{\varpi-(\varpi-1)\norm{A}},
		\end{split}
		\end{align}
		in which 
		\begin{align}\label{k_seq_detect2}
		%\begin{split}
		j_i^*=&\arg\min_{j\in\mathcal{\hat{N}}_i\cup \mathcal{\hat S}_i(T_i)} |j-i|,\nonumber\\
		\tilde Q_i=&\frac{(\epsilon+\bar\mu)(2L+1-b)+(b-| \mathcal{\hat  S}^a_i(T_i)|)\beta}{2L},\nonumber\\
		\tilde  m_i=&1-\frac{|\mathcal{\hat S}_{i,1}(T_i)|+(2L+1-b-|\mathcal{\hat S}_{i,1}(T_i)|) k_i^*}{2L},\\
		k_i^*=&\frac{\beta}{\norm{A}q+\epsilon+\bar\mu},\nonumber\\
		\mathcal{\hat S}_{i,1}(T_i)=&\mathcal{\hat S}_i(T_i)\cap \left(\mathcal{N}_i\cup \{i\}\right).\nonumber
		%\end{split}
		\end{align}
		%where $\mathcal{\hat S}_{i,1}(T_i)=\mathcal{\hat S}(T_i)\bigcap \left(\mathcal{N}_i\bigcup \{i\}\right).$ 
	\end{theorem}
	\begin{pf}
		See Appendix \ref{pf_thm_estimation2}.
	\end{pf}
		 Theorem \ref{thm_estimation2} is     based on the available information at  some time $T_i\geq 0$. If $T_i=0$, $\mathcal{\hat S}_i(T_i)=\mathcal{\hat S}^a_{i}(T_i)=0$, the corresponding bound is the worst bound which can be offline obtained. With the increase of $T_i$, 
		  $|\mathcal{\hat S}_i(T_i)|$ and $|\mathcal{\hat S}^a_{i}(T_i)|$ are non-decreasing. As a result,  the  error bound is non-increasing. Thus, it motivates us to design effective detector to enlarge the sets $\mathcal{\hat S}_i(T_i)$ and $\mathcal{\hat S}^a_{i}(T_i)$.

	In the following proposition, we study the feasibility of the condition on $\omega$ in Theorem \ref{thm_estimation2}.
	\begin{proposition}\label{prop_feasibility}
		A necessary condition of the condition that 
		there is a scalar $\omega\in(0,1)$, such that $0<\bar\beta_1(\omega)<\bar\beta_2(\omega)$, is 
		\begin{align*}
		b\leq L,
		\end{align*}
		where $\bar\beta_1(\omega)$ and $\bar\beta_2(\omega)$ are introduced in \eqref{eq_beta}.
		It is also a sufficient condition, if
		%	 the time step $T>0$ (i.e., $\norm{A}>1$), the disturbance bounds $\epsilon\geq 0$  and $\mu\geq 0$ are sufficiently small and
		there exists a scalar $\omega_0\in(0,1)$,  such that 
		\begin{align}\label{condi_prop}
		\begin{split}
		\frac{2L+1-b}{b}&> \frac{\omega_0 q+f_2}{\omega_0 q-f_1}>0\\
		\frac{2L+1-b}{2L}&> \frac{\omega_0+\norm{A}-1}{\norm{A}}
		\end{split}
		\end{align}
		where $f_1=\frac{\left(\epsilon+\bar\mu\right)\left(2L+1-b\right)}{2L}$, and $f_2=\frac{\omega_0(\epsilon+\bar\mu)+\left(\norm{A}-1\right)\beta_0}{\norm{A}}$.
		%	\begin{align*}
		%	f_1&=\frac{\left(\epsilon+\bar\mu\right)\left(2L+1-b\right)}{2L}\\
		%	f_2&=f_1+\frac{\omega_0(\epsilon+\bar\mu)+\left(\norm{A}-1\right)\beta_0}{\norm{A}}.
		%	\end{align*}
	\end{proposition}
	\begin{pf}
		See   Appendix \ref{pf_prop_feasibility}.
	\end{pf}
	\begin{remark}
%		Recalling $\beta_0=\norm{A}q+\epsilon+\bar\mu$ from \eqref{eq_beta},   for $q>0$,
	It can be proved that when $b\leq L$,	if the  time step $T$ is sufficiently small, such that 
	$\norm{A}<1+\frac{(2L+1-b)(2L+1-2b)}{2bL+(2L+1-2b)(b-1)},$ then one can find  a  scalar $\omega_0\in (0,1)$ and  scalars $q,\epsilon,\mu$  satisfying Assumption \ref{ass_noise} such that 
%		 (i.e., $\norm{A}$ is sufficiently close to one) and $\epsilon+\bar\mu$ is  sufficiently small, then there is always one $\omega_0\in(0,1)$ such that 
		 the conditions in \eqref{condi_prop} are satisfied. 	  
		%$\epsilon\geq 0,\mu\geq 0,T>0$ are sufficiently small, then $f_1\geq 0$, $f_2\geq 0$, and $\norm{A}>1$ are sufficiently small.
		% As a result, the inequalities in \eqref{condi_prop} hold when   
	\end{remark}
	\begin{remark}
		The maximum number of the attacked vehicle sensors that the proposed architecture can tolerate is   $b=L=\lceil N/2\rceil -1$, which is the most general condition. Because  the sparse observability \cite{shoukry2016event} shows that if  half or more than half  vehicle sensors  are attacked, it is infeasible to recover the  states of all vehicles. 
	\end{remark}

	\subsection{Observer property with adaptive threshold}\label{subsec:beta_varying}
		In this subsection, we design the observer threshold    $\beta_j(t)$ in the following way: for $t\geq 1$,
	\begin{align}\label{eq_beta_varying}
	\begin{split}
	\beta_j(t)&=k_{j,0}\left(\norm{A}\rho_j(t-1)+\epsilon+\bar\mu\right), \quad j\in \mathcal{V}_1,
	%\rho_i(t)&=a_{21}\rho_i(t-1)+a_{22},
	\end{split}
	\end{align}
	where $\rho_j(\cdot)$ is introduced in \eqref{sequ_detect}, $\bar \mu$  is in \eqref{new_bound_noise}, and $k_{j,0}=\frac{\beta_{j,0}}{\norm{A}q+\epsilon+\bar\mu},$ in which  $\beta_{j,0}$ is a positive scalar designed in the following theorem. 
	%\begin{align*}
	%%a_{11}&=k_0\norm{A}\\
	%%a_{12}&=k_0(\epsilon+\bar\mu)\\
	%%a_{21}&=\left(1-\frac{2L+1-2b}{2L}k_0\right)\norm{A}\\
	%%a_{22}&=(\epsilon+\bar\mu)\frac{2L+1-2b+bk_0}{2L})\\
	%k_0&=\frac{\beta_0}{\norm{A}q+\epsilon+\bar\mu}.
	%\end{align*}
	\begin{theorem}\label{thm_estimation22}
		%	Under the same conditions as in Theorem \ref{thm_estimation2}, if 
		Consider the observer in  Algorithm \ref{alg:obser} for the system \eqref{eq_local_state}--\eqref{eq_system3} satisfying Assumptions \ref{ass_attack}--\ref{ass_detection}. 
		%		If there is a scalar $\omega\in(0,1)$, such that $0<\bar\beta_1(\omega)<\bar\beta_2(\omega)$,  then for $\forall \beta_{i,0}\in (\bar\beta_1(\omega),\bar\beta_2(\omega))$, $j\in\mathcal{V}_1$, given the sets $\mathcal{\hat S}_i(T_i)$ and $\mathcal{\hat  S}^a_i(T_i)$ at time $T_i$,  $i\in\mathcal{V}$, 
		Given the sets $\mathcal{\hat S}_i(T_i)$ and $\mathcal{\hat  S}^a_i(T_i)$ at time $T_i\geq 0$ for any $i\in\mathcal{V}$, 
		if there is a scalar $\omega\in(0,1)$, such that $0<\bar\beta_1(\omega)<\bar\beta_2(\omega)$, 	
		then the design of $\beta_j(t)$  in \eqref{eq_beta_varying} with $\beta_{j,0}\in (\bar\beta_1(\omega),\bar\beta_2(\omega))$ and   $j\in\mathcal{V}_1$  ensures that  the estimation error of vehicle $i$ is asymptotically upper bounded, i.e.,
		\begin{align*}
		&\limsup\limits_{t\rightarrow \infty}\norm{\hat x_i(t)-x_i(t)}\leq 
		\begin{cases}
		\bar\alpha_1, &\text{if }  i\in\mathcal{V}_1,\\
		\bar\alpha_2, & \text{if }i\in\mathcal{V}_2\bigcap \mathcal{\hat S}_i(T_i),\\
		\bar\alpha_3, &\text{if }  i\in\mathcal{V}_2\setminus\mathcal{\hat S}_i(T_i),
		\end{cases}
		\end{align*}
		where $\bar\beta_1(\omega)$ and $\bar\beta_2(\omega)$ are defined in \eqref{eq_beta}, and
		\begin{align}\label{bounds2}
		\begin{split}
		\bar\alpha_1&=\frac{a_{i,2}(T_i)}{1-a_{i,1}(T_i)\norm{A}}\\
		\bar\alpha_2&=\tilde\alpha_2\\
		\bar\alpha_3&=\frac{\epsilon\varpi+\mu|j_i^*-i|+\norm{A}\max \{\bar\alpha_1,\bar\alpha_2\}}{\varpi-(\varpi-1)\norm{A}}.
		\end{split}
		\end{align}
		in which 
		\begin{align*}
		a_{i,1}(T_i)=&1-\frac{|\mathcal{\hat S}_{i,1}(T_i)|+( \bar L-b+| \mathcal{\hat  S}^a_i(T_i)|-|\mathcal{\hat S}_{i,1}(T_i)|)k_{i,0}}{2L},\nonumber\\
		% \bar k_i(t)=&\min\{1,\frac{\beta_i(t)}{\norm{A}\rho_i(t-1)+\epsilon+\bar\mu}\},\\
		a_{i,2}(T_i)=&\frac{ \bar L+(b-| \mathcal{\hat  S}^a_i(T_i)|)k_{i,0}}{2L}(\epsilon+\bar\mu),\nonumber\\
		k_{i,0}=&\frac{\beta_{i,0}}{\norm{A}q+\epsilon+\bar\mu},\nonumber
		\end{align*}
		where $ \bar L=2L+1-b,$ the scalar $j_i^*$ and the set $\mathcal{\hat S}_{i,1}(T_i)$ are the same as in \eqref{k_seq_detect2}, and the scalar  $\tilde\alpha_2$ is in \eqref{bounds}.
		%	\begin{align}\label{k_seq_detect22}
		%	%\begin{split}
		%%	j_i^*=&\arg\min_{j\in\mathcal{\hat{N}}_i\cup \mathcal{\hat S}_i(T_i)} |j-i|,\nonumber\\
		%%	\tilde Q_i=&\frac{(\epsilon+\bar\mu)(2L+1-b)+(b-| \mathcal{\hat  S}^a_i(T_i)|)\beta}{2L},\nonumber\\
		%%	\tilde  m_i=&1-\frac{|\mathcal{\hat S}_{i,1}(T_i)|+(2L+1-b-|\mathcal{\hat S}_{i,1}(T_i)|) k_i^*}{2L},\\
		%%	k_{i,0}=&\frac{\beta_{i,0}}{\norm{A}q+\epsilon+\bar\mu},\\
		%%	\mathcal{\hat S}_{i,1}(T_i)=&\mathcal{\hat S}_i(T_i)\cap \left(\mathcal{N}_i\cup \{i\}\right).\nonumber
		%	%\end{split}
		%	\end{align}
		%where $\mathcal{\hat S}_{i,1}(T_i)=\mathcal{\hat S}(T_i)\bigcap \left(\mathcal{N}_i\bigcup \{i\}\right).$ 
	\end{theorem}
	\begin{pf}
		See Appendix \ref{pf_thm_estimation22}.	
	\end{pf}
	%From Theorem \ref{thm_estimation22}, it is straightforward to obtain the following conclusion for the system without disturbances.
	%\begin{corollary}
	%	Under the same conditions as in Theorem \ref{thm_estimation22}, if the system is disturbance-free, i.e., $\epsilon=\mu=0$, then 
	%$	\limsup\limits_{t\rightarrow \infty}\norm{\hat x_i(t)-x_i(t)}=0, \forall i\in\mathcal{V}.$
	%\end{corollary}
	\begin{remark}
		In comparison with Theorems \ref{thm_estimation2} under the same conditions,
		Theorems \ref{thm_estimation22} shows that  the adaptive design of $\beta_i(t)$   achieves better estimation performance than the static design  in the sense of providing a smaller error bound.
%		, while, on the other hand, the resource for the online calculation of the sequence $\{\rho_i(t)\}$ is required. 
	\end{remark}

	\section{Detector Design}\label{sec:detection}
	In this section, we design an attack detector algorithm  and then study when all  attacked and attack-free vehicle sensors can be identified by the detector in finite time.
	%to update the three sets  $ \{\mathcal{\hat S}_i(t),\mathcal{\hat  S}_i^a(t),\mathcal{\hat S}_i^s(t)\}$ introduced in \eqref{message}, and  study when the two sets $\mathcal{\hat S}_i(t)$ and $\mathcal{\hat  S}_i^a(t)$ converge to the true sets $\mathcal{S} $ and $\mathcal{S}^a$ in finite time.
	
	% develop two local attack detectors to find out whether it is under attack.
	
	\subsection{Detector algorithm}
	%In last subsection, a saturation based  estimation method is provided to estimate the state of each vehicle under potentially attacked measurements. Although this method can be employed directly, to improve the estimation performance and find out which vehicle is under attack, next we will study the attack detection problem. 
%	In this subsection, we design the detector algorithm. 
	Based on the relative measurements between two neighbor vehicles, 
	we consider the following detection condition:	
%	the following detector is  to identify the suspiciously attacked vehicle sensors so as to update the set $\mathcal{\hat S}^s_i(t).$
%	\begin{detector}\label{detector1}
%		For vehicle $i\in\{2,\dots,N\}$, if there is a time $t$ such that
		\begin{align}\label{item1}
		\norm{y_{i-1,i}(t)+y_{i-1,i-1}(t)-y_{i,i}(t)}>3\mu.
		\end{align}
		%	where $f_{i,i-1}(t)=y_{i-1,i}(t)+y_{i-1}(t)-y_i(t)$, 
		This condition \eqref{item1} is to infer whether either sensor $i$ or   $i-1$ is attacked under the bounded measurement noise.
%		then we claim that at least one vehicle sensor between vehicles  $i$ and $i-1$ is under attack, and  the sensors of vehicles  $i$ and $i-1$ are suspected to be attacked.
		%	 and we let $\mathcal{\hat S}^s_i(t)=\mathcal{\hat S}^s_i(t)\bigcup\{i-1,i\}$.
		%			\item 
		%			if there is a time $t$ such that \ref{item2}) holds, either vehicle $i$ or $i+1$ is under attack, and we let $\mathcal{\hat S}^s_i(t)=\mathcal{\hat S}^s_i(t)\bigcup\{i,i+1\}$;\vskip 5pt
		%			\item if there are two times under which \ref{item1}) and \ref{item2}) hold respectively, then vehicle $i$ is claimed to be under attack and we let $\mathcal{\hat  S}^a_i(t)=\{i\}$;
		%	\item if there are two times under which \ref{item1}) and \ref{item3}) hold respectively, vehicle $i-1$ is claimed to be under attack and we let $\mathcal{\hat  S}^a_i(t)=\{i-1\}$.
		%	\item if \ref{item1}) holds but \ref{item2}) and \ref{item3}) do not hold, each least
%	\end{detector}
	%	We note that  under Assumptions \ref{ass_noise}--\ref{ass_attack}, the claim in Detector \ref{detector1} holds according to   Lemma \ref{lem_relative}.
	%\begin{footnotesize}
	\begin{algorithm}
						\caption{Online Attack Detector}
							\label{alg:detec}
		\small
				\begin{algorithmic}[1]
%		\SetAlgoLined
%		\SetKwInOut{Input}{init.}
%		\SetKwInOut{Output}{output}
		\STATE{\textbf{Initialization}: Initial estimate for attacked vehicle sensor set $\mathcal{\hat  S}^a_i(0)=\emptyset,$ initial estimate for suspicious vehicle set $\mathcal{\hat S}^s_i(0)=\emptyset$, and initial estimate for attack-free  vehicle set $\mathcal{\hat S}_i(0)=\emptyset,$ $i\in\mathcal{V}$.}
		\STATE{\textbf{Output}: Sets $\mathcal{\hat  S}^a_i(t),$ $\mathcal{\hat S}^s_i(t)$, and $\mathcal{\hat S}_i(t)$}
		\FOR{$t\geq 0$}
		\STATE{
					\textbf{Communications between neighboring vehicles:}  Vehicle $i$ sends out $\mathcal{M}_i$ defined in \eqref{message}\;
					Each vehicle $i$ fuses the sets from its neighbors: 
					$\mathcal{\hat  S}^a_i(t)=\cup_{j\in\mathcal{N}_i}\mathcal{\hat  S}^a_j(t-1)\cup \mathcal{\hat  S}^a_i(t-1),$ $\mathcal{\hat S}^s_i(t)=\cup_{j\in\mathcal{N}_i}\mathcal{\hat S}^s_j(t-1)\cup \mathcal{\hat S}^s_i(t-1)$, $\mathcal{\hat S}_i(t)=\cup_{j\in\mathcal{N}_i}\mathcal{\hat S}_j(t-1)\cup \mathcal{\hat S}_i(t-1)$\;\\
%							\textbf{Detection based on Sub-detectors 1 and 2:}\\
							\IF{$i\geq 2$, \text{and} $i\notin \mathcal{\hat  S}^a_i(t)$, \text{and} $i-1\notin \mathcal{\hat  S}^a_i(t)$}
%							\STATE{
%								Run Sub-detector 1\;
%\vskip -5pt
								\IF{\eqref{item1} holds}
%								\STATE{
										\IF{$i\in \mathcal{\hat S}_i(t)$}
									\STATE{let $\mathcal{\hat  S}^a_i(t)=\mathcal{\hat  S}^a_i(t)\cup\{i-1\}$}
									\ELSIF{$i-1\in \mathcal{\hat S}_i(t)$}
									\STATE{let $\mathcal{\hat  S}^a_i(t)=\mathcal{\hat  S}^a_i(t)\cup\{i\}$}
									\ELSE
									\STATE{let $\mathcal{\hat S}^s_i(t)=\mathcal{\hat S}^s_i(t)\cup \{i-1,i\}$}
									\ENDIF
%								}
								\ENDIF								
%							}
							\ENDIF  
							\IF{$i\notin \mathcal{\hat  S}^a_i(t)$ and $i\notin \mathcal{\hat  S}_i(t)$}
%							\STATE{
%								Run Sub-detector 2\;
								\IF{\eqref{cond_detector3} holds}
								\STATE{let $\mathcal{\hat  S}^a_i(t)=\mathcal{\hat  S}^a_i(t)\cup \{i\}$}
								\ENDIF
%							}
							\ENDIF
							\IF{\eqref{eq_detec1_good} holds}
							\STATE{$\mathcal{\hat S}_i(t)=\mathcal{\hat S}_i(t)\cup\left(\mathcal{V}-\mathcal{\hat S}^s_i(t)-\mathcal{\hat  S}^a_i(t)\right)$}
							\ENDIF
							\IF{$|\mathcal{\hat  S}^a_i(t)|=b$}
							\STATE{$\mathcal{\hat S}_i(t)=\mathcal{V}-\mathcal{\hat  S}^a_i(t)$}
							\ENDIF
		}
		\ENDFOR
			\end{algorithmic}
	\end{algorithm}	
	%\end{footnotesize}	
	%The idea of the second detector is that for an attack-free vehicle, due to $a_{i}(t)=0,$ its measurement innovation is supposed to be upper bounded by a sequence related to the estimation error. Otherwise, this vehicle is under attack. 
	%For vehicle $i\notin \mathcal{\hat  S}^a_i(t)$, 
	%by comparing the  norm of the measurement innovation with a time-varying threshold   built based on the sequences $\{\rho_i(t)\}$ and $\{\tau_i(t)\}$ in \eqref{sequ_detect} and \eqref{tau_detect}, 
	Moreover, in order to find out whether sensor $i$ is under attack, we also consider the following detection condition:
	%For $i\notin \mathcal{\hat  S}^a_i(t)\bigcup \mathcal{\hat S}(t)$, we run the following detector to detect whether it is under attack.
%	\begin{detector}\label{detector3}
%		We claim vehicle sensor $i$ is  under attack, if 
		\begin{align}\label{cond_detector3}
		\begin{split}
		&\norm{y_{i,i}(t)-A\hat x_{i}(t-1)}>g_i(t),
		\end{split}
		\end{align}
		where $g_i(t)=	\epsilon+\mu+\norm{A}\rho_i(t-1)$ if $i\in\mathcal{V}_1$, otherwise, $g_i(t)=	\epsilon+\mu+\norm{A}\tau_i(t-1)$, in which  
		$\rho_i(t-1)$ and $\tau_i(t-1)$ are generated through \eqref{sequ_detect} and \eqref{tau_detect}, respectively.
%	\end{detector}	

	The two conditions in \eqref{item1}--\eqref{cond_detector3} will be used to update the two sets $\mathcal{\hat  S}^a_i(t)$ and $\mathcal{\hat S}^s_i(t)$.
%	the estimate of the attacked vehicle sensor set, i.e., $\mathcal{\hat  S}^a_i(t)$, the suspicious attacked vehicle sensor set, i.e., $\mathcal{\hat S}^s_i(t)$, and   the estimate of the attack-free vehicle sensor set, i.e., $\mathcal{\hat S}_i(t)$, at each time $t$, $i\in\mathcal{V}$. 
	%These sets provide identity information on the vehicles, which are    used in the observer by Algorithm \ref{alg:obser} to improve the estimation precision.
	%, but also can  enable decision-makings in the vehicle network, such as emergent repair or new data encryption of GPS receivers \cite{braasch1999gps}.
	Denote $\overline{\mathcal{\hat S}^s}_i(t):=\mathcal{\hat S}^s_i(t)\bigcup \mathcal{\hat  S}^a_i(t)$, which  includes the sensors under attack or suspected to be under attack. Then  we   analyze the minimal number of   attacked   sensors in the set $\overline{\mathcal{\hat S}^s}_i(t)$ as follows. 
	Split $\overline{\mathcal{\hat S}^s}_i(t)$  into multiple  subsets comprising of successive   sensor labels, i.e., $\overline{\mathcal{\hat S}^s}_{i,j}(t),j=1,2,\dots,l_i$, where $\bigcup_{j=1}^{l_i}\overline{\mathcal{\hat S}^s}_{i,j}(t)=\overline{\mathcal{\hat S}^s}_i(t)$.  It is to be proved in Lemma \ref{prop_ass} that the minimal number of    attacked   sensors in the set $\overline{\mathcal{\hat S}^s}_i(t)$ is $\sum_{j=1}^{l_i}\lceil  |\overline{\mathcal{\hat S}^s}_{i,j}(t)|/3 \rceil$, if the set $\overline{\mathcal{\hat S}^s}_i(t)$ is fault-free.	
	For instance, if $\mathcal{\hat S}^s_i(t)=\{1,2,3,9,10,11,12\}$ and $\mathcal{\hat  S}^a_i(t)=\{2,6,15\}$, then $\overline{\mathcal{\hat S}^s}_i(t)=\{1,2,3,6,9,10,11,12,15\}$. By  splitting  $\overline{\mathcal{\hat S}^s}_i(t)$, we have $\overline{\mathcal{\hat S}^s}_{i,1}(t)=\{1,2,3\}$, $\overline{\mathcal{\hat S}^s}_{i,2}(t)=\{6\}$,  $\overline{\mathcal{\hat S}^s}_{i,3}(t)=\{9,10,11,12\}$, and $\overline{\mathcal{\hat S}^s}_{i,4}(t)=\{15\}$. We conclude that at least five attacked  sensors are in the set $\overline{\mathcal{\hat S}^s}_i(t)$. Because $\overline{\mathcal{\hat S}^s}_{i,1}(t)$ has at least one, $\overline{\mathcal{\hat S}^s}_{i,2}(t)$ has one, $\overline{\mathcal{\hat S}^s}_{i,3}(t)$ has at least two, and $\overline{\mathcal{\hat S}^s}_{i,4}(t)$ has  one.
	Then we consider the following detection condition:
		\begin{align}\label{eq_detec1_good}
		\sum_{j=1}^{l_i}\lceil  |\overline{\mathcal{\hat S}^s}_{i,j}(t)|/3 \rceil=b.
% \text{ and } |\overline{\mathcal{\hat S}^s}_i(t)|<N,
		\end{align}
	The condition \eqref{eq_detec1_good} is to infer whether the number of  sensors under attack and detected by vehicle $i$ reaches the known maximum  number of attacked sensors.

	Based on the observer in Algorithm \ref{alg:obser} and the detection conditions \eqref{item1}--\eqref{eq_detec1_good}, an online distributed attack detector is provided in Algorithm \ref{alg:detec}, which  is able to update the three sets: $\mathcal{\hat  S}^a_i(t)$,  $\mathcal{\hat S}^s_i(t)$, and $\mathcal{\hat S}_i(t)$, $i\in\mathcal{V}$. 	
%%	Based on the definition on $\overline{\mathcal{\hat S}^s}_{i,j}(t)$, we find the following properties of Algorithm \ref{alg:detec}.
%	\begin{lemma}\label{prop_relative}
%		Under Assumptions \ref{ass_noise}--\ref{ass_attack}, denote $\overline{\mathcal{\hat S}^s}_i(t)=\mathcal{\hat S}^s_i(t)\bigcup \mathcal{\hat  S}^a_i(t)$,  then the following properties hold.
%		\begin{enumerate}[label=\roman*)]
%%			\item The claim in Sub-detector \ref{detector1} holds;
%            \item If the detection condition \eqref{item1} is satisfied,  either  vehicle $i$ or vehicle $i-1$ is attacked.
%            \item If the detection condition \eqref{cond_detector3} is satisfied,  vehicle $i$ is attacked.
%			\item The minimal number of the  attacked vehicle sensors in the set $\overline{\mathcal{\hat S}^s}_i(t)$ is $\sum_{j=1}^{j_i}\lceil  |\overline{\mathcal{\hat S}^s}_{i,j}(t)|/3 \rceil$.
%			\item If 
%		
%			then the vehicles in the set $\mathcal{V}-\overline{\mathcal{\hat S}^s}_i(t)$ are all attack-free for sure,
%		\end{enumerate}		
%		where $\overline{\mathcal{\hat S}^s}_{i,j}(t),j=1,2,\dots,j_i$ are the sets with successive  elements such that  $\bigcup_{j=1}^{j_i}\overline{\mathcal{\hat S}^s}_{i,j}(t)=\overline{\mathcal{\hat S}^s}_i(t)$.
%		%		all the three claims hold.
%		%		%    \begin{itemize}
%		%		%    	\item   all the three claims hold;
%		%		%%    	\item 	 the inequalities in \ref{item2}) and \ref{item3})  can not be satisfied simultaneously.
%		%		%    \end{itemize}
%	\end{lemma}
%	\begin{pf}
%		See Appendix \ref{pf_prop_relative}.	
%	\end{pf}
	
		\subsection{Detector properties}
	%The next proposition shows that Assumption \ref{ass_detection} can be removed in the framework of this paper. the detection results for the vehicles in the sets $\mathcal{\hat S}_i(t)$ and $\mathcal{\hat  S}_i^a(t)$ are fault-free
	\begin{lemma}\label{prop_ass}
		The observer in  Algorithm  \ref{alg:obser} and the detector in Algorithm \ref{alg:detec} for the system \eqref{eq_local_state}--\eqref{eq_system3}  under Assumptions \ref{ass_attack}--\ref{ass_noise}  satisfy
		Assumption  \ref{ass_detection}.
	\end{lemma}
		\begin{pf}
			See Appendix \ref{pf_prop_ass}.	
		\end{pf}
	Lemma \ref{prop_ass} states that  the two sets $\mathcal{\hat S}_i(t)$ and $\mathcal{\hat  S}_i^a(t)$ are fault-free, which differs from the existing results of false alarms (e.g., \cite{baras2019trust}) since we study bounded noise. The following proposition studies the finite-time convergence of the detection sets $\mathcal{\hat  S}^a_i(t)$ and $\mathcal{\hat  S}_i(t)$.

	\begin{theorem}\label{prop_detect}
		Consider the observer in  Algorithm  \ref{alg:obser} and the detector in Algorithm \ref{alg:detec}  for the system \eqref{eq_local_state}--\eqref{eq_system3}  under Assumptions \ref{ass_attack}--\ref{ass_noise}. If there is a time $T_j$ and a vehicle $j\in\mathcal{V}$, such that the number of the attacked vehicle sensors estimated by vehicle $j$ equals to its upper bound    in Assumption~\ref{ass_attack}, i.e., $|\mathcal{\hat  S}^a_j(T_j)|=b$, then there exists a time $T_*$, such that for  $t\geq T_*$, the sets of attacked and attack-free vehicle sensors estimated by each vehicle $i\in\mathcal{V}$ equals  the true sets, i.e., 
		%all the attacked vehicle sensors and attack-free vehicle sensors are identified, i.e.,
		\begin{align*}
		\mathcal{\hat  S}^a_i(t)&=\mathcal{S}^a, \quad 
		\mathcal{\hat  S}_i(t)=\mathcal{S}.
		\end{align*}
	\end{theorem}
	\begin{pf}
		By  Algorithm \ref{alg:detec}, when there is a time $T_j$ and a vehicle $j\in\mathcal{V}$, such that $|\mathcal{\hat  S}^a_j(T_j)|=b$, then $\mathcal{\hat  S}^a_j(T_j)=\mathcal{S}^a$ and $\mathcal{\hat  S}_j(T_j)=\mathcal{S}.$ Since  both $|\mathcal{\hat  S}^a_i(t)|$ and $|\mathcal{\hat S}_i(t)|$ are non-decreasing and the vehicle network is finite, there is a time at which all vehicles   update their set estimates to the true sets.
	\end{pf}
	Theorem \ref{prop_detect}  holds under the condition that  the attacker compromises $b$ sensors with aggressive attack signals, which is possible when the attacker has no  knowledge of the detector. Otherwise, the attacker can inject  stealthy signals  making the attacked sensors undetectable.

%	 i) The number of the attacked vehicle sensors reaches the bound known by the system designer (i..e, $b$). Otherwise, there is the possibility that at least one unknown sensor is under attack; and ii) The attacker injects aggressive attack signals such that the   $b$ attacked  vehicle sensors are able to be identified by the detector.  
%	% Note that  the convergence conclusion in  Theorem \ref{prop_detect} will not affect the boundedness of the overall performance function $\varphi(t)$ (see Theorem \ref{thm_control}).

\section{Controller Design}\label{sec:control}
In this section,  we design an observer-based distributed controller algorithm, and then analyze the boundedness of the overall performance function of the architecture consisting of the observer in Algorithm \ref{alg:obser}, the detector in Algorithm \ref{alg:detec}, and the distributed controller. 
%The attack resilience related  notions introduced in Definitions \ref{defn_stability}--\ref{defn_stability_distur} are studied. 
%Then the unified framework, integrated from the resilient observer, the attack detector, and the distributed controller, is applied to vehicle platooning.
% by employing Algorithm \ref{alg:obser} and the estimates from adjacent vehicles.  
%observer estimates from neighboring vehicles to achieve the formation control defined in Subsection \ref{sec:form_problem}, and apply the proposed framework to  vehicle platooning.

%{\color{red} I think it is good to investigate the string stability of the platoon for the proposed distributed controller here. We will also put some string stability reference here, such as \cite{besselink2017string, herman2016scaling, somarakis2019risk}, and discuss a bit about their approaches.}

\subsection{Controller algorithm}
%\textbf{Estimate communication for controller design:}
%Next, we design the controller $u_i(t)$ based on the  estimates  of two neighboring vehicles 
%and the desired relative position distance  between two neighboring vehicles. 
Denote $\bar{\mathcal{N}}_i$ the set of vehicle(s) nearest  to vehicle $i$,  $i=0,1,\dots,N$, i.e., 
%\begin{small}
	\begin{align}\label{eq_measurement_notation}
	\bar{\mathcal{N}}_i=
	\begin{cases}
	\{1\}, &\text{if } i=0\\
	\{i-1,i+1\},&\text{if } i\in\{1,2,\dots,N-1\}\\
	%\{0,2\}, &\text{if } i=1\\
	\{N-1\}, &\text{if } i=N,
	\end{cases}
	\end{align}
%\end{small}
where vehicle $0$, which is virtual and introduced for convenience, stands for the reference state of the leader vehicle 1. 
Assume  $\hat s_{i}(t)$ and  $\bar s_{i}(t)$   are the estimate and predicted value of  $s_{i}(t)$, and  $\hat v_{i}(t)$ and $\hat v_{i}(t)$ are the estimate and predicted value of  $v_{i}(t)$. 
%Let $t_i^*$ be the start time for the control input. For $0\leq t< t_i^*$, we let $u_{i}(t)=0.$ 
Then, we propose a distributed observer-based   controller in Algorithm \ref{alg:control}, where $\Delta x_{i-1,i}^s(t)$ and $\Delta x_{i-1,i}^v(t)$ are the desired relative position and velocity between vehicles $i-1$ and $i$, and $g_s>0$,  $g_v>0$  are parameters to be determined.
% Algorithm \ref{alg:control}    is one of the controllers discussed in \eqref{eq_control}.
%	Regarding the acceleration $u_i(t)$ in  \eqref{eq_pred}, by employing the predicted estimates $\bar x_j(t)=[\bar s_{j}(t),\bar v_{j}(t)]^{\sf T}$ from the vehicle $j\in\mathcal{N}_i$, vehicle $i$ is equipped with the following acceleration input, for $t\geq t_i^*\geq 0$,  
%	\begin{align}\label{eq_control}
%	\begin{split}
%	u_{i}(t)=&\sum_{j\in\mathcal{N}_i}\big(g_s(\bar s_{j}(t)-\hat s_{i}(t)+\Delta_{i,j})\\
%	&+g_v(\bar v_{j}(t)-\hat v_{i}(t))\big), 2\leq i\leq N,
%	\end{split}
%	\end{align}
%	where $g_s$ and $g_v$ are positive scalars to be determined.

%Note that if the initial estimates are very accurate, we can choose $t_i^*=0$. Otherwise, setting a larger $t^*$ can lead to better estimates but need more time to achieve the platooning of vehicles. 

% $\hat u_{i,j}(t)=[0,\Delta t\tilde u_{i,j}(t)]^{\sf T}$.

\begin{algorithm}
			\caption{Distributed Controller}
			\label{alg:control}
%	\small
		\begin{algorithmic}[1]
%	\SetAlgoLined
%	\SetKwInOut{Input}{init.}
%	\SetKwInOut{Output}{output}
	\STATE{\textbf{Initialization}: Control parameter $g_s$ and $g_v$, desired relative position and velocity between vehicles $i-1$ and $i$, i.e., $\Delta x_{i-1,i}^s(t)$ and $\Delta x_{i-1,i}^v(t)$, $i=1,2,\dots,N$}
	\STATE{\textbf{Output}: Control input $u_{i}(t)$}
	%		\KwResult{Write here the result }
	%		initialization\;
	\FOR{$t\geq 0$}
	\STATE{
		%		\textbf{Communications between neighboring vehicles:}  Vehicle $i$ receives the predicted estimate $\bar x_j(t)=[\bar s_{j}(t),\bar v_{j}(t)]^{\sf T}$ from the vehicle $j\in\bar{\mathcal{N}}_i$\;
		\textbf{Communications between neighboring vehicles:}  Vehicle $i$ sends out $\mathcal{M}_i$ defined in \eqref{message}\;\\
		\textbf{Distributed controller}
		\begin{align*}
		\begin{split}
		u_{i}(t)=&\sum_{j\in\bar{\mathcal{N}}_i}\big(g_s(\bar s_{j}(t)-\hat s_{i}(t)+\Delta x_{j,i}^s(t))\\
		&+g_v(\bar v_{j}(t)-\hat v_{i}(t)+\Delta x_{j,i}^v(t))\big), \quad \\
		\text{where } &[\bar s_{0}(t),\bar v_{0}(t)]^{\sf T}=:x_0(t).								
		\end{split}
		\end{align*}
		%			instructions\;
		%			\eIf{condition}{
		%				instructions1\;
		%				instructions2\;
		%			}{
		%			instructions3\;
		%		}
	}
	\ENDFOR
				\end{algorithmic}
\end{algorithm}

\begin{remark}\label{rem_relative}
	The relative state measurements in \eqref{eq_system3} are not directly used in the controller  but the estimates, because: i) The relative measurements are noisy. ii) There is no sensor of the leader vehicle to measure the relative state to the reference state (i.e., $x_1(t)-x_0(t)$).
	%	For the leader vehicle,  its relative state to the reference state $x_0$ can not be directly measured. 
\end{remark}
%Let 
%\begin{align}\label{eq_roots}
%s^2+(\lambda_lTg_v-2)s+\lambda_l T^2g_s-\lambda_lTg_v+1=0,
%\end{align}
%where $s\in \mathbb{C}.$

\subsection{Closed-loop property}
The following lemma, proved in \cite{he2020secure}, is useful in the following analysis.
\begin{lemma}\label{lem_stability}
	Consider the  linear dynamical system 
	$	x(t+1)=Fx(t)+G(t),$
	where $F\in\mathbb{R}^{n\times n}$ is a Schur stable matrix. If $\limsup\limits_{t\rightarrow \infty}\norm{G(t)}\leq \varsigma$,   the equation $F^{\sf T}PF-P=-I_n$ has a solution $P\succ 0$ such that 
	 $\limsup\limits_{t\rightarrow \infty}\norm{x(t)}\leq \sqrt{\frac{2\theta\varsigma^2\lambda_{\max}(P)}{\lambda_{\min}(P)}},$  where  	$\theta=\norm{P}+2\norm{PF}^2$. 		
	%		there is a positive definite matrix $P,$ satisfying $F^{\sf T}PF-P=-I_n$, such that $		\norm{x(t)}^2\leq \frac{\lambda^{\sf T}\lambda_{\max}(P)\norm{x(0)}^2}{\lambda_{\min}(P)}+\frac{\beta}{\lambda_{\min}(P)}\sum_{l=0}^{t-1}\lambda^{t-1-l}\norm{G(l)}^2,$
	%		%		\begin{align*}%\label{eq_x}
	%		%		\norm{x(t)}^2\leq& \frac{\lambda^{\sf T}\lambda_{\max}(P)\norm{x(0)}^2}{\lambda_{\min}(P)}+\frac{\beta}{\lambda_{\min}(P)}\sum_{l=0}^{t-1}\lambda^{t-1-l}\norm{G(l)}^2.
	%		%		\end{align*}
	%		
	%		Furthermore, if $\sup_{t\geq 0}\norm{G(t)}\leq \alpha_1$, then $\limsup\limits_{t\rightarrow \infty}\norm{x(t)}^2\leq \frac{2\beta\alpha_1^2\lambda_{\max}(P)}{\lambda_{\min}(P)};$ 
\end{lemma}
Let $\mathcal{L}\in\mathbb{R}^{(N+1)\times (N+1)}$ be the graph Laplacian matrix \cite{xie2012consensus} corresponding to the neighbor sets in \eqref{eq_measurement_notation}.		
Denote  $\mathcal{L}_g\in\mathbb{R}^{N\times N}$  the grounded graph Laplacian matrix with respect to the nodes $\{1,2,3,\dots,N\}$, which is obtained by removing the first row and first column of   Laplacian matrix $\mathcal{L}$.
%Let $\lambda_l$ be the $l$-th eigenvalue of $\mathcal{L}_g$, $l=1,2,\dots,N$. 
\begin{assumption}\label{ass_g}
	The parameters $g_s$ and $g_v$ of the controller in Algorithm \ref{alg:control} are subject to  
	$g_v>Tg_s>0$ and $
	T^2g_s-2Tg_v>-\frac{4}{\lambda_{\max}(\mathcal{L}_g)}$.
%	\begin{align}\label{design_parameters}
%	g_v>Tg_s>0, \quad 
%	T^2g_s-2Tg_v>-\frac{4}{\lambda_{\max}(\mathcal{L}_g)}.
%	\end{align}
\end{assumption}
Assumption \ref{ass_g} can be satisfied for any positive $g_s$ and $g_v$ if the time step $T>0$ is sufficiently small.
%From \eqref{design_parameters}, one with the knowledge of the graph topology $\mathcal{G}$ can 
%design the parameters  $g_s$ and $g_v$ offline.
%\begin{remark}
%%	The condition  \eqref{design_parameters} is a necessary condition of the parameters $g_s$ and $g_v$ to ensure the formation control of the vehicle network. If the estimates of the vehicles are asymptotically convergent to the true states, it is also a sufficient condition. 
%\end{remark} 
In the following theorem, the  closed-loop performance function  $\varphi(t)$ in  \eqref{eq_function} is studied.
%control performance when the observer is with the static observer threshold~$\beta_i$,~$i\in\mathcal{V}_1$.		
\begin{theorem}\label{thm_control}
	Consider the observer in  Algorithm  \ref{alg:obser}, the detector in Algorithm \ref{alg:detec}, and the controller in Algorithm \ref{alg:control}   satisfying 
	Assumption \ref{ass_g} for the system \eqref{eq_local_state}--\eqref{eq_system3}.  Then the following properties hold:
	\begin{enumerate}[label=\roman*)]
		\item 		If  the observer threshold is   static and the   conditions   in Theorem \ref{thm_estimation2} are satisfied,    the performance function $\varphi(t)$ in  \eqref{eq_function} is asymptotically upper bounded, i.e.,
		\begin{align*}
		\limsup_{t\rightarrow \infty}\varphi(t)\leq \hat \alpha+\eta\xi;
		\end{align*} 
		\item If  the observer threshold is   adaptive and the   conditions    in Theorem \ref{thm_estimation22}  are satisfied,   $\varphi(t)$  is asymptotically upper bounded, i.e.,
		\begin{align*}
		\limsup_{t\rightarrow \infty}\varphi(t)\leq \bar{\hat \alpha}+\bar \eta\xi;
		\end{align*} 
	\end{enumerate}
	%					\begin{align}\label{design_parameters}
	%					\lambda_{\max}(\mathcal{L}_g)<\frac{4g_s}{g_v^2}, \quad	T<\frac{g_v}{g_s}
	%					\end{align}
	%					then the position tracking error, i.e., $\tilde s_i=s_i(t)- s_{i-1}(t)+\Delta_{i,i-1}$ and the speed tracking error, i.e., $\tilde v_i=v_i(t)-v_1(t)$,  are such that
	
	%					\begin{align}\label{eq_bound}
	%					\limsup\limits_{t\rightarrow \infty}\left(\sum_{i=1}^{N}\left(\tilde s_i^2(t)+\tilde v_i^2(t)\right)\right)\leq \frac{2\kappa\eta^2\lambda_{\max}(M)}{\lambda_{\min}(M)},
	%					\end{align}
	where   
	\begin{align}\label{thm2_notations}
	\xi&=\sqrt{\frac{2\kappa\lambda_{\max}(M)}{\lambda_{\min}(M)}},\quad
	M=\sum_{i=0}^{\infty}(P^i)^{\sf T}P^i, F =\begin{pmatrix}
	0&0\\
	T g_s&	T g_v
	\end{pmatrix}\nonumber\\ 
	P&=I_{N}\otimes A-\mathcal{L}_g\otimes F,\quad \kappa=\norm{M}+2\norm{MP}^2\nonumber\\
	\eta&=2\sqrt{N}T\hat \alpha\left(g_s(\norm{A}+1)+2 g_v\right)+\sqrt{N}\epsilon,\\
	\bar \eta&=2\sqrt{N}T\bar{\hat \alpha}\left(g_s(\norm{A}+1)+2 g_v\right)+\sqrt{N}\epsilon\nonumber\\
	\hat \alpha&=\max\{\tilde\alpha_1,\tilde\alpha_2,\tilde\alpha_3\},\quad \bar{\hat \alpha}=\max\{\bar\alpha_1,\bar\alpha_2,\bar\alpha_3\},\nonumber
	\end{align}			
	in which $\tilde\alpha_i $ and $\bar\alpha_i$, for $i=1,2,3$,   are  introduced in Theorems  \ref{thm_estimation2}  and \ref{thm_estimation22}, respectively.
	%			\begin{align*}
	%			&\limsup\limits_{t\rightarrow \infty}|v_i(t)-v_1(t))|\leq \mathcal{\hat  S}^a_1,\\
	%			&\limsup\limits_{t\rightarrow \infty}|s_i(t)- s_{i-1}(t)+\Delta_{i,i-1}|\leq \mathcal{\hat  S}^a_2.
	%			\end{align*}
\end{theorem}
%		\begin{theorem}\label{thm_control}
%			Under the same conditions as Theorem \ref{thm_estimation}, 
%			if 
%			\begin{align}\label{design_parameters}
%			\lambda_{\max}(\mathcal{L}_g)<\frac{4g_s}{g_v^2}, \quad	T<\frac{g_v}{g_s}
%			\end{align}
%			then the position tracking error, i.e., $\tilde s_i=s_i(t)- s_{i-1}(t)+\Delta_{i,i-1}$ and the speed tracking error, i.e., $\tilde v_i=v_i(t)-v_1(t)$,  are such that
%			\begin{align}\label{eq_bound}
%			\limsup\limits_{t\rightarrow \infty}\left(\sum_{i=2}^{N}\left(\tilde s_i^2(t)+\tilde v_i^2(t)\right)\right)\leq \frac{2\kappa\eta^2\lambda_{\max}(M)}{\lambda_{\min}(M)},
%			\end{align}
%			where   
%			\begin{align}\label{thm2_notations}
%			\begin{split}
%						M&=\sum_{i=0}^{\infty}(P^i)^{\sf T}P^i,\\
%			\kappa&=\norm{M}+2\norm{MP}^2,\\
%			\eta&=2\sqrt{N}T\hat \alpha\left(g_s(\norm{A}+1)+2 g_v\right)+\sqrt{N}\epsilon,\\ \hat \alpha&=\max\{\alpha_0,\alpha_i,i\in\mathcal{V}_2\},
%			\end{split}
%			\end{align}			
%			where $\alpha_0,\alpha_i$ are the estimation error bounds given in \eqref{thm_estimation}.
%%			\begin{align*}
%%			&\limsup\limits_{t\rightarrow \infty}|v_i(t)-v_1(t))|\leq \mathcal{\hat  S}^a_1,\\
%%			&\limsup\limits_{t\rightarrow \infty}|s_i(t)- s_{i-1}(t)+\Delta_{i,i-1}|\leq \mathcal{\hat  S}^a_2.
%%			\end{align*}
%		\end{theorem}
\begin{pf}
	See Appendix \ref{pf_thm_control}.			
\end{pf}
\begin{remark}
	It follows from Theorems \ref{thm_estimation2}--\ref{thm_estimation22} that under the same condition, 	the upper bounds in Theorem \ref{thm_control} fulfill $\bar{\hat \alpha}+\bar \eta\xi\leq \hat \alpha+\eta\xi$, because  the design of the adaptive observer threshold can employ the measurements more effectively and help to detect more attacked sensors. This illustrates the advantage of using an adaptive threshold instead of a static one in the observer.
\end{remark}

\begin{figure*}[t]
	\centering
	\subfigure[State estimation error of each vehicle]{
		\includegraphics[height=5.1cm, width=5.7cm]{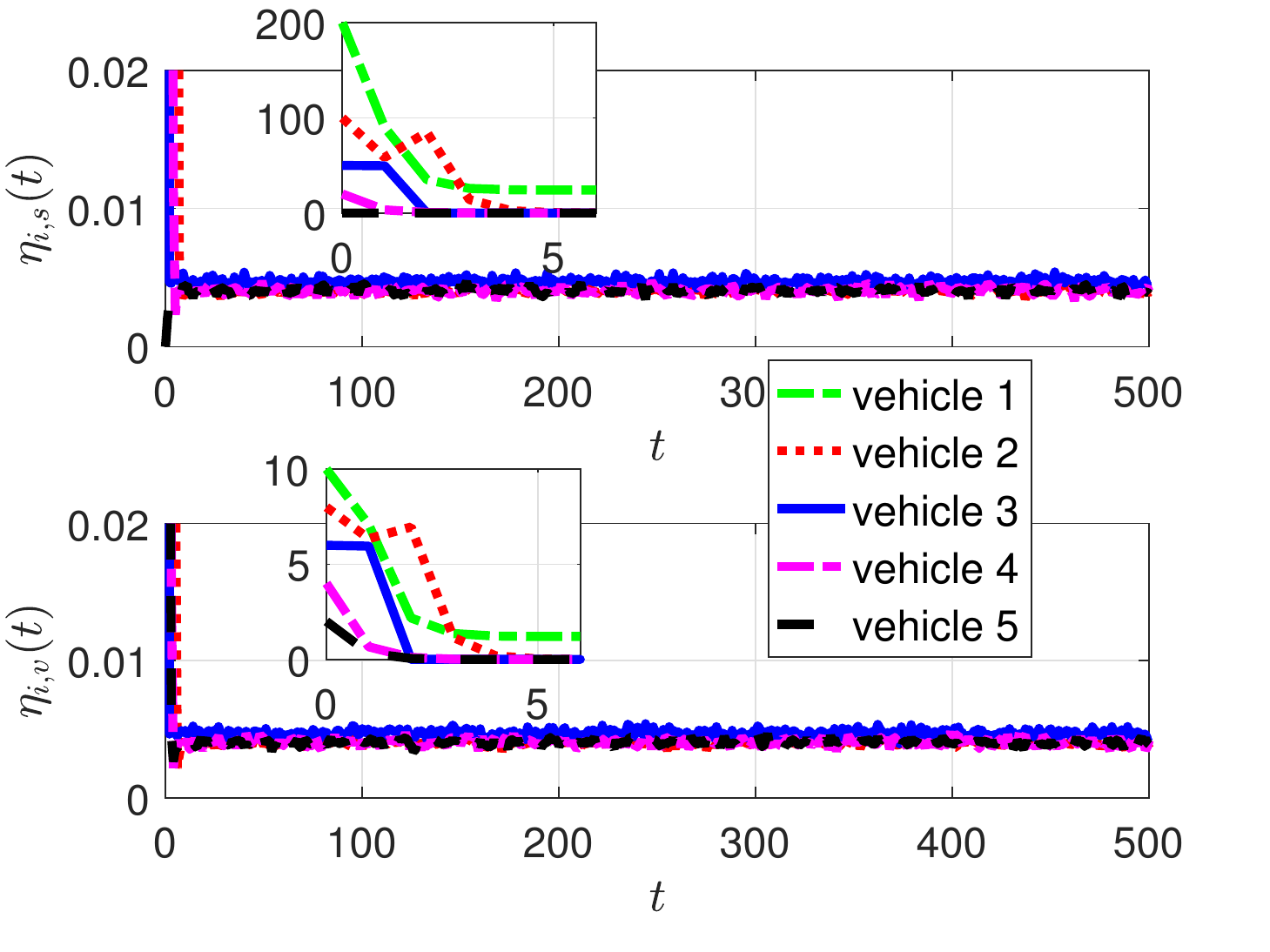}}
	\subfigure[\label{inertia2}Estimation error bounds of each vehicle]{
		\includegraphics[height=5.1cm, width=5.7cm]{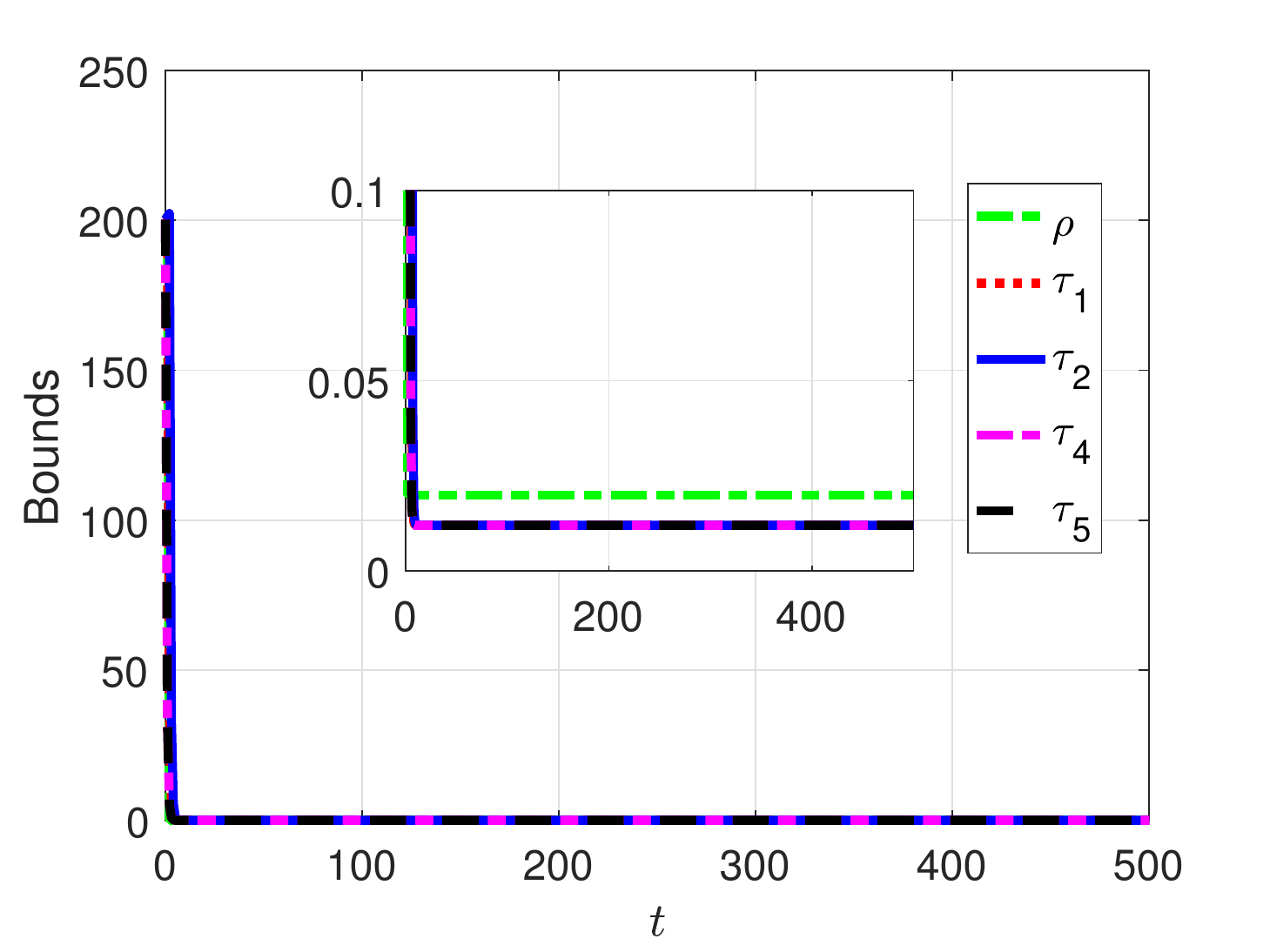}}
	\subfigure[\label{asyn2}Relative states between reference state and the  vehicles]{
		\includegraphics[height=5.1cm, width=5.7cm]{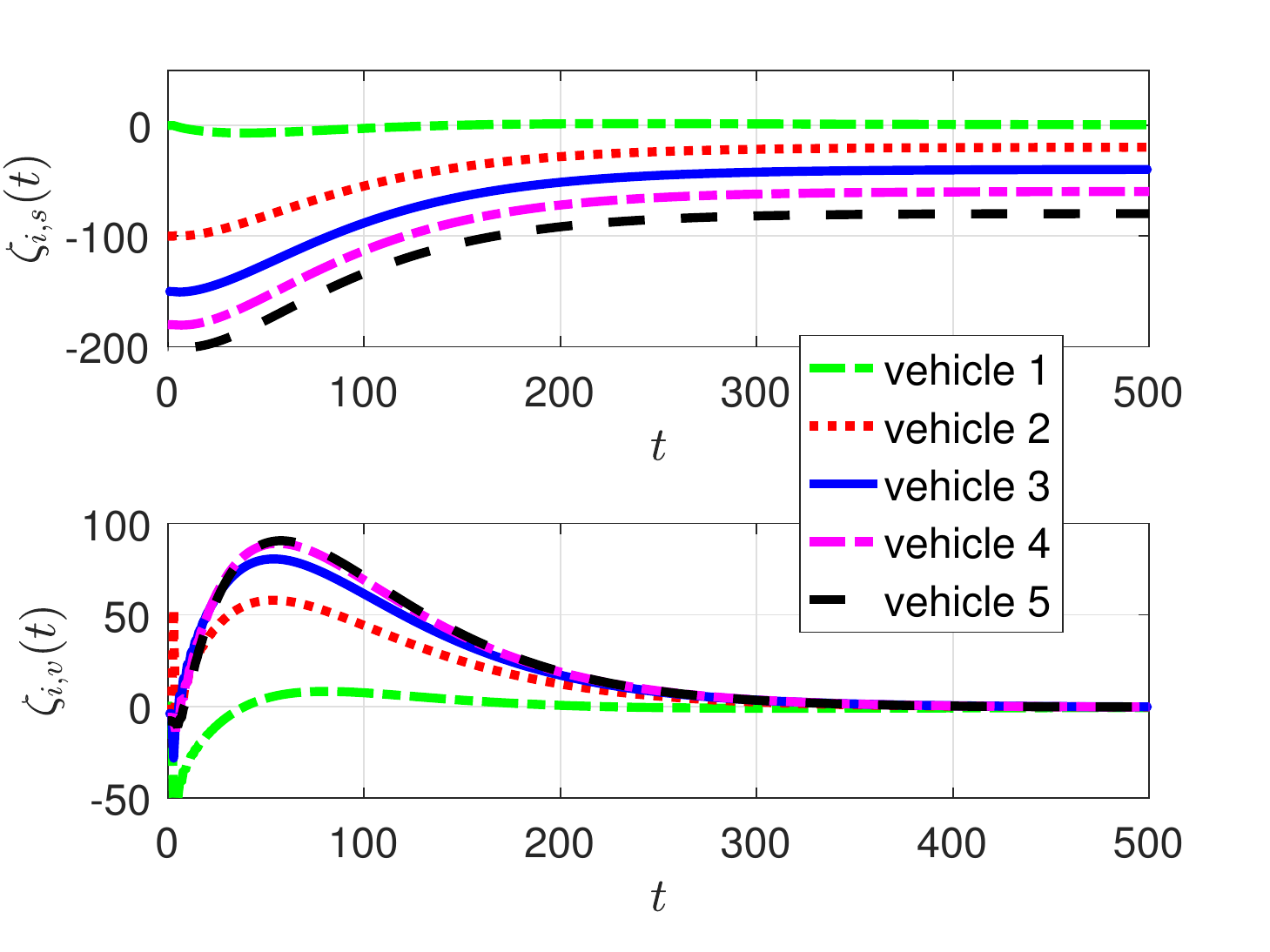}}
	\caption{Estimation and platooning error of Algorithms \ref{alg:obser}--\ref{alg:control}. In (a), the state estimation errors of each vehicle in position and velocity are provided. Corresponding to Proposition \ref{lem_detec22}, the dynamics of the online estimation error bounds $\rho(t)$ and $\lambda_i(t)$, $i\in\mathcal{V}_2\bigcap \mathcal{\hat S}(t)=\{1,2,4,5\}$, are provided in (b). In (c), the relative state (i.e., relative position and velocity) between the  vehicles $1,2,3,4,5$ and the reference state is shown. }\label{fig:test1}
\end{figure*}
\begin{figure*}[t]
	\centering
	\subfigure[The error function $\varphi(t)$ with different noise magnitudes ]{
		\includegraphics[height=5.1cm, width=5.7cm]{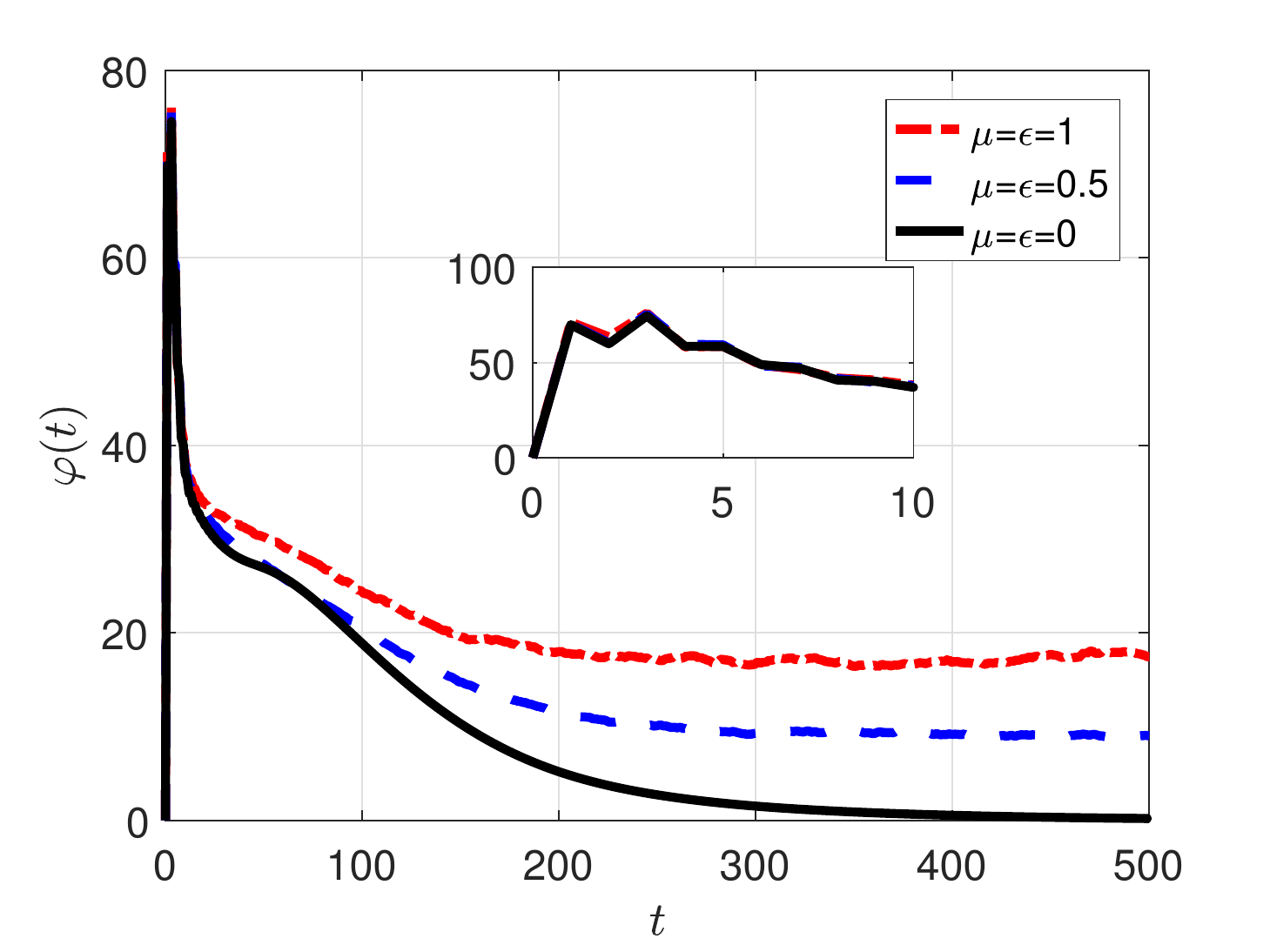}}
	%			\subfigure[The error function $\varphi(t)$  with different $|\mathcal{S}^a|$ ]{
	%				\includegraphics[height=5.1cm, width=5.8cm]{attack_number.eps}}
	\subfigure[ The error function $\varphi(t)$ under different attacks ]{
		\includegraphics[height=5.1cm, width=5.7cm]{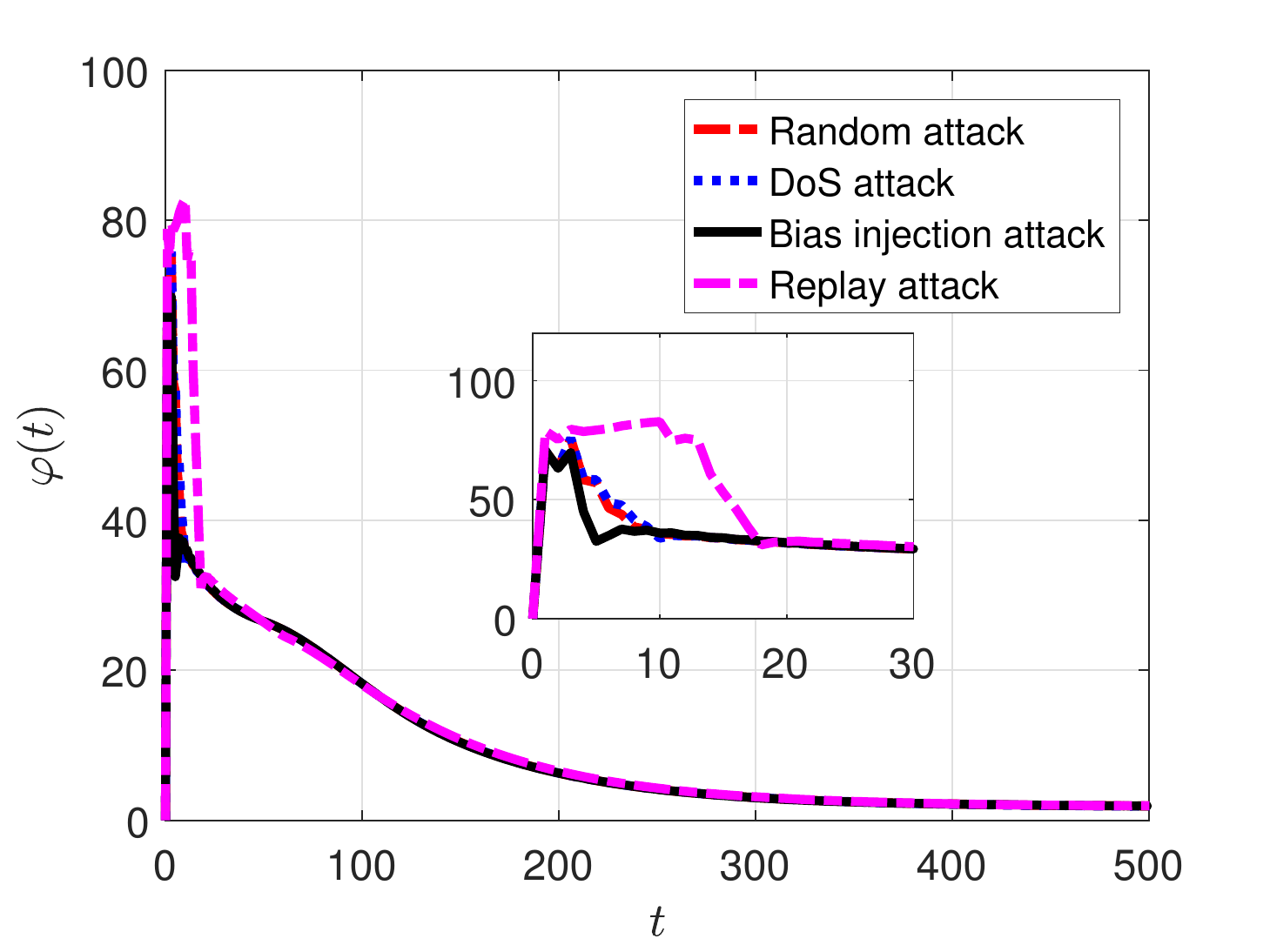}}
	\subfigure[ Comparison of five  algorithms in   platooning error. ]{
		\includegraphics[height=5.1cm, width=5.7cm]{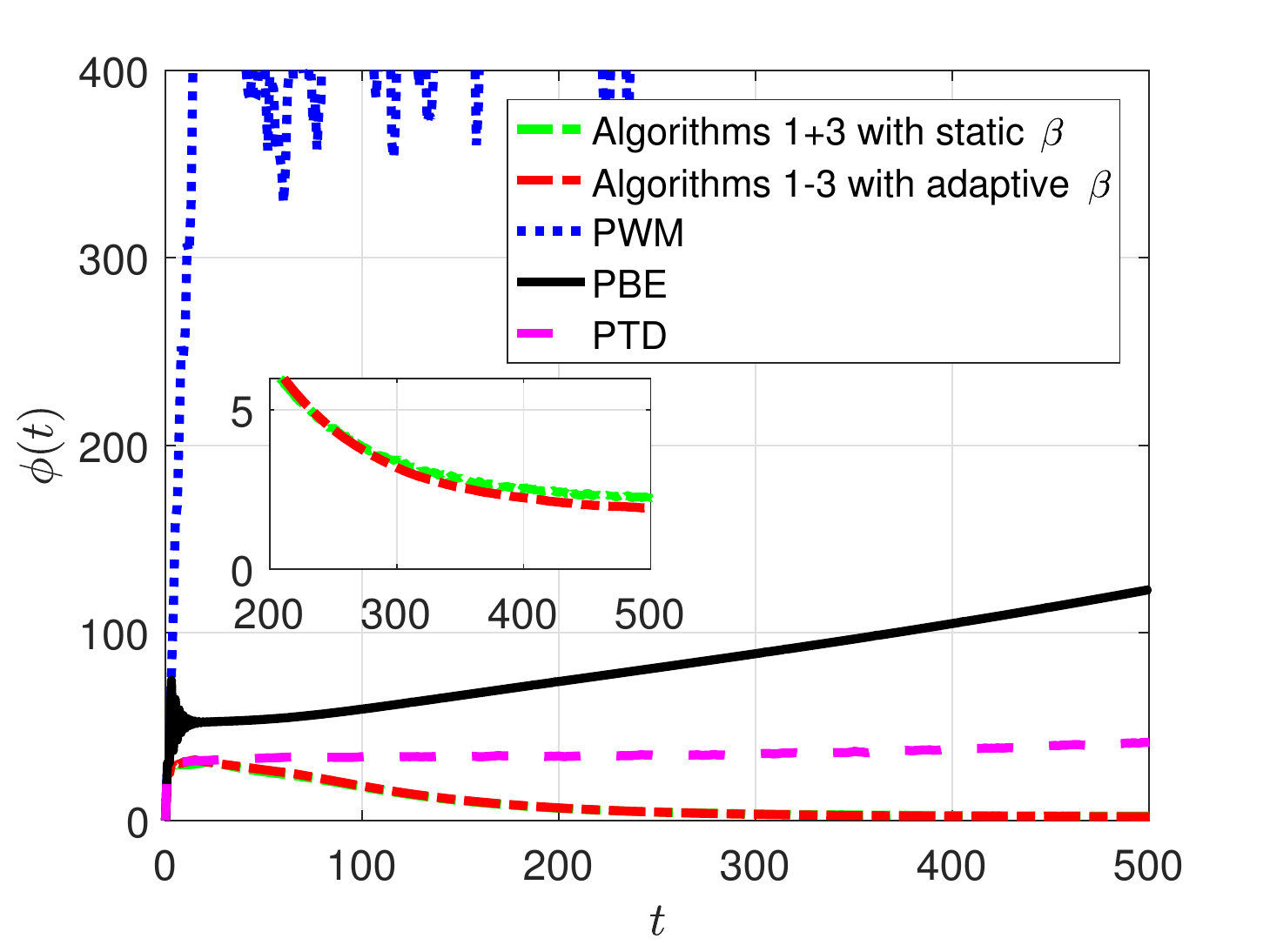}}
	\caption{The influence of some essential variables to the performance of Algorithms \ref{alg:obser}--\ref{alg:control}, and a comparison of five algorithms.
		%				, where PWM is obtained from Algorithm \ref{alg:control} by replacing the estimates by measurements, PBE is  obtained from Algorithm \ref{alg:control} by using the estimates from Byzantine strategy \cite{mitra2019byzantine}, and PTD is from \cite{lin2009consensus}.
	}\label{fig:test21}
\end{figure*}

Theorem  \ref{thm_control} and the following corollary  provide the solution to the formulated problem in Section \ref{sec:form_problem}.
\begin{corollary}\label{coro_overall1}
	Consider the observer in  Algorithm  \ref{alg:obser}, the detector in Algorithm \ref{alg:detec}, and the controller in Algorithm \ref{alg:control}   satisfying 
	Assumption \ref{ass_g} for the system \eqref{eq_local_state}--\eqref{eq_system3}.  
	Then   the performance function $\varphi(t)$  tends to zero, i.e.,
	\begin{align*}
	\limsup_{t\rightarrow \infty}\varphi(t)=0,
	\end{align*} 
	if   the system is known to be noise-free, i.e., $\mu=\epsilon=0,$  and one of the following two conditions is satisfied
	\begin{enumerate}[label=\roman*)]
		\item 		the observer threshold is   static,  the   conditions   in Theorem \ref{thm_estimation2} hold, and   there is a vehicle sensor $i$ at some $T_i<\infty$, such that $| \mathcal{\hat  S}^a_i(T_i)|=b$;
		\item  the observer threshold is   adaptive, and the   conditions    in Theorem \ref{thm_estimation22} hold.
	\end{enumerate}
\end{corollary}
\begin{pf}
	The proof follows from Theorems \ref{thm_estimation2}--\ref{thm_control}.
\end{pf}
%\begin{corollary}\label{coro_overall2}
%	Under the same conditions as in 1) of Theorem \ref{thm_control22}, 
%	if the system is known to be disturbance-free, i.e., $\mu=\epsilon=0,$   then $\limsup_{t\rightarrow \infty}\varphi(t)= 0.$
%\end{corollary}
%\begin{pf}
%	The conclusions hold according to Theorems \ref{thm_estimation22} and~\ref{thm_control22}.
%\end{pf}	
%Theorem \ref{thm_control2} shows that given an adaptive $\beta_i(t)$ in \eqref{eq_beta_varying}, the convergence of the estimation error and platooning error  is independent of the detection results.
\begin{remark}
Corollary \ref{coro_overall1} shows the improvement of performance achieved in the noise-free case in comparison to the noisy case Theorem \ref{thm_control}.
Note that the first conclusion of Corollary \ref{coro_overall1}  means that there is one vehicle that has detected the maximal number of attacked sensors. This makes it possible to conclude that there can be no other attacked sensors, so the mitigation mechanism of the observer can fully compensate for the attack. The second conclusion of Corollary \ref{coro_overall1} means that 
whatever the detection results, the observer with the adaptive threshold makes the  space of stealthy attacks diminish to an empty set asymptotically.
% Because in   $\lim\limits_{t\rightarrow}\max\{\beta_{i}(t),\tau_i(t)\}=0$,  $\forall i\in\mathcal{V}$, which, together with the detection condition \eqref{cond_detector3}, makes 
%the space of stealthy attacks diminish to an empty set asymptotically.
\end{remark}

\section{Simulations}\label{sec:simulation}
In this section, the effectiveness of the proposed methods is  evaluated through simulations by an application to vehicle platooning.

%				\begin{figure*}[t]
%					\centering
%					\subfigure[ The error function $\varphi(t)$ with the observer parameter $\beta=200$  ]{
%						\includegraphics[height=5.1cm, width=5.8cm]{compare_large_beta.eps}}
%					\subfigure[ The error function $\varphi(t)$ with the observer parameter $\beta=5$  ]{
%						\includegraphics[height=5.1cm, width=5.8cm]{compare_middle_beta.eps}}
%					\subfigure[ The error function $\varphi(t)$ with the observer parameter $\beta=0.1$]{
%						\includegraphics[height=5.1cm, width=5.8cm]{compare_small_beta.eps}}
%					\caption{The error function $\varphi(t)$ of Algorithms \ref{alg:obser}--\ref{alg:control} with different $\beta$, where the algorithms with adaptive observer parameter use this setting in the initialization.}\label{fig:test2}
%				\end{figure*}

%			\begin{figure}[t]
%				\centering
%				%					\subfigure[Performance with different $\mathcal{S}^a$]{
%				%						\includegraphics[height=5.1cm, width=5.8cm]{attack_number.eps}}
%				%						\subfigure[Comparison of algorithms]{
%				\includegraphics[width=1\linewidth]{compare.eps}
%				%							}
%				\caption{Comparison of five  algorithms in   platooning error. Here,  PWM is obtained from Algorithm \ref{alg:control} by replacing the estimates by measurements, PBE is  obtained from Algorithm \ref{alg:control} by using the estimates from Byzantine strategy \cite{mitra2019byzantine}, and PTD is from \cite{lin2009consensus}.} \label{fig:test22}
%			\end{figure}	
Suppose there are five vehicles, i.e., $N=5$, 
with time step  $T=0.01$ and time range $t=0,1,\dots,500.$ 
All elements of the process noise  $d_i(t)$ and measurement noise $n_{i,j}(t)$, $j\in\mathcal{N}_i\cup \{i\}$, $i=1,\dots,5$,   follow the uniform distribution between $(0,\mu_0/\sqrt{2})$, where $\mu_0=0.1$. The bounds in Assumption \ref{ass_noise} are assumed to be $\mu=\epsilon=\mu_0$ and $q=300.$ The initial state is $x_1(0)=(200, 10)^{\sf T},$ $x_2(0)=(100, 8)^{\sf T},$ $x_3(0)=(50, 6)^{\sf T},$ $x_4(0)=(20, 4)^{\sf T},$ $x_5(0)=(0, 2)^{\sf T}$, whose observer estimates are all $0^{2\times 1}$. The required position distance between vehicles $i$ and $i+1$ is $|\Delta_{i,i+1}|=20$, $i=1,2,\dots,N-1$. The control gains in Algorithm \ref{alg:control} are   $g_s=g_v=50$, and  the communication range $L=2$. Suppose   the reference position and the reference velocity of the leader vehicle are $s_{0}(t+1)=s_{0}(t)+v_0T$ and $v_0=10$, where $s_{0}(0)=200$. In the following, we assume all vehicles share the same observer threshold $\beta(\cdot).$
%	Assume vehicle 3 is under spoofing attack and the communication range $L=2$.

We conduct a Monte Carlo experiment with $100$ runs. 
Define the average estimation error in position and velocity by $\eta_{i,s}(t)$ and $\eta_{i,v}(t)$, respectively,  and
define the relative position and velocity between vehicle $i\in\{1,2,3,4,5\}$ and the leader vehicle $0$ by $\zeta_{i,s}(t)$ and $\zeta_{i,v}(t)$, respectively, i.e.,
\begin{align}
\eta_{i,s}(t)&=\frac{1}{100}\sum_{j=1}^{100}|e_{i,s}^j(t)|,\zeta_{i,s}(t)=\frac{1}{100}\sum_{j=1}^{100}(s_{i}^j(t)-s_{0}(t)),\nonumber\\
\eta_{i,v}(t)&=\frac{1}{100}\sum_{j=1}^{100}|e_{i,v}^j(t)|,	\zeta_{i,v}(t)=\frac{1}{100}\sum_{j=1}^{100}(v_{i}^j(t)-v_{0}),\nonumber
%\eta(t)&=\frac{1}{100}\sum_{j=1}^{100}\max_{i\in\{1,\dots,30\}}\norm{e_i^j(t)},
\end{align}
where $e_{i,s}^j(t)$ and $e_{i,v}^j(t)$ are the state estimation errors of vehicle $i$ in position and velocity, respectively, at time $t$ in the $j$-th run,  and $s_{i}^j(t)$ and $v_{i}^j(t)$ are the position and velocity of vehicle $i$, respectively, at time $t$ in the $j$-th run.

First, we study the performance of Algorithms \ref{alg:obser}--\ref{alg:control} with the adaptive observer parameter $\beta(t)$ designed in \eqref{eq_beta_varying}. For one vehicle $i$ under FDI sensor attacks, assume that the measurements would be compromised by the random attack signal  $a_i(t)=w_i(t)x_i(t)$, where $w_i(t)$ is drawn from the standard normal distribution.
For the case of the attacked vehicle sensor set $\mathcal{S}^a=\{3\}$, the state estimation error, estimation error bounds, and vehicle platooning error   are provided in Fig. \ref{fig:test1}.  Fig. \ref{fig:test1}--(a) shows that   the estimation errors in position and velocity are convergent to small neighborhoods of zero rapidly. Fig. \ref{fig:test1}--(b) shows that the offline bounds of the estimation errors are  convergent to  small neighborhoods of zero. It is shown in Fig. \ref{fig:test1}--(c) that  the speeds of all vehicles   converge to the reference velocity, and the relative positions between two neighbor vehicles tend to the desired one, i.e., 20. 
%Since Algorithm \ref{alg:platoon} also contains Algorithm \ref{alg:detec} apart from  Algorithms \ref{alg:obser}--\ref{alg:control}, it provides better estimation performance   as shown in   Fig. \ref{fig:test12}  and \ref{fig:test1}, and shares good platooning performance.  
 We study the  performance function $\varphi(t)$  of Algorithms \ref{alg:obser}--\ref{alg:control}  with $\mathcal{S}^a=\{2,3\}$
  under different noise magnitudes (i.e., $\epsilon$ and $\mu$) and under different types of attacks   in  (a)   and   (b) of Fig. \ref{fig:test21}, respectively.   Fig. \ref{fig:test21}--(a) shows that $\varphi(t)$ decreases as the noise magnitudes decrease. 
  In Fig. \ref{fig:test21}--(b), we study four typical attack types, including random attack, DoS attack, bias injection attack, and replay attack \cite{teixeira2015secure}. It  shows that Algorithms \ref{alg:obser}--\ref{alg:control} with adaptive observer parameter is able to deal with multiple kinds of attacks.
%The reason that why  few attacked vehicles can not have better performance is due to the

%Second, we study four typical combinations, namely, Algorithms 1+3 with static observer parameter $\beta$,  Algorithms 1+3 adaptive observer parameter $\beta(t)$, Algorithms 1-3 with static observer parameter $\beta_i$, and  Algorithms 1-3 adaptive observer parameter $\beta_i(t)$, where  Algorithms 1+3 indicates that it has no detection scheme.
%Through using the above algorithms, the results are provided in Fig. \ref{fig:test2}. The results show that Algorithms 1-3 outweigh Algorithms 1+3, which is due to the detection scheme in Algorithm 2. Moreover, the algorithms with adaptive observer parameter $\beta(t)$ outperform the ones with static observer parameter $\beta$.

Then, we compare the proposed methods, i.e., Algorithms \ref{alg:obser}+\ref{alg:control} (\ref{alg:obser} and \ref{alg:control}) with static observer parameter $\beta$, Algorithms \ref{alg:obser}--\ref{alg:control} with adaptive observer parameter $\beta(t)$,
%abbreviated by  RLP and RLPDF, respectively, 
with  PWM, which is obtained from Algorithm \ref{alg:control} by replacing the estimates by measurements, and with PBE, which is  obtained from Algorithm \ref{alg:control} by using the estimates following Byzantine strategy \cite{mitra2019byzantine}, as well as   PTD  \cite{lin2009consensus}. 
To evaluate the platooning error of each algorithm, we use the performance function $\phi(t)$:
$	\phi(t)=	\frac{1}{N}\sum_{i=1}^N\norm{x_{i}(t)-x_{i}^*(t)}$.
The algorithm comparison result is provided in  Fig. \ref{fig:test21}--(c), which shows that our  algorithms  outperform the other three algorithms, and Algorithms \ref{alg:obser}--\ref{alg:control} achieves best platooning performance among the five algorithms. In  Fig. \ref{fig:test21}--(c), PWM is divergent since the compromised measurements  directly affect  the platooning.

\section{Conclusion and Future Work}\label{sec_conclusion}
This paper studied  how to design a secure observer-based distributed controller 	such 	that  a group of vehicles   can achieve accurate state estimates and formation control under the case that a static subset of   vehicle sensors are  compromised by a malicious attacker.	
We proposed an architecture consisting of  a resilient observer, an online attack detector, and a distributed  controller. Some important properties of the observer, detector, and controller were analyzed. An application of the proposed architecture to vehicle platooning was investigated in  numerical simulations.

There are some directions of  future work. One is to extend the architecture to the  attack detection on actuators of vehicles in platoon. Another is to study more general  models of vehicles and  sensors. It is also promising to extend the   methods from the string vehicle topology to more complex vehicle topologies with higher dimensions and more leaders.

	\footnotesize
	\bibliography{All_references}

\begin{thebibliography}{10}

\bibitem{shoukry2018smt}
Y.~Shoukry, M.~Chong, M.~Wakaiki, P.~Nuzzo, A.~Sangiovanni-Vincentelli, S.~A.
  Seshia, J.~P. Hespanha, and P.~Tabuada, ``\protect{SMT}-based observer design
  for cyber-physical systems under sensor attacks,'' {\em ACM Transactions on
  Cyber-Physical Systems}, vol.~2, no.~1, pp.~1--27, 2018.

\bibitem{baras2019trust}
J.~S. Baras and X.~Liu, ``Trust is the cure to distributed consensus with
  adversaries,'' in {\em Mediterranean Conference on Control and Automation},
  pp.~195--202, 2019.

\bibitem{pasqualetti2013attack}
F.~Pasqualetti, F.~D{\"o}rfler, and F.~Bullo, ``Attack detection and
  identification in cyber-physical systems,'' {\em IEEE Transactions on
  Automatic Control}, vol.~58, no.~11, pp.~2715--2729, 2013.

\bibitem{RN18}
Z.~H. Tang, M.~Kuijper, M.~S. Chong, I.~Mareels, and C.~Leckie, ``Linear system
  security-detection and correction of adversarial sensor attacks in the
  noise-free case,'' {\em Automatica}, vol.~101, pp.~53--59, 2019.

\bibitem{gallo2020distributed}
A.~J. Gallo, M.~S. Turan, F.~Boem, T.~Parisini, and G.~Ferrari-Trecate, ``A
  distributed cyber-attack detection scheme with application to \protect{DC}
  microgrids,'' {\em IEEE Transactions on Automatic Control}, vol.~65, no.~9,
  pp.~3800--3815, 2020.

\bibitem{RN25}
X.~H. Ge, Q.~L. Han, M.~Y. Zhong, and X.~M. Zhang, ``Distributed
  \protect{Krein} space-based attack detection over sensor networks under
  deception attacks,'' {\em Automatica}, vol.~109, 2019.

\bibitem{deghat2019detection}
M.~Deghat, V.~Ugrinovskii, I.~Shames, and C.~Langbort, ``Detection and
  mitigation of biasing attacks on distributed estimation networks,'' {\em
  Automatica}, vol.~99, pp.~369--381, 2019.

\bibitem{forti2018distributed}
N.~Forti, G.~Battistelli, L.~Chisci, S.~Li, B.~Wang, and B.~Sinopoli,
  ``Distributed joint attack detection and secure state estimation,'' {\em IEEE
  Transactions on Signal and Information Processing over Networks}, vol.~4,
  no.~1, pp.~96--110, 2018.

\bibitem{chowdhury2020observer}
N.~R. Chowdhury, J.~Belikov, D.~Baimel, and Y.~Levron, ``Observer-based
  detection and identification of sensor attacks in networked \protect{CPSs},''
  {\em Automatica}, vol.~121, p.~109166, 2020.

\bibitem{kim2018detection}
J.~Kim, C.~Lee, H.~Shim, Y.~Eun, and J.~H. Seo, ``Detection of sensor attack
  and resilient state estimation for uniformly observable nonlinear systems
  having redundant sensors,'' {\em IEEE Transactions on Automatic Control},
  vol.~64, no.~3, pp.~1162--1169, 2018.

\bibitem{yang2020multi}
T.~Yang, C.~Murguia, M.~Kuijper, and D.~Ne{\v{s}}i{\'c}, ``A multi-observer
  based estimation framework for nonlinear systems under sensor attacks,'' {\em
  Automatica}, vol.~119, p.~109043, 2020.

\bibitem{RN89}
T.~Shinohara, T.~Namerikawa, and Z.~H. Qu, ``Resilient reinforcement in secure
  state estimation against sensor attacks with a priori information,'' {\em
  IEEE Transactions on Automatic Control}, vol.~64, no.~12, pp.~5024--5038,
  2019.

\bibitem{fawzi2014secure}
H.~Fawzi, P.~Tabuada, and S.~Diggavi, ``Secure estimation and control for
  cyber-physical systems under adversarial attacks,'' {\em IEEE Transactions on
  Automatic control}, vol.~59, no.~6, pp.~1454--1467, 2014.

\bibitem{pajic2017attack}
M.~Pajic, I.~Lee, and G.~J. Pappas, ``Attack-resilient state estimation for
  noisy dynamical systems,'' {\em IEEE Transactions on Control of Network
  Systems}, vol.~4, no.~1, pp.~82--92, 2017.

\bibitem{shoukry2017secure}
Y.~Shoukry, P.~Nuzzo, A.~Puggelli, A.~L. Sangiovanni-Vincentelli, S.~A. Seshia,
  and P.~Tabuada, ``Secure state estimation for cyber-physical systems under
  sensor attacks: A satisfiability modulo theory approach,'' {\em IEEE
  Transactions on Automatic Control}, vol.~62, no.~10, pp.~4917--4932, 2017.

\bibitem{RN92}
A.~Y. Lu and G.~H. Yang, ``Secure switched observers for cyber-physical systems
  under sparse sensor attacks: A set cover approach,'' {\em IEEE Transactions
  on Automatic Control}, vol.~64, no.~9, pp.~3949--3955, 2019.

\bibitem{RN13}
Y.~B. Gao, G.~H. Sun, J.~X. Liu, Y.~Shi, and L.~G. Wu, ``State estimation and
  self-triggered control of \protect{CPSs} against joint sensor and actuator
  attacks,'' {\em Automatica}, vol.~113, 2020.

\bibitem{su2019finite}
L.~Su and S.~Shahrampour, ``Finite-time guarantees for
  \protect{Byzantine}-resilient distributed state estimation with noisy
  measurements,'' {\em IEEE Transactions on Automatic Control}, vol.~65, no.~9,
  pp.~3758--3771, 2020.

\bibitem{ren2020secure}
X.~Ren, Y.~Mo, J.~Chen, and K.~H. Johansson, ``Secure state estimation with
  \protect{Byzantine} sensors: A probabilistic approach,'' {\em IEEE
  Transactions on Automatic Control}, vol.~65, no.~9, pp.~3742--3757, 2020.

\bibitem{mitra2019byzantine}
A.~Mitra and S.~Sundaram, ``\protect{Byzantine}-resilient distributed observers
  for \protect{LTI} systems,'' {\em Automatica}, vol.~108, p.~108487, 2019.

\bibitem{mitra2019resilient}
A.~Mitra, J.~A. Richards, S.~Bagchi, and S.~Sundaram, ``Resilient distributed
  state estimation with mobile agents: overcoming \protect{Byzantine}
  adversaries, communication losses, and intermittent measurements,'' {\em
  Autonomous Robots}, vol.~43, no.~3, pp.~743--768, 2019.

\bibitem{lee2020fully}
J.~G. Lee, J.~Kim, and H.~Shim, ``Fully distributed resilient state estimation
  based on distributed median solver,'' {\em IEEE Transactions on Automatic
  Control}, vol.~65, no.~9, pp.~3935--3942, 2020.

\bibitem{chen2018resilient}
Y.~Chen, S.~Kar, and J.~M. Moura, ``Resilient distributed estimation: Sensor
  attacks,'' {\em IEEE Transactions on Automatic Control}, vol.~64, no.~9,
  pp.~3772--3779, 2019.

\bibitem{he2020secure_journal}
X.~He, X.~Ren, H.~Sandberg, and K.~H. Johansson, ``How to secure distributed
  filters under sensor attacks?,'' {\em arXiv preprint arXiv:2004.05409}, 2020.

\bibitem{zhu2013distributed}
M.~Zhu and S.~Mart{\'\i}nez, ``On distributed constrained formation control in
  operator--vehicle adversarial networks,'' {\em Automatica}, vol.~49, no.~12,
  pp.~3571--3582, 2013.

\bibitem{RN65}
Y.~Z. Zhu and W.~X. Zheng, ``Observer-based control for cyber-physical systems
  with periodic \protect{DoS} attacks via a cyclic switching strategy,'' {\em
  IEEE Transactions on Automatic Control}, vol.~65, no.~8, pp.~3714--3721,
  2020.

\bibitem{RN1}
D.~Zhao, Z.~D. Wang, G.~L. Wei, and Q.~L. Han, ``A dynamic event-triggered
  approach to observer-based \protect{ PID} security control subject to
  deception attacks,'' {\em Automatica}, vol.~120, 2020.

\bibitem{feng2017distributed}
Z.~Feng, G.~Wen, and G.~Hu, ``Distributed secure coordinated control for
  multiagent systems under strategic attacks,'' {\em IEEE Transactions on
  Cybernetics}, vol.~47, no.~5, pp.~1273--1284, 2017.

\bibitem{weerakkody2016graph}
S.~Weerakkody, X.~Liu, S.~H. Son, and B.~Sinopoli, ``A graph-theoretic
  characterization of perfect attackability for secure design of distributed
  control systems,'' {\em IEEE Transactions on Control of Network Systems},
  vol.~4, no.~1, pp.~60--70, 2016.

\bibitem{an2019distributed}
L.~An and G.-H. Yang, ``Distributed secure state estimation for cyber--physical
  systems under sensor attacks,'' {\em Automatica}, vol.~107, pp.~526--538,
  2019.

\bibitem{teixeira2015secure}
A.~Teixeira, I.~Shames, H.~Sandberg, and K.~H. Johansson, ``A secure control
  framework for resource-limited adversaries,'' {\em Automatica}, vol.~51,
  pp.~135--148, 2015.

\bibitem{shoukry2016event}
Y.~Shoukry and P.~Tabuada, ``Event-triggered state observers for sparse sensor
  noise/attacks,'' {\em IEEE Transactions on Automatic Control}, vol.~61,
  no.~8, pp.~2079--2091, 2016.

\bibitem{he2020secure}
X.~He, E.~Hashemi, and K.~H. Johansson, ``Secure platooning of autonomous
  vehicles under attacked \protect{GPS} data,'' {\em arXiv preprint
  arXiv:2003.12975}, 2020.

\bibitem{xie2012consensus}
D.~Xie and S.~Wang, ``Consensus of second-order discrete-time multi-agent
  systems with fixed topology,'' {\em Journal of Mathematical Analysis and
  Applications}, vol.~387, no.~1, pp.~8--16, 2012.

\bibitem{lin2009consensus}
P.~Lin and Y.~Jia, ``Consensus of second-order discrete-time multi-agent
  systems with nonuniform time-delays and dynamically changing topologies,''
  {\em Automatica}, vol.~45, no.~9, pp.~2154--2158, 2009.

\bibitem{hao2010effect}
H.~Hao, P.~Barooah, and J.~Veerman, ``Effect of network structure on the
  stability margin of large vehicle formation with distributed control,'' in
  {\em IEEE Conference on Decision and Control}, pp.~4783--4788, 2010.

\end{thebibliography}
	\bibliographystyle{ieeetr}	
	
%\appendix
%\section{A summary of Latin grammar}    % Each appendix must have a short title.
%\section{Some Latin vocabulary}         % Sections and subsections are supported
%                                        % in the appendices.

\normalsize
\appendix
{\centering \section*{Appendix}}
\section{Proof of Proposition \ref{lem_detec22}}\label{pf_lem_detec22}
Denote the estimation error by $e_i(t)=\hat x_i(t)-x_i(t)$, the prediction error by $\bar e_i(t)=\bar x_i(t)-x_i(t)$, $i\in\mathcal{V}$. For notational convenience, we let $\lambda_i(t)=\tau_i(t)$, $t\leq T_i$, where $T_i$ is the time after which vehicle $i$ is attack-free by detection, i.e.,  $i\in\mathcal{\hat S}_i(t)$, $t\geq T_i+1.$ 
We use an inductive method for proof. At the initial time, due to $\rho_i(0)=\lambda_i(0)=\tau_i(0)=q$, according to Assumption \ref{ass_noise}, the conclusion holds. 	
Assume at time $t-1\geq 0$, the conclusion holds.								
In the following, we consider the case at time $t\geq 1$. 

First, we consider each vehicle sensor $i\in\mathcal{V}_1,$ which has at least $2L+1-b$ attack-free vehicle sensors as   neighbors. 
Suppose $\mathcal{J}$ is the set of these $2L+1-b$ sensors, i.e., $\mathcal{J}\subseteq \mathcal{S}$ with $|\mathcal{J}|=2L+1-b$,  which is unknown to  vehicles but useful for the following analysis.		
Let $\mathcal{J}^a=\mathcal{N}_i\cup \{i\}-\mathcal{J}$. It holds that  $|\mathcal{J}^a|=b$ and   the sensors in the set $\mathcal{\hat  S}^a_i(t)\subseteq \mathcal{J}^a$ are surely attacked under Assumption \ref{ass_detection}. 
%		It means that  the measurements of at most $b-|\mathcal{\hat  S}^a_i(t-1)|$ vehicles will be used at time $t$.  
%		Since the conclusion holds at time $t-1$, we have 
%		$		\norm{y_{j,j}(t)-A\hat x_{j}(t-1)}\leq \epsilon+\bar\mu+\norm{A}\bar\rho_j(t-1)$,  $j\in \mathcal{J}.$		
%		Under Assumption \ref{ass_attack}, there are at most $b$  vehicles under attack. 
%		Since each vehicle can obtain the measurements from  $2L+1$ vehicles, i.e., the measurements of itself and its $2L$ neighbor vehicles, and $L\geq b$, then there are at least $2L+1-b$ attack-free vehicles whose measurements can be obtained by vehicle $i$. 
%		Suppose $\mathcal{J}$ is the set of these $2L+1-b$ vehicles, i.e., $\mathcal{J}\subset \mathcal{S}$ with $|\mathcal{J}|=2L+1-b$, which is unknown to all vehicles but useful for the following analysis.	
Denote 
%\begin{small}
%\begin{align*}
%\bar K_{i,\mathcal{J}}(t)&=\diag\bigg\{	k_{i,m_s}(t)\mathbb I_{m_s\in \mathcal{J}}I_2
%\bigg\}_{s=1}^{2L+1}\in\mathbb{R}^{(4L+2)\times (4L+2)},\nonumber
%\end{align*}
$	\bar K_{i,\mathcal{J}}(t)=\diag\bigg\{	k_{i,m_s}(t)\mathbb I_{m_s\in \mathcal{J}}I_2
\bigg\}_{s=1}^{2L+1}\in\mathbb{R}^{(4L+2)\times (4L+2)}$
%\end{small}	
where $k_{i,m_s}(t)$ is introduced in \eqref{eq_K2}.
%	\begin{align*}
%	\bar K_{i,\mathcal{J}}(t)&=\diag\bigg\{k_{i,j_1}(t)\mathbb I_{j_1\in \mathcal{J}}, 	k_{i,j_2}(t)\mathbb I_{j_2\in \mathcal{J}},\quad \dots,\\
%	&\qquad\quad k_{i,j_{2L+1}}(t)\mathbb I_{j_{2L+1}\in \mathcal{J}}
%	\bigg\}\otimes I_2\in\mathbb{R}^{(4L+2)\times (4L+2)}.\nonumber
%	\end{align*}	
Let $\bar K_{i}^{[j]}(t)$ be the $j\text{-th}$ diagonal element of $\bar K_{i,\mathcal{J}}(t)$, $j=1,\dots,4L+2$,  $\boldsymbol{n}_i^{[j]}(t)$ be the $j$-th element of $\boldsymbol{n}_i(t)$ in \eqref{eq_system5}, and
%$\hat  K_{i}(t)=\diag\left\{\sum\limits_{j=1,3,\dots,4L+1}\bar K_{i}^{[j]}(t),\sum\limits_{j=2,4,\dots,4L+2}\bar K_{i}^{[j]}(t)\right \}$ and $W_i(t)=\sum\limits_{j=1,3,\dots,4L+1}\left(\begin{smallmatrix}
%\bar K_{i}^{[j]}(t)\boldsymbol{n}_i^{[j]}(t)\\
%\bar K_{i}^{[j+1]}(t)\boldsymbol{n}_i^{[j+1]}(t)\end{smallmatrix}\right),$
\begin{align*}
\hat  K_{i}(t)&=\diag\left\{\sum\limits_{j=1,3,\dots,4L+1}\bar K_{i}^{[j]}(t),\sum\limits_{j=2,4,\dots,4L+2}\bar K_{i}^{[j]}(t)\right \}\\
W_i(t)&=\sum\limits_{j=1,3,\dots,4L+1}\begin{pmatrix}
\bar K_{i}^{[j]}(t)\boldsymbol{n}_i^{[j]}(t)\\
\bar K_{i}^{[j+1]}(t)\boldsymbol{n}_i^{[j+1]}(t)\end{pmatrix},
\end{align*}
through which we have $\hat  K_{i}(t)\in\mathbb{R}^{2\times 2}$ and $W_i(t)\in\mathbb{R}^{2}.$
By Algorithm \ref{alg:obser},  we have
\begin{align*}
e_i(t)
%		=&Ae_i(t-1)+\frac{1}{2L} C^{\sf T} K_{i}(t)(z_i(t)-C\bar x_i(t))\\
=&(I_2-\frac{1}{2L}\hat  K_{i}(t))Ae_i(t-1)+\frac{1}{2L}\hat  K_{i}(t)d_i(t-1)\\
&+\frac{1}{2L}W_i(t)+\frac{1}{2L} C^{\sf T}\bar K_{i,\mathcal{J}^a}(t)(z_i(t)-C\bar x_i(t)),
\end{align*}
where $	\bar K_{i,\mathcal{J}^a}(t)=	 K_{i}(t)-\bar K_{i,\mathcal{J}}(t).$
%		As we see, $\bar K_{i,\mathcal{J}}(t)$ is diagonal and its diagonal  elements are zeros if the corresponding vehicles are in the set $\mathcal{J}$. 
%		Then we define the complementary of $\bar K_{i,\mathcal{J}}(t)$ with respect to $K_{i}(t)$  in \eqref{eq_notation} as follows: 
%		\begin{align}\label{eq_comp}
%	
%		\end{align}
According to  \eqref{eq_K2}, the measurement update of sensor $i$ at time $t$ will be affected by 
at most $b-|\mathcal{\hat  S}^a_i(t)|$ attacked vehicle sensors, which remain stealthy till  time $t$. The measurements of these vehicles will be used at time $t$. 
Under the noise bound in equation \eqref{new_bound_noise} and the saturation operation in equation \ref{eq_K2}, taking the norm  of  $e_i(t)$ yields
\begin{align*} 
\norm{e_i(t)}\leq& \norm{(I_2-\frac{1}{2L}\hat  K_{i}(t))A}\norm{e_i(t-1)}\nonumber\\
&+|\mathcal{J}|\frac{\epsilon+\bar\mu}{2L}+(b-|\mathcal{\hat  S}^a_i(t)|)\frac{\beta_i(t)}{2L}\nonumber\\
%	\leq &\norm{(I_2-\frac{1}{2L}\hat  K_{i}(t))A}\norm{e_i(t-1)}+\bar Q_i(t)\nonumber\\
\leq&\rho_i(t),
\end{align*}
where 
%	 $\bar Q_i(t)=\frac{(\epsilon+\bar\mu)(2L+1-b)+(b-| \mathcal{\hat  S}^a_i(t)|)\beta_i(t)}{2L}$. 
the last inequality is obtained because: 1)  In the set $\mathcal{J}$, there  are    $|\mathcal{\hat S}_{i,1}(t)|$ attack-free vehicles    whose measurements have been fully utilized in the update at time $t$ (i.e., without saturation), where $\mathcal{\hat S}_{i,1}(t)$ is defined in \eqref{eq_normal};	2) There are $2L+1-b-|\mathcal{\hat S}_{i,1}(t)|$ attack-free  vehicles, whose measurement innovations is saturated with the corresponding gain satisfying $\hat  K_{i}^{[j]}(t)\geq  \bar k_i(t)=\min\{1,\frac{\beta_i(t)}{\norm{A}\rho_i(t-1)+\epsilon+\bar\mu}\}.$

Second, for   vehicle $i\in\mathcal{V}_2\bigcap \mathcal{\hat S}(t)$, according to \eqref{update_2} and Assumption \ref{ass_noise},  it is straightforward to prove that the estimation error is upper bounded by $\lambda_i(t)$.	
Third, for   vehicle $i\in\mathcal{V}_2-\mathcal{\hat S}(t)$,
By   Algorithm \ref{alg:obser}, we have  
\begin{align*}
e_i(t)=\frac{(\varpi-1)A}{\varpi}e_i(t-1)-\frac{(\varpi-1)d_i(t-1)}{\varpi}+\frac{\hat n_{i|j_i(t)}(t)}{\varpi},
\end{align*}
%	where $\hat n_{i|j_i}(t)$ is given in \eqref{pf_d}, and $\varpi>\norm{A}$. 
Regarding $\hat n_{i|j_i(t)}(t)$  in \eqref{pf_d}, 
according to  Assumption \ref{ass_noise}, the definition $j_i(t)=\arg\min_{j\in\mathcal{\hat{N}}_i\cup \mathcal{\hat S}_i(t)} |j-i|$, and $\norm{	\bar e_{j_i(t)}(t)}\leq \norm{A}s_i(t-1)+\epsilon$, we have $\norm{\hat n_{i|j_i(t)}(t)}\leq \mu|j_i(t)-i|+\norm{A}s_i(t-1)+\epsilon$, where $s_i(t-1)=	\rho_{j_i}(t-1),$ if $j_i(t)\in \mathcal{V}_1$, otherwise $	s_i(t-1)=\lambda_{j_i}(t-1)$.
%	\begin{align*}
%%\norm{n_i(t-1)}&\leq \frac{\epsilon}{2}\\
%\norm{\hat n_{i|j_i}(t)}&\leq \mu|j_i-i|+\norm{A}\rho(t-1)+\epsilon.
%	\end{align*}
Taking norm of both sides of  $	e_i(t)$,
we have $\norm{e_i(t)}\leq \tau_i(t)$.

\section{Proof of Theorem \ref{thm_estimation2}}\label{pf_thm_estimation2}
At time $T_i\geq 0$, the estimate of the attacked vehicle sensor set is $\mathcal{\hat  S}^a_i(T_i)$ and the estimate of the attack-free  vehicle set is $\mathcal{\hat S}(T_i)$. By Assumption \ref{ass_detection}, both $|\mathcal{\hat  S}^a_i(t)|$ and $|\mathcal{\hat S}_i(t)|$ are non-decreasing, thus $|\mathcal{\hat  S}^a_i(t)|\geq |\mathcal{\hat  S}^a_i(T_i)|$ and $|\mathcal{\hat S}_i(t)|\geq |\mathcal{\hat S}_i(T_i)|$, for any $t\geq T_i$.
%	 Since the estimation error is upper bounded by $\rho_i(t)$, $\lambda_i(t)$ $\tau_i(t)$. In the following, we prove the boundedness of these sequences.
Instead of proving the upper boundedness of the estimation error, in the following we prove the upper boundedness of  $\rho_i(t)$, $\lambda_i(t)$, and $\tau_i(t)$, which are upper bounds of the estimation error according to Proposition \ref{lem_detec22}.

First, we consider the case for $ i\in\mathcal{V}_1$.
By choosing  $\forall \beta_i\in (\bar\beta_1(\omega),\bar\beta_2(\omega))$,  where $\bar\beta_1(\omega)$ and $\bar\beta_2(\omega)$ are in  \eqref{eq_beta}, we directly have 
\begin{align}
\beta_i&<\beta_0\label{pf_thm_1}\\
\beta_i&<\frac{2L}{b}\left(\omega q-\frac{\left(\epsilon+\bar\mu\right)\left(2L+1-b\right)}{2L}\right)\label{pf_thm_2}\\
\beta_i&>\frac{2L}{2L+1-b}\frac{\left(\omega+\norm{A}-1\right)\beta_0}{\norm{A}}.\label{pf_thm_3}
\end{align}
It follows from   \eqref{pf_thm_1} that $k_i^*:=\frac{\beta_i}{\norm{A}q+\epsilon+\bar\mu}<1$. 
Then according to \eqref{pf_thm_3},  it is derived that $
(1-L_0k_i^*)\norm{A}q<(1-\omega )q,$ where $L_0=\frac{2L+1-b}{2L}$. Since the inequality in \eqref{pf_thm_2} is equivalent to 
$\frac{(\epsilon+\bar\mu)(2L+1-b)+b\beta_i}{2L}<\omega q$, we have 
\begin{align}\label{eq_condition}
(1-L_0k_i^*)\norm{A}q+\frac{(\epsilon+\bar\mu)(2L+1-b)+b\beta_i}{2L}<q.
\end{align}	
From \eqref{eq_condition} and Proposition \ref{lem_detec22}, by using an inductive method, we are able to obtain that $\rho_i(t)< q$, for $t\geq 1$, which, together with  \eqref{sequ_detect}, ensures that 
%	For vehicle $i\in\mathcal{V}_1$, according to  the proof of Theorem \ref{thm_estimation} and Proposition \ref{lem_detec22}, we have $\bar\rho_i(t)\leq \rho(t)\leq q$.
%	for $t\geq T_i$,  we have
\begin{align}\label{sequ_detect2}
\rho_i(t+1)\leq \tilde  m_i\norm{A}	\rho_i(t)+\tilde Q_i, \quad t\geq T_i
\end{align}
where $\tilde  m_i$ and $\tilde Q_i$ are given in \eqref{k_seq_detect2}.	
According to \eqref{eq_condition}, we have $(1-L_0k_i^*)\norm{A}<1$, which, together with  $0< \tilde  m_i\leq 1-L_0k_i^*$, leads to
$\tilde  m_i\norm{A}\in (0,1)$. Thus, it follows from \eqref{sequ_detect2} that $\limsup\limits_{t\rightarrow \infty}\rho_i(t)\leq  \tilde\alpha_1,$ where $\tilde\alpha_1$ is in \eqref{bounds}.

Second, for vehicle $ i\in\mathcal{V}_2\bigcap\mathcal{\hat S}_i(T_i)$, according to   \eqref{lambada} and $\frac{(\varpi-1)\norm{A}}{\varpi}\in (0,1)$, we have 
$\limsup\limits_{t\rightarrow \infty}\lambda_i(t)\leq \tilde\alpha_2,$
where $\tilde\alpha_2$ is in \eqref{bounds}.

Third, for vehicle $ i\in\mathcal{V}_2-\mathcal{\hat S}_i(T_i)$, since $\mathcal{\hat S}_i(t)$ is non-decreasing,  
we have $|j_i(t)-i|\leq |j_i^*-i|$, where $j_i^*$ is in \eqref{k_seq_detect2}, and 
$	j_i(t)=\arg\min_{j\in\mathcal{\hat{N}}_i\cup \mathcal{\hat S}_i(t)} |j-i|,\quad t\geq T_i.$
From \eqref{tau_detect} and $\limsup\limits_{t\rightarrow \infty}s_i(t)\leq \max \{\tilde\alpha_1,\tilde\alpha_2\}$, we obtain $\limsup\limits_{t\rightarrow \infty}\tau_i(t)\leq \tilde\alpha_3, $ where $\tilde\alpha_3$ is in \eqref{bounds}.

\section{Proof of Proposition \ref{prop_feasibility}}\label{pf_prop_feasibility}
Necessity: We assume $b>L$ for the proof by contradiction.
Then $2L+1-b\leq b$, which leads to $\frac{2L}{2L+1-b}\geq \frac{2L}{b}$. 
It is known from $\bar\beta_1(\omega)>0$ that $2L+1>b.$ Given   $\omega\in (0,1)$,
due to $\norm{A}>1$,  we have 
$\frac{\left(\omega+\norm{A}-1\right)\beta_0}{\norm{A}}>  \left(\omega q-\frac{\left(\epsilon+\bar\mu\right)\left(2L+1-b\right)}{2L}\right),$
where $\beta_0=\norm{A}q+\epsilon+\bar\mu$. Thus, $\bar\beta_1(\omega)>\bar\beta_2(\omega)$. The assumption $b>L$ does not hold.

Sufficiency: We will prove that if the inequalities in \eqref{condi_prop} are satisfied, the scalar $\omega_0$ is such that  $\bar \beta_1(\omega_0)<\bar \beta_2(\omega_0)$.

According to \eqref{eq_beta} and the first inequality in \eqref{condi_prop},  
$	\bar \beta_1(\omega_0)<\frac{2L}{b}\left(\omega_0 q-\frac{\left(\epsilon+\bar\mu\right)\left(2L+1-b\right)}{2L}\right).$
If the second inequality in \eqref{condi_prop} holds, then 
$	\frac{2L}{2L+1-b}\frac{\left(\omega_0+\norm{A}-1\right)}{\norm{A}}< 1.$
Multiplying both sides of the inequality by $\beta_0$ in \eqref{eq_beta} leads to $\bar \beta_1(\omega_0)<\beta_0. $  Therefore, $\bar \beta_1(\omega_0)<\bar \beta_2(\omega_0).$ Due to $\norm{A}>1$ and $L\geq b$, $\bar \beta_1(\omega_0)>0.$

\section{Proof of Theorem \ref{thm_estimation22}}\label{pf_thm_estimation22}
According to Proposition \ref{lem_detec22}, we prove the boundedness of the three sequences $\rho_i(t)$, $\lambda_i(t)$ $\tau_i(t)$ for the case that $\beta_i(t)$ is  designed as in \eqref{eq_beta_varying}.
Denote $ \bar L=2L+1-b.$

First, we consider the case for vehicle $ i\in\mathcal{V}_1$.  Since $\beta_{i,0}$ satisfies the same condition as $\beta_i$ in Theorem \ref{thm_estimation2}, according to the proof of Theorem \ref{thm_estimation2}, we have $k_{i,0}:=\frac{\beta_{i,0}}{\norm{A}q+\epsilon+\bar\mu}<1$ and 
\begin{align}\label{eq_condition2}
(1-\frac{\bar L}{2L}k_{i,0})\norm{A}q+\frac{(\epsilon+\bar\mu)\bar L+b\beta_{i,0}}{2L}<q.
\end{align}
which corresponds to \eqref{eq_condition}. From \eqref{eq_condition2} and $\beta_{i,0}=k_{i,0}(\norm{A}q+\epsilon+\bar\mu),$   we are able to obtain
\begin{align}\label{pf_condition}
\left(1-\frac{\bar L-b}{2L}k_{i,0}\right)\norm{A}<1.
\end{align}
Submitting $\beta_{i}(t)$ in \eqref{eq_beta_varying} into \eqref{sequ_detect} yields
\begin{align}\label{sequ_detect22}
\rho_i(t)=a_{i,1}(t)\norm{A}	\rho_i(t-1)+a_{i,2}(t),
\end{align}
where 
\begin{align*}
a_{i,1}(t)=&1-\frac{|\mathcal{\hat S}_{i,1}(t)|+(\bar L-b+| \mathcal{\hat  S}^a_i(t)|-|\mathcal{\hat S}_{i,1}(t)|)k_{i,0}}{2L},\nonumber\\
% \bar k_i(t)=&\min\{1,\frac{\beta_i(t)}{\norm{A}\rho_i(t-1)+\epsilon+\bar\mu}\},\\
a_{i,2}(t)=&\frac{\bar L+(b-| \mathcal{\hat  S}^a_i(t)|)k_{i,0}}{2L}(\epsilon+\bar\mu),\nonumber
\end{align*}
By Assumption \ref{ass_detection}, both $|\mathcal{\hat  S}^a_i(t)|$ and $|\mathcal{\hat S}_i(t)|$ are non-decreasing, thus $|\mathcal{\hat  S}^a_i(t)|\geq |\mathcal{\hat  S}^a_i(T_i)|$ and $|\mathcal{\hat S}_i(t)|\geq |\mathcal{\hat S}_i(T_i)|$, for any $t\geq T_i$.  Due to $k_{i,0}<1$, we have $\sup_{t\geq T_i}a_{i,1}(t)\leq a_{i,1}(T_i)\leq  1-\frac{\bar L-b}{2L}k_{i,0}$ and $\sup_{t\geq T_i}a_{i,2}(t)\leq  a_{i,2}(T_i),$
%\begin{align*}
%\sup_{t\geq T_i}a_{i,1}(t)&\leq a_{i,1}(T_i)\leq  1-\frac{\bar L-b}{2L}k_{i,0}\\
%\sup_{t\geq T_i}a_{i,2}(t)&\leq  a_{i,2}(T_i),
%\end{align*}
which, together with \eqref{pf_condition}--\eqref{sequ_detect22}, leads to 
$\limsup\limits_{t\rightarrow \infty} \rho_i(t)\leq \frac{a_{i,2}(T_i)}{1-a_{i,1}(T_i)\norm{A}}$.

The proofs for  vehicle $ i\in\mathcal{V}_2\bigcap\mathcal{\hat S}_i(T_i)$ and for vehicle $ i\in\mathcal{V}_2-\mathcal{\hat S}_i(T_i)$ are similar to the proofs in Theorem \ref{thm_estimation2}.

\section{Proof of Lemma \ref{prop_ass}}\label{pf_prop_ass}
We use an inductive method to prove the conclusion. At the initial time, Assumption \ref{ass_detection} holds trivially. Assume at time $t-1$, Assumption \ref{ass_detection} is satisfied. Then, we consider the case at time $t$.		
First, we aim to prove the following conclusions corresponding to lines 7, 20, and 24 of Algorithm \ref{alg:detec} under the  preconditions in lines 5 and 18:
\begin{enumerate}[label=\roman*)]
	\item If the detection condition \eqref{item1} is satisfied,  either  sensor $i$ or sensor $i-1$ is attacked.
	\item If the detection condition \eqref{cond_detector3} is satisfied,  sensor $i$ is attacked.
	\item If the detection condition \eqref{eq_detec1_good} is satisfied, the sensors in the set $\mathcal{V}\setminus (\mathcal{\hat S}^s_i(t)\cup\mathcal{\hat  S}^a_i(t))$ are attack-free.
	%		  The minimal number of the  attacked vehicle sensors in the set $\overline{\mathcal{\hat S}^s}_i(t)$ is $\sum_{j=1}^{j_i}\lceil  |\overline{\mathcal{\hat S}^s}_{i,j}(t)|/3 \rceil$.
\end{enumerate}

Proof of i): 
%			To prove the conclusion, we  make a conjecture that	for each attack-free vehicle $i\geq 2$ and $i-1$, i.e., $i\in\{2,\dots,N\}\cap\mathcal{S}$,  
%			$			\norm{f_{i,i-1}(t)}\leq 3\mu.$
By equation \eqref{eq_system2}, for two attack-free   sensors $i-1$ and $i$,  due to $a_i=a_{i-1}=0$, it holds that $				y_{i,i}(t)-y_{i-1,i-1}(t)
=x_i(t)-x_{i-1}(t)+n_{i,i}(t)-n_{i-1,i-1}(t),$
%		\begin{equation*}
%			\begin{split}
%				&y_{i,i}(t)-y_{i-1,i-1}(t)\\
%				=&x_i(t)-x_{i-1}(t)+n_{i,i}(t)-n_{i-1,i-1}(t),
%				%y_{i-1,i-1}(t)&=x_{i-1}(t)+n_{i-1,i-1}(t).
%			\end{split}
%		\end{equation*}
which, together with   \eqref{eq_system3}, leads to $	y_{i-1,i}(t)+y_{i-1,i-1}(t)-y_{i,i}(t)=n_{i-1,i}(t)+n_{i-1,i-1}(t)-n_{i,i}(t).$
%		\begin{equation}\label{pf_lem3}
%			\begin{split}
%				f_{i,i-1}(t)&=n_{i-1,i}(t)+n_{i-1,i-1}(t)+n_{i,i}(t).
%			\end{split}
%		\end{equation}
Under Assumption \ref{ass_noise}, taking the norm of its both sides  yields the conclusion.	 The conclusion  ii) is satisfied according to Proposition \ref{lem_detec22} by noting that $ i\notin \mathcal{\hat  S}_i(t)$.
Proof of iii):  Since $\bigcup_{j=1}^{j_i}\overline{\mathcal{\hat S}^s}_{i,j}(t)=\overline{\mathcal{\hat S}^s}_i(t)$ and each set $\overline{\mathcal{\hat S}^s}_{i,j}(t)$ contains successive sensor labels, the minimal number of the  attacked   sensors is no smaller than the sum of the minimal attacked   sensor number in each $\overline{\mathcal{\hat S}^s}_{i,j}(t)$. One attacked sensor can lead to at most  three suspicious sensors comprising of itself and its two neighbor sensors, hence, each $\overline{\mathcal{\hat S}^s}_{i,j}(t)$ contains $\lceil  |\overline{\mathcal{\hat S}^s}_{i,j}(t)|/3 \rceil$ attacked sensors at least.
%, which leads to    3)  by summing the subset number $j_i$.  Proof of 4):
%Owing to $|\overline{\mathcal{\hat S}^s}_i(t)|<N$, 
%Then  
%$\mathcal{V}-\overline{\mathcal{\hat S}^s}_i(t)$ is non-empty. Then the conclusion in 3) follows from the fact that given $\sum_{j=1}^{j_i}\lceil  |\overline{\mathcal{\hat S}^s}_{i,j}(t)|/3 \rceil=b$, 
Given the detection condition \eqref{eq_detec1_good},
the conclusion of iii) is obtained by noting that the set  $\overline{\mathcal{\hat S}^s}_i(t)=\mathcal{\hat S}^s_i(t)\bigcup \mathcal{\hat  S}^a_i(t)$ contains all attacked   sensors.  
%See Appendix \ref{pf_prop_relative}.
%	Under Assumption \ref{ass_attack}, there is one vehicle under attack, thus the second conclusion holds.		

Based on i)--iii), Algorithms \ref{alg:obser}--\ref{alg:detec}
%		Since the detection conditions \eqref{item1}--\eqref{cond_detector3} of the detector in  Algorithm \ref{alg:detec} are designed in a deterministic way, which, together with the estimation error bound sequences in Section \ref{sec:observer},  
ensures that the  sets $\mathcal{\hat  S}^a_i(t),$ $\mathcal{\hat S}^s_i(t)$, and $\mathcal{\hat S}_i(t)$ are all fault-free. The updates of the three sets  in Algorithm \ref{alg:detec} ensures that $\mathcal{\hat S}_i(t)$ and $\mathcal{\hat  S}^a_i(t)$ are monotonically non-decreasing. Therefore, Assumption \ref{ass_detection} is satisfied at time $t.$

\section{Proof of Theorem \ref{thm_control}}\label{pf_thm_control}
%\begin{equation}\label{state_indi}
%\begin{split}
%s_i(t+1)&=s_i(t)+\Delta tv_i(t)+n_{s,i}(t)\\
%v_i(t+1)&=v_i(t)+Tu_i(t)+n_{v,i}(t),i=1,\dots,N,\\
%\end{split}
%\end{equation}
%Proof of 1): 
Recall from \eqref{eq_function} that $x_i^*(t)=[s_i^*(t),v_i^*(t)]^{\sf T}$ is the desired state of vehicle $i$, $0\leq i\leq N$, which is such that $s_i^*(t)=s_{j}^*(t)+\Delta x_{j,i}^s(t)$ and $v_i^*(t)=v_{j}^*(t)+\Delta x_{j,i}^v(t)$, $j\in\mathcal{\bar N}_i$, then we denote $\tilde e_i(t)=x_i(t)-x_i^*(t)=[\tilde s_i(t),\tilde v_i(t)]^{\sf T} $  the tracking error of vehicle $i$.			Since the virtual reference vehicle 0 is in its desired state, then   $\tilde s_{0}(t)=\tilde v_{0}(t)=0.$ For $1\leq i\leq N$,   it holds that
\begin{align}\label{system_control}
\begin{split}
\tilde e_i(t+1)&=A\tilde e_i(t)+[0,T	\tilde  u_i(t)]^{\sf T}+\delta_i(t)\\
\delta_i(t)&=[0,T\hat u_i(t)]^{\sf T}+d_{i}(t)
\end{split}
\end{align}
where
\begin{align}\label{eq_u_error}
\begin{split}
\tilde u_{i}(t)=&\sum_{j\in\mathcal{\bar N}_i}\big(g_s(	\tilde s_{j}(t)-	\tilde s_{i}(t))\\
&+g_v( 	\tilde v_{j}(t)-	\tilde v_{i}(t))\big), 0\leq i, j\leq N,\\
\hat u_i(t)=&\sum_{j\in\mathcal{\bar N}_i}\big(g_s(	(\bar s_{j}(t)-s_{j}(t))-	(\hat s_{i}(t)- s_{i}(t)))\\
&+g_v( 	(\bar v_{j}(t)-v_{j}(t))-	(\hat v_{i}(t)- v_{i}(t)))\big).
\end{split}
\end{align}
From \eqref{system_control} and \eqref{eq_u_error}, we have
\begin{align}\label{eq_Error}
\tilde E(t+1)=P\tilde E(t)+\delta(t).
\end{align}
where $P$ is in \eqref{thm2_notations},  $\tilde E(t)=[\tilde e_1(t)^{\sf T},\dots,\tilde e_N(t)^{\sf T}]^{\sf T}$, and $\delta(t)=[\delta_1(t)^{\sf T},\dots,\delta_N(t)^{\sf T}]^{\sf T}$. 
%			where $P=I_{N-1}\otimes A-\mathcal{L}_g\otimes F$, $F=\begin{pmatrix}
%			0&0\\
%			T g_s&	T g_v
%			\end{pmatrix}$ and $\mathcal{L}_g$ is the ground Laplacian matrix with respect to the nodes $\{2,3,\dots,N\}$ obtained by removing the first row and first column of the Laplacian matrix $\mathcal{L}$. 			
By Theorem \ref{thm_estimation2}, $\sup_{t\geq 0}\norm{\delta(t)}<\infty$. Based on the BIBO stability principle, the  asymptotic stability of $\tilde E(t)$ in \eqref{eq_Error} is determined by the eigenvalues of  $P$. According to \cite{hao2010effect}, the spectrum of $P$ is 
$			\sigma(P)=\bigcup_{\lambda_{l}\in \sigma(\mathcal{L}_g)}\{A-\lambda_{l}F\}=\bigcup_{\lambda_{l}\in \sigma(\mathcal{L}_g)}Q_{l},$
where $\sigma(\cdot)$ is the set of distinct  eigenvalues, 
and 
$Q_l=\left(\begin{smallmatrix}
1&T\\
-\lambda_l Tg_s& 1-\lambda_lTg_v
\end{smallmatrix}\right)$, $l=1,2,\dots,N$. From \cite{hao2010effect}, all eigenvalues of $\mathcal{L}_g$ are real-valued and positive, i.e., $\lambda_l>0$. 	
Denote the eigenvalues of $Q_l$ by $s$, which  are the roots of  $\phi(s)=0$, where
%			\begin{align}\label{charastic}
%			s^2+(\lambda_lTg_v-2)s+\lambda_l T^2g_s-\lambda_lTg_v+1=0.
%			\end{align}		
%	Define
%	\begin{align}\label{S_set}
%	\mathbb{S}_{l}:=\left\{(g_s, g_v)\in\mathbb{R}^2|	\text{\eqref{eq_roots} holds, and } |s|<1, s\in \mathbb{C}\right\},
%	\end{align}
%	where 
$	\phi(s)=s^2+(\lambda_lTg_v-2)s+\lambda_l T^2g_s-\lambda_lTg_v+1.$
To prove the Schur stability of $P$, in the following, we aim to prove for each $\lambda_l$, $l=1,2,\dots,N$, $s$ 
falls into the open unit disk, i.e., $|s|<1$.		
By applying bilinear transformation	to $\phi(s)$, we can transfer the Schur stability of $\phi(s)$ into the Hurwitz stability of a continuous-time
system. Then we are able to prove that $s$ falls into the open unit disk, i.e., $|s|<1$, if and only if  $g_v>Tg_s>0$ and $T^2g_s-2Tg_v>-\frac{4}{\lambda_l}$. 
We refer  to \cite{xie2012consensus} for a similar proof.   Thus, when $(g_s,g_v)$ are chosen as in Assumption \ref{ass_g},  $P$ is Schur stable.			

From Theorem \ref{thm_estimation2}, \eqref{system_control}, and \eqref{eq_u_error}, we have 
$	\limsup\limits_{t\rightarrow \infty}\norm{\delta(t)}\leq \eta,$
where $\eta$ is given in \eqref{thm2_notations}.
Since $P$ is Schur stable, we use Lemma \ref{lem_stability} with respect to  \eqref{eq_Error}. Due to $\norm{\tilde e_i(t)}\leq \norm{\tilde E(t)}$, from the definition of the overall function $\varphi(t)$ in \eqref{eq_function} and Theorem \ref{thm_estimation2},  the conclusion in 1) is obtained. 
The proof of 2) is the same as  the proof of 1) but using Theorem \ref{thm_estimation22} in the evaluation of the estimation error instead of using Theorem~\ref{thm_estimation2}.
%			\begin{align}
%			\limsup\limits_{t\rightarrow \infty}\norm{\tilde X(t)}\leq \frac{\kappa\eta^2}{\lambda_{\min}(M)(1-\lambda)}
%			\end{align}
%			where $\lambda=1-\frac{1}{2\lambda_{\max}(M)}\in(0,1)$,  	$ \kappa=\norm{M}+2\norm{MP}^2,$  and $M=\sum_{i=0}^{\infty}(P^i)^{\sf T}P^i$.
%			
%	  $	\limsup\limits_{t\rightarrow \infty}\sqrt{\sum_{i=2}^{N}\left(\tilde s_i^2(t)+\tilde v_i^2(t)\right)}=\limsup\limits_{t\rightarrow \infty}\norm{\tilde X(t)}. $

\end{document}